\documentclass[twocolumn]{aastex63}
\usepackage{amsmath}
\usepackage{multirow}
\usepackage{mathtools}
\usepackage[thinc]{esdiff}

\usepackage[T1]{fontenc}
\usepackage{ae,aecompl}

\usepackage{graphicx}	
\usepackage{amssymb}	
\newcommand{\bz}{$\langle B_z \rangle$}

\received{----}
\revised{----}
\accepted{----}
\submitjournal{ApJ}

\shorttitle{CU\,Vir at wideband}
\shortauthors{Das \& Chandra}
\begin{document}

\title{Ultra-Wideband, Multi-epoch Radio Study of the First Discovered `Main sequence Radio Pulse emitter' CU\,Vir}

\correspondingauthor{Barnali Das}
\email{barnali@ncra.tifr.res.in}

\author[0000-0001-8704-1822]{Barnali Das}
\affil{National Centre for Radio Astrophysics, Tata Institute of Fundamental Research,  Pune University Campus, Pune-411007, India}

\author[0000-0002-0844-6563]{Poonam Chandra}
\affil{National Centre for Radio Astrophysics, Tata Institute of Fundamental Research,  Pune University Campus, Pune-411007, India}




\begin{abstract}
Presence of large-scale surface magnetic field in early-type stars leads to several unique electromagnetic phenomena producing radiation over X-ray to radio bands. 
Among them, the rarest type of emission is electron cyclotron maser emission (ECME) observed as periodic, circularly polarized radio pulses.
The phenomenon was first discovered in the hot magnetic star CU\,Vir. Past observations of this star led to the consensus that the star produces only right circularly polarized ECME, suggesting that only one magnetic hemisphere takes part in the phenomenon.
Here we present the first ultra-wideband (0.4--4 GHz) study of this star using the upgraded Giant Metrewave Radio telescope and the Karl G. Jansky Very Large Array, which led to the surprising discovery of ECME of both circular polarizations up to around 1.5 GHz. 
The GHz observations also allowed us to infer that the upper ECME cut-off frequency is at $\gtrsim 5\,\mathrm{GHz}$. 
The sub-GHz observation led to the unexpected observation of more than two pairs of ECME pulses per rotation cycle. In addition, we report the discovery of a `giant pulse', and transient enhancements, which are potentially the first observational evidence of `centrifugal breakout' of plasma from the innermost part of the stellar magnetosphere. 
The stark contrast between the star's behavior at GHz and sub-GHz frequencies could either be due to propagation effects, a manifestation of varying magnetic field topology as a function of height, or a signature of an additional `ECME engine'.
\end{abstract}




\section{Introduction}\label{sec:intro}
A wide varieties of stellar objects produce coherent radio emission by the process of electron cyclotron maser emission (ECME), like magnetic early-type stars, planets, ultracool dwarfs (UCDs) etc. \citep[e.g.][]{trigilio2000,hallinan2006}. ECME gives rise to highly directed, circularly polarized radiation which are seen as pulses at certain stellar rotational phases. In case of the magnetic early-type stars (spectral type A or B), the observed emission is highly regular in terms of the rotational phases of arrival. Because of the similarity of the phenomenon with the pulsar radio emission, the hot magnetic stars emitting ECME has been termed as main sequence pulsars \citep[e.g.][]{krticka2019}. However, as the corresponding acronym `MSP' is already in use for millisecond pulsars, we propose to call this class of objects `Main-sequence Radio Pulse emitter' (MRP). Currently seven MRPs are known: CU\,Vir \citep{trigilio2000}, HD\,133880 \citep{chandra2015,das2018}, HD\,142990 \citep{lenc2018,das2019a}, HD\,142301 \citep{leto2019}, HD\,35298 \citep{das2019b}, $\rho\,\mathrm{Oph\,A}$ \citep{leto2020} and $\rho\,\mathrm{Oph\,C}$ \citep{leto2020b}.

The reason that the pulses produced by MRPs are highly regular is that these objects have stable large-scale (mostly dipolar) magnetic fields \citep[e.g.][]{kochukhov2014}. 
The interaction between the magnetic field and the radiatively driven stellar wind forms a co-rotating magnetosphere around these stars \citep[e.g.][]{andre1988}. Upto a certain distance (defined by the Alfv\'en radius $R_\mathrm{A}$), the magnetic field dominates over the stellar wind kinetic energy and the wind plasma is forced to follow the magnetic field lines. This region around the star where all the magnetic field lines are closed is called `inner magnetosphere'. Inside the inner magnetosphere, plasma is trapped and the mass-loss is prohibited, except via `centrifugal breakout' or `leakage' mechanism \citep[e.g.][]{shultz2020,owocki2020}. Beyond a certain distance away from the star, the stellar wind dominates over the magnetic field and the field lines now become open. This part of the magnetosphere is called the `outer magnetosphere'. The region that marks the transition from inner to outer magnetosphere is called the `middle magnetosphere'. This region consists of magnetic field lines that produces a current sheet at the magnetic equator \citep[see Figure 1 of ][]{trigilio2004}. The current sheet is the site of acceleration of electrons. These electrons, while traveling back towards the stellar surface following the magnetic field lines, experience magnetic mirroring effect due to the converging field lines. It results into developing a `loss-cone distribution' where, over a certain velocity range, population inversion is established \citep[e.g.][]{treumann2006,trigilio2004}. This unstable distribution leads to the generation of ECME, provided the plasma density is small enough so that the local plasma frequency $\nu_\mathrm{p}$ is smaller than the local electron gyrofrequency $\nu_\mathrm{B}$. The latter is related to the local magnetic field strength as $\nu_\mathrm{B}\approx 2.8B$, where $\nu_\mathrm{B}$ is in MHz and the magnetic field strength $B$ is in gauss units \citep[e.g.][]{melrose1982,dulk1985}. The emission happens at frequencies very close to the harmonics of $\nu_\mathrm{B}$ with a bandwidth of only a few percents \citep[e.g.][]{melrose1982,sharma1984,dulk1985}. However in most astrophysical systems, the observed radiation is comprised of radiation generated at sites with different magnetic field strengths (e.g. in case of the MRPs) so that effectively it becomes a broadband phenomenon. Two of its distinguishing properties are the high directivity and circular polarization. For mildly relativistic electrons, the emission is directed nearly perpendicular to the magnetic field direction. The angular width of the beam $\Delta \theta$ is related to the angle of emission $\theta_0$ w.r.t. the magnetic field direction as $\Delta\theta\approx\cos\theta_0$ \citep{melrose1982}, i.e. larger the angle $\theta_0$, more directed will be the emission.

\begin{figure}
    \centering
    \includegraphics[trim={0.2cm 0cm 4cm .8cm}, clip, width=0.45\textwidth]{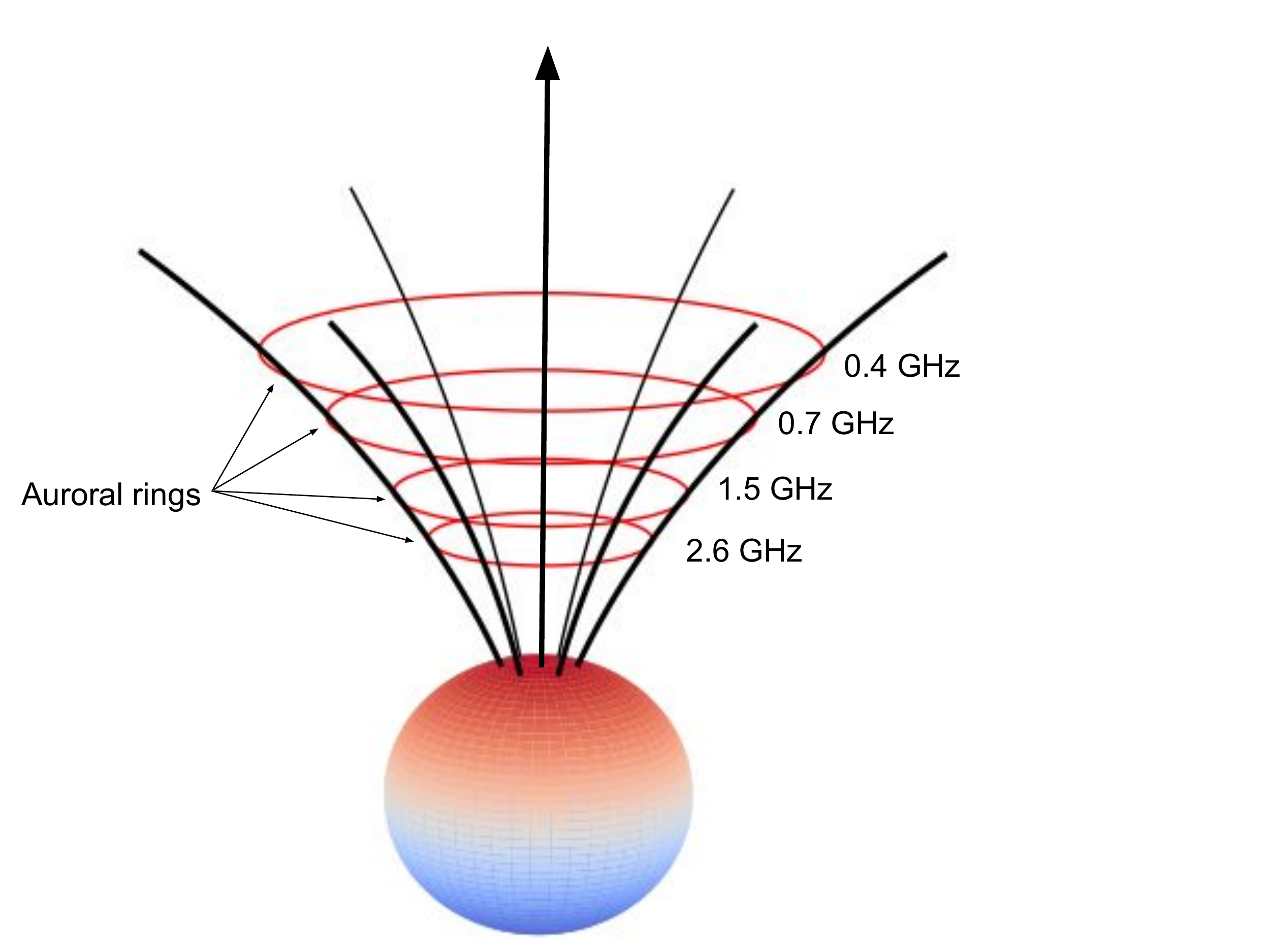}
    \caption{The different sites of origin of ECME at different frequencies in a star with a dipolar magnetic field. The arrow represents the magnetic dipole axis. Here we show the sites for only the north magnetic hemispheres. Since, the frequency of emission is proportional to the magnetic field strength (see \S\ref{sec:intro}), the sites producing ECME at a given frequency corresponds to sites with a given magnetic field value. For a dipolar magnetic field, such sites constitute ring-shaped regions, known as auroral rings. Here, we have assumed a dipolar magnetic field of strength 4 kG at the poles. The auroral rings shown are the sites of origin of ECME at the frequencies mentioned in the figure, assuming emission at the second harmonic of the local electron gyrofrequency $\nu_\mathrm{B}$. }
    \label{fig:auroral_rings}
\end{figure}

Inspired by the angular beaming characteristics of the auroral kilometric radiation \citep[e.g.][]{mutel2008}, \citet{trigilio2011} proposed that ECME beam pattern from MRPs also follows the `tangent plane beaming model', i.e. the emission is directed nearly perpendicular to the local magnetic field line and parallel to the magnetic equatorial surface. As a result, they are seen close to the rotational phases corresponding to the zeros of the stellar longitudinal magnetic field \bz~\citep{lo2012,leto2016}. Since, the frequency of ECME is proportional to the local magnetic field strength, higher frequency arises closer to the star and vice-versa (Figure \ref{fig:auroral_rings}). Also, radiation produced at opposite magnetic poles are oppositely circularly polarized. In case of the extra-ordinary (X) mode of emission, the radiation produced near the north magnetic pole is right circularly polarized (RCP) and those produced near the south magnetic pole is left circularly polarized (LCP). For ordinary (O) mode, the sense of circular polarization is just the opposite. Assuming that ECME is produced near both magnetic poles, it is expected that there will be a pair of oppositely circularly polarized ECME pulses per stellar rotation cycle, observable near each magnetic null \citep{leto2016}. Under certain circumstances, the oppositely circularly polarized pulses may overlap in their rotational phases of arrival leading to very low circular polarization in the observed radio pulses \citep{leto2016,das2020a}.

The first MRP discovered is CU\,Vir \citep{trigilio2000}. Naturally, it has been observed several times thereafter, not only in radio bands, but also in X-ray and optical bands \citep[e.g.][etc.]{ravi2010,lo2012,krticka2019,robrade2018}. 
The X-ray observations showed that the star produces very hard X-ray \citep[dominated by hot plasma at 25 MK for multi-thermal model,][]{robrade2018} which led to the speculation of non-thermal origin of the hard X-ray emission as first proposed by \citet{leto2017} for the magnetic B star HD\,182180.
The published radio measurements so far revealed that CU\,Vir produces only RCP ECME. In case of the rest of the MRPs, both LCP and RCP pulses are observed \citep[e.g.][]{leto2019}. Since oppositely circularly polarized pulses correspond to activities near the two magnetic poles, the absence of LCP pulse indicates that ECME is not produced near one of the magnetic poles of CU\,Vir \citep{trigilio2000}. It was speculated that this could be a result of deviation of the stellar magnetic field from dipolar topology \citep{trigilio2000}.

In all the reported radio observations of CU\,Vir, which are aimed at studying ECME, one but all has observing frequency $\geq 1\,\mathrm{GHz}$. The only observation below 1 GHz was the one at 610 MHz by \citet{stevens2010} using the legacy Giant Metrewave Radio Telescope (GMRT). They covered a rotational phase range of $\approx 0.25$ cycle. The authors claimed to have detected low frequency counterpart of the ECME pulses seen at higher frequencies ($>1\,\mathrm{GHz}$), however this study was not pursued further.

In order to explore the sub-GHz properties of this well-known star, we observed CU\,Vir over 0.3--0.8 GHz using the upgraded GMRT (uGMRT) for one full stellar rotation cycle and at two epochs. We also acquired near-simultaneous data of this star covering the frequency range of 1--4 GHz using the Karl G. Jansky Very Large Array (VLA). In this paper, we present the results obtained from our ultra-wideband observation and discuss possible scenarios that can explain the several surprising results that we obtained.


This paper is structured as follows: in the next section (\S\ref{sec:obs_data}), we describe the details of the observations and data analysis. We then summarize the results obtained from our higher frequency ($\geq 1\,\mathrm{GHz}$) observations (\S\ref{sec:high_frequenecy_cuv}). Following that, we present the results of the sub-GHz observation of CU\,Vir which are the highlights of this paper (\S\ref{sec:results_low_freq}). We then connect the results obtain at different radio bands (\S\ref{sec:combined_results}) and discuss a scenario to consistently explain the behaviour over the entire frequency range (\S\ref{sec:all_data_combined}). The penultimate section (\S\ref{sec:transients}) is dedicated to the transient phenomena that we observed at sub-GHz frequencies. We finally present our conclusions and summarize the paper in \S\ref{sec:summary}.

\section{Observations and Data analysis}\label{sec:obs_data}
We observed the star in band 3 (300--500 MHz) and band 4 (550--900 MHz) of the uGMRT in January 2019 (epoch 1) and June 2020 (epoch 2); and in L (1--2 GHz) and S (2--4 GHz) of the VLA in June and July 2019. At all the frequency bands, the star was observed for one full rotation cycle. 

In case of the uGMRT observations (sub-GHz frequency), we observed the star at two epochs so as to examine the persistency of the emission over time (the same was not needed at higher radio frequencies since the star was already observed several times above 1 GHz). To reduce the required observing time (and also to compare the results obtained in the two bands without including any time-dependent effect), the data at the second epoch were acquired in subarray mode, i.e. data over band 3 and band 4 were recorded simultaneously by using half of the antennas for each frequency band. The time resolution for each uGMRT observation was 8 seconds. At epoch 1, the original frequency resolution in band 3 and band 4 were respectively 97.7 kHz and 195.3 kHz, and at epoch 2, it was 195.3 kHz for both bands. Details of our uGMRT observations are provided in Table \ref{tab:obs}. 
The standard calibrators 3C286 and 3C147 were used for calibrating the absolute flux density scale and also for bandpass calibration, and J1445+099 was used as the phase calibrator. The data were analysed using the method described in \citet{das2019b}. The final frequency resolution in band 4 at both epochs was 586 kHz. In band 3, it was 293 kHz at epoch 1 and 391 kHz at epoch 2.

\begin{deluxetable*}{cccc|cccc|cccc}
\scriptsize
\tablecaption{Log of uGMRT observations of CU\,Vir showing the dates and duration of observations at different wavebands and the effective observing frequency ranges (Eff. band) for each band on different days. The hours here refer to total observation hour including the overheads. Also shown are the rotational phase ranges ($\phi_\mathrm{rot}$) covered on different days. The rotational phases are calculated using the ephemeris proposed by \citet{mikulasek2011}.\label{tab:obs}} 
\footnotesize
\tablehead{
\hline
\multicolumn{8}{c|}{Epoch 1}       & \multicolumn{4}{c}{Epoch 2}\\
\hline
\multicolumn{4}{c|}{band 3} & \multicolumn{4}{c|}{band 4} &\multicolumn{4}{c}{band3+band4}\\
Date & Hours & Eff. band & $\Delta \phi_\mathrm{rot}$ & Date & Hours & Eff. band &$\Delta \phi_\mathrm{rot}$ & Date & Hours & Eff. band & $\Delta \phi_\mathrm{rot}$\\
& & MHz & &  & & MHz & &  & & MHz &
}
\startdata
\hline
2019-01-11 & 7 &  333--358, & 0.25--0.75 & 2019-01-04 & 5 & 570--804 & 0.11--0.46 & 2020-06-07 & 8 & 279--461, & 0.69--1.19\\
 &  & 378--461  & & 2019-01-06 & 5 & 570--804 & 0.88--1.23 & & & 570--804 &\\
2019-01-18 & 7 & 334--461 & 0.69--1.21 & 2019-01-07 & 5 & 570--804  & 0.62--0.99 & 2020-06-13 & 8 & 279--461, & 0.18--0.71\\
& & & & 2019-01-13 & 5 & 570--804 & 0.43--0.75 &  & & 570--804 &\\
\hline
\enddata
\end{deluxetable*}

In case of the VLA data, the full stellar rotation cycle was covered by observing on 2019 June 12 and 2019 July 23, with seven hours of observing time on each day. The rotational phase covered on 2019 June 12 was 0.26--0.78 and that on 2019 July 23 was 0.80--1.32. We used subarray mode that enabled us to acquire simultaneous data at L (1--2 GHz) and S (2--4 GHz) bands. In both bands, there were 16 spectral windows (SPWs), each containing 64 channels. Each SPW has a bandwidth of 64 MHz in L band and 128 MHz in S band. The time resolution was 3 seconds. 3C286 was used as the absolute flux density calibrator and J1354-0206 was used as the phase calibrator.
The data were calibrated using the VLA Scripted Calibration Pipeline\footnote{\url{https://science.nrao.edu/facilities/vla/data-processing/pipeline/scripted-pipeline}} which is based on the Common Astronomy Software Application \citep[\textsc{casa},][]{mcmullin2007}.
The calibrated data for the target were further inspected for bad data and flagged if required. The L band data were selfcalibrated without masking the target using the \textsc{casa} task `tclean'. The sources other then the target were then subtracted using the task `uvsub' and the residual was imaged with 2 minute time resolution. A similar approach was followed for the S band data except that no selfcalibration was done.

In one case (for band 3 data), we extracted the dynamic spectrum from the self-calibrated uGMRT data. For that, we first subtracted the contribution of all the sources in the field of view excluding the target from the self-calibrated visibilities. We then averaged over a few channels (to enhance the signal to noise ratio). Further averaging was done over all the baselines for each time-frequency bin. The resulting visibilities (for each polarization) were then plotted on the time-frequency plane to obtain the dynamic spectrum for the given polarization.

The data were phased using the ephemeris provided by \citet{mikulasek2011}. 
According to this ephemeris, the rotational phase is given by $v-\mathrm{int}(v)$ ($\mathrm{int}(v)$ is the integer part of $v$); the quantity $v$ is defined as below:

\begin{align}
v&\cong v_0-\frac{B}{P_0}\left(\frac{3}{2}\Theta^2-\Theta^4\right);
v_0=\frac{t-M_0}{P_0};
\Theta=\frac{t-T_0}{\Pi}\label{eq:cuv_ephem}\\
B&=\frac{A\Pi}{P_0}\nonumber
\end{align}

Here $t$ corresponds to the time (Heliocentric Julian Day HJD) when we want to clculate the rotational phase, $M_0=2446730.4447$, $T_0=2446636(24)\,\mathrm{d}$, $P_0=0.52069415(8)\,\mathrm{d}$, $A=1.915(10)\,\mathrm{s}$ and $\Pi=13260(70)\,\mathrm{d}$.
Although a more recent ephemeris is available \citep{mikulasek2019}, we found that the older ephemeris is much more effective in terms of aligning pulses observed at different epochs. Note that neither work providing the ephemeris \citep{mikulasek2011,mikulasek2019}, includes data acquired in the year 2019 or latter.


Recently it was discovered that the circular polarization convention of the uGMRT in band 3 and band 4 is opposite to the IAU/IEEE convention \citep{das2020,das2020b}. To maintain a uniform convention for all our observations, we swapped the circular polarization manually for the uGMRT data during the post-processing phase.
Throughout the paper, we will use the color `red' for RCP and `blue' for LCP (unless stated otherwise), both of which are in accordance with the IAU/IEEE convention.

The flux density measurements at the four frequency bands are given in Tables \ref{tab:band3_data}, \ref{tab:band4_data}, \ref{tab:Lband_data} and \ref{tab:Sband_data}.

\section{Radio emission from CU\,Vir over 1--4 GHz}\label{sec:high_frequenecy_cuv}
\begin{figure}
    \centering
    \includegraphics[width=0.48\textwidth]{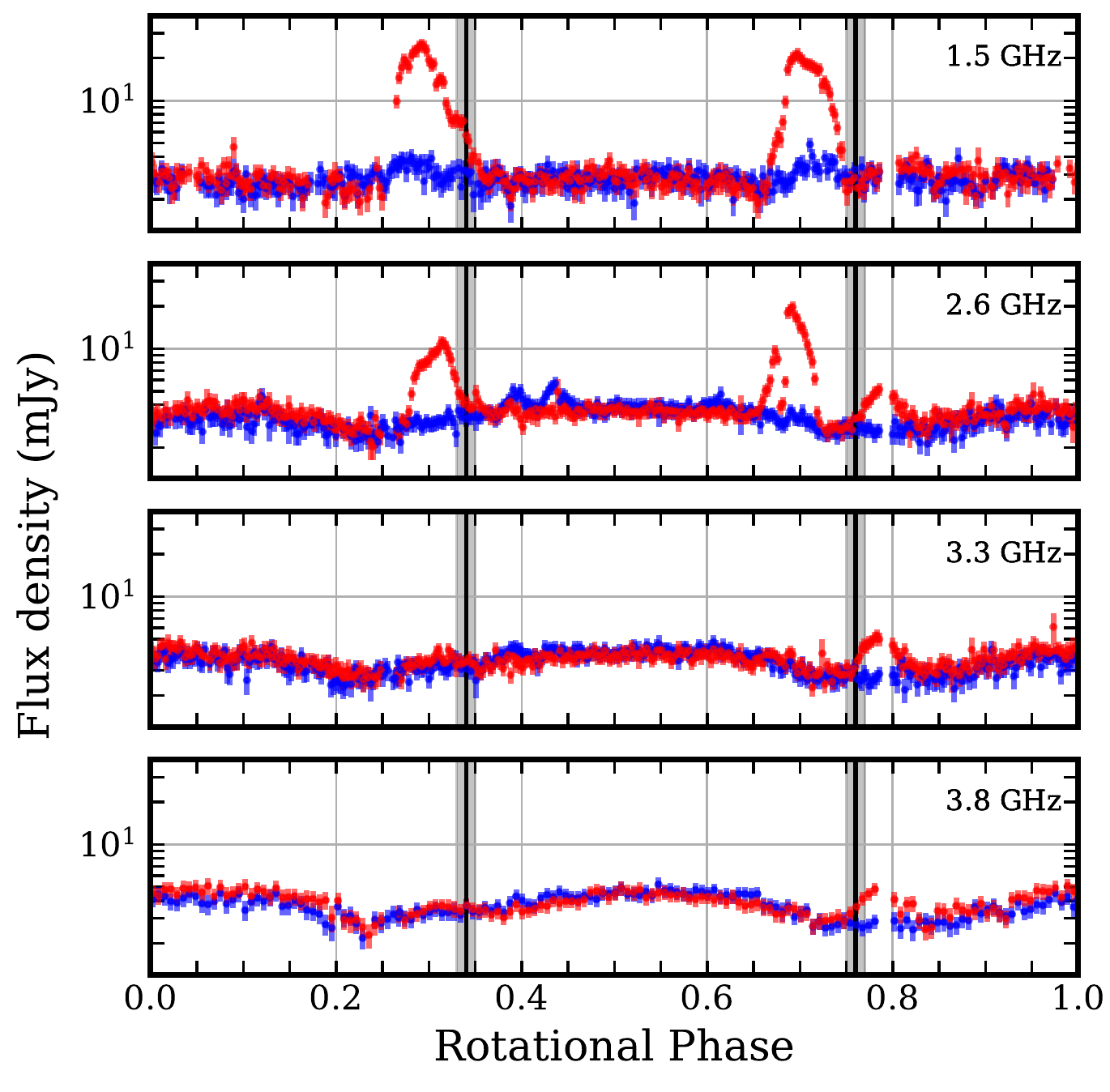}
    \caption{The lightcurves of CU\,Vir over 1--4 GHz obtained from data taken on 2019 June 12 and 2019 July 23. Red and blue markers represent right and left circular polarization respectively. The data were phased using the ephemeris of \citet{mikulasek2011}. The vertical lines mark the rotational phases of arrival of ECME pulses (RCP) at 1.4 GHz (using the same ephemeris) observed  over the years 1998 and 2010. \citep{trigilio2000,trigilio2011,kochukhov2014}. The surrounding grey shaded regions represent the associated uncertainties.}
    \label{fig:L_S_lightcurves}
\end{figure}

As mentioned already, ECME from CU\,Vir was discovered by \citet{trigilio2000} at 1.4 GHz. Subsequent observations established that this is a stable phenomenon that gives rise to exclusively right circularly polarized (RCP, according to IAU/IEEE convention) radio emission that are visible twice every rotation period \citep[e.g.][etc.]{trigilio2008,ravi2010,trigilio2011,lo2012}. It was also found that CU\,Vir exhibits ECME at least over the frequency range of 1.4--2.5 GHz, but is absent at frequencies $\geq5$ GHz. The multi-epoch observations also helped in diagnosing rotation period evolution of the star \citep{trigilio2008}. 

Here we report observation of this star over a continuous frequency range of 1--4 GHz. 
In Figure \ref{fig:L_S_lightcurves}, we show the lightcurves at 1.5 GHz (1--2 GHz), 2.6 GHz (2.05--3.05 GHz), 3.3 GHz (3.2--3.4 GHz) and 3.8 GHz (3.6--3.9 GHz). 
The vertical lines and the surrounding grey shaded regions in the panels mark the rotational phases $0.34\pm0.01$ and $0.76\pm0.01$, which are the rotational phases of arrival of the 1.4 GHz ECME pulses using the same ephemeris as the one used in this work, but using data acquired over the years 1998--2010 \citep{trigilio2000,trigilio2011,kochukhov2014}.
In case of the 1.5 GHz pulses reported here (data acquired in the year 2019), the rotational phases corresponding to the peak of the pulses are $0.292\pm0.001$ and $0.697\pm0.001$, which are offset by $0.05\pm0.01$ and $0.06\pm0.01$ rotation cycles respectively from those reported in the past (Figure \ref{fig:L_S_lightcurves}). This suggests that the extrapolated rotation period, obtained from the rotational period evolution model of CU\,Vir, reported by \citet{mikulasek2011} is incorrect, at least to phase the radio data. Note that the need for a more precise model to explain the rotation period evolution of CU\,Vir has already been mentioned by \citet{mikulasek2019}.

The new data provide several important results. Probably the most important one is the discovery of LCP enhancements in these frequencies, though at a significantly weaker level than that for the RCP. The second one is the discovery of a relatively weak RCP pulse near phase 0.8,  at frequencies above 2 GHz, that has previously gone unnoticed. Interestingly, this pulse seems to be present even at 5 GHz. We provide estimation/constraint on the upper cut-off frequencies for all these pulses, and also the lower cut-off frequency for the newly discovered RCP pulse. In addition, we present the rotational modulation of the higher frequency (gyrosynchrotron) lightcurves.


\subsection{Discovery of LCP enhancement from CU\,Vir at GHz frequencies}\label{subsec:lcp_at_GHz}
\begin{figure}
    \centering
    \includegraphics[width=0.45\textwidth]{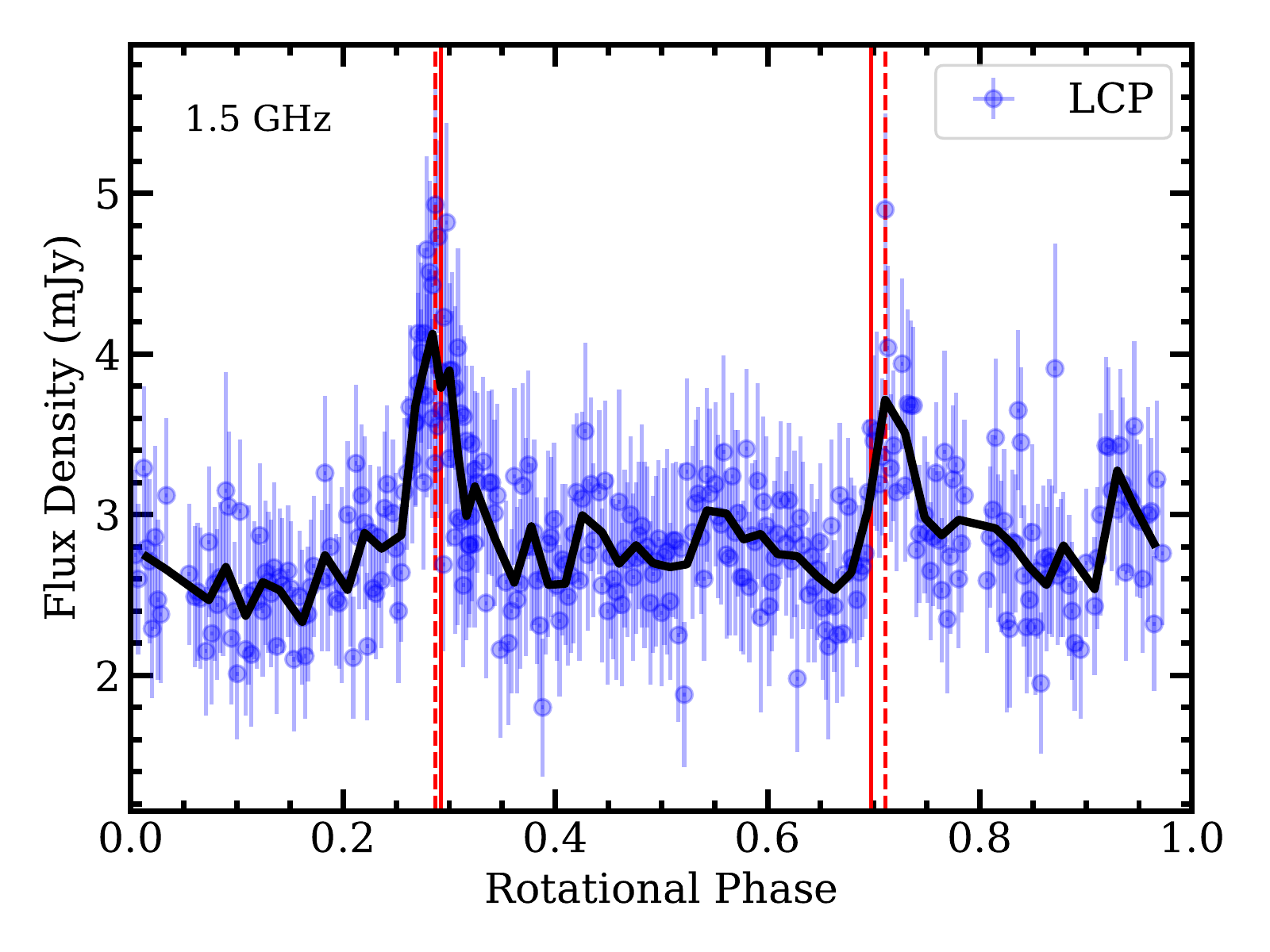}
    \caption{The weak LCP enhancement observed at 1.5 GHz from CU\,Vir. The vertical solid lines mark the maxima of the RCP ECME pulses, and the vertical dashed lines represent the same for the LCP enhancements shown here. The blue markers are the flux density measurements with 2 minutes time resolution (same as the top panel of Figure \ref{fig:L_S_lightcurves}). The black thick curve shows the smoothed lightcurve with 12 minutes time resolution. Note that this is the first time that LCP enhancement is detected above 1 GHz.}
    \label{fig:L_band_LCP}
\end{figure}

From the top panel of Figure \ref{fig:L_S_lightcurves}, that shows the lightcurves at 1.5 GHz, we hardly see any enhancement in LCP. However, a careful investigation (with the aid of a highly sensitive instrument: VLA) revealed LCP enhancements close to those observed at RCP, though not perfectly aligned with the former (Figure \ref{fig:L_band_LCP}). The latter fact allows us to rule out any instrumental leakage causing the enhancements. Comparing this figure with the top panel of Figure \ref{fig:L_S_lightcurves}, one can easily see that the prime reason for not noticing the LCP enhancements until now is that these pulses are extremely weak as compared to those in RCP adjacent to the former. 

We do not find any hint of LCP enhancement near the RCP enhancements in S band within our sensitivity limit, suggesting that the upper cut-off frequency of the LCP pulses lies in 1.5--2.6 GHz. In \S\ref{subsec:upper_cut_off}, a more detailed exercise to locate the upper cut-off frequency is described for both RCP and LCP ECME pulses.

\subsection{Discovery of a RCP pulse present only at S band}\label{subsec:additional_RCP_pulse}
In Figure \ref{fig:L_S_lightcurves}, we find an enhancement in the RCP between 0.75--0.85 rotation cycle which is not detectable at 1.5 GHz, but present at 2.6 GHz (2.05--3.05 GHz, Figure \ref{fig:L_S_lightcurves}) up to 3.8 GHz, where the `regular' ECME pulses are absent. Between 2.6--3.8 GHz, it exhibits similar shape and strength. This feature is actually covered partially on the two days of our observation (rising part on 2019 June 12 and the falling part on 2019 July 23), which implies that the enhancement is persistent. 
As can be seen from the bottom panel of Figure \ref{fig:gyrosynchrotron}, it is $\approx25\%$ right circularly polarized at 3.8 GHz, much higher than that in case of the gyrosynchrotron emission around the primary maxima at the same frequency. This raises the question of whether this enhancement is indeed a part of the gyrosynchrotron lightcurve. In the literature, there is no mention of observation of such a structure to the best of our knowledge, but a hint of such a feature is visible in the 2.5 GHz lightcurve in Stokes V ((RCP--LCP)/2) presented in \citet{trigilio2008} (see their Figure 5, upper right panel) and also in the 5 GHz lightcurves reported by \citet{trigilio2000,leto2006}. Combining all these observations, we find the following properties of this pulse:

\begin{enumerate}
\item This pulse is visible at frequencies higher than 2 GHz and at least up to 5 GHz. 
\item This pulse is right circularly polarized throughout its bandwidth.
\item The percentage circular polarization appears to decrease between 4 and 5 GHz \citep[by $\approx 10\%$, based on measurements of this work and those of ][]{leto2006}.
\item The rotational phases of arrival of the pulse at different frequencies appear to be the same (within our time resolution)\footnote{Though the rotational phase of arrival of the pulse at 5 GHz is apparently offset by $\approx 0.05$ cycle \citep{leto2006} from that observed at 3.8 GHz, the rotational phases corresponding to the primary maxima are also offset by similar amount, showing that the offsets are spurious. These offsets very likely arise due to the same reason as that behind the offsets observed between ECME pulses observed in the past and those reported here.}
\end{enumerate}

The top three characteristics clearly set this pulse apart from the rest of the lightcurve (consisting of gyrosynchrotron emission). Once we rule out gyrosynchrotron, the only explanation left is that this pulse is also produced by ECME. Note that this inference extends the highest frequency of ECME detection from MRPs to 5 GHz. Possible reasons for the appearance of this pulse include contribution from higher order magnetic field close to the stellar surface, propagation effects in the magnetospheric plasma and ECM emission at a higher harmonic than that for the known pulses.

\subsection{The cut-off frequency of ECME}\label{subsec:upper_cut_off}
\begin{figure*}
\centering
\includegraphics[width=0.3\textwidth]{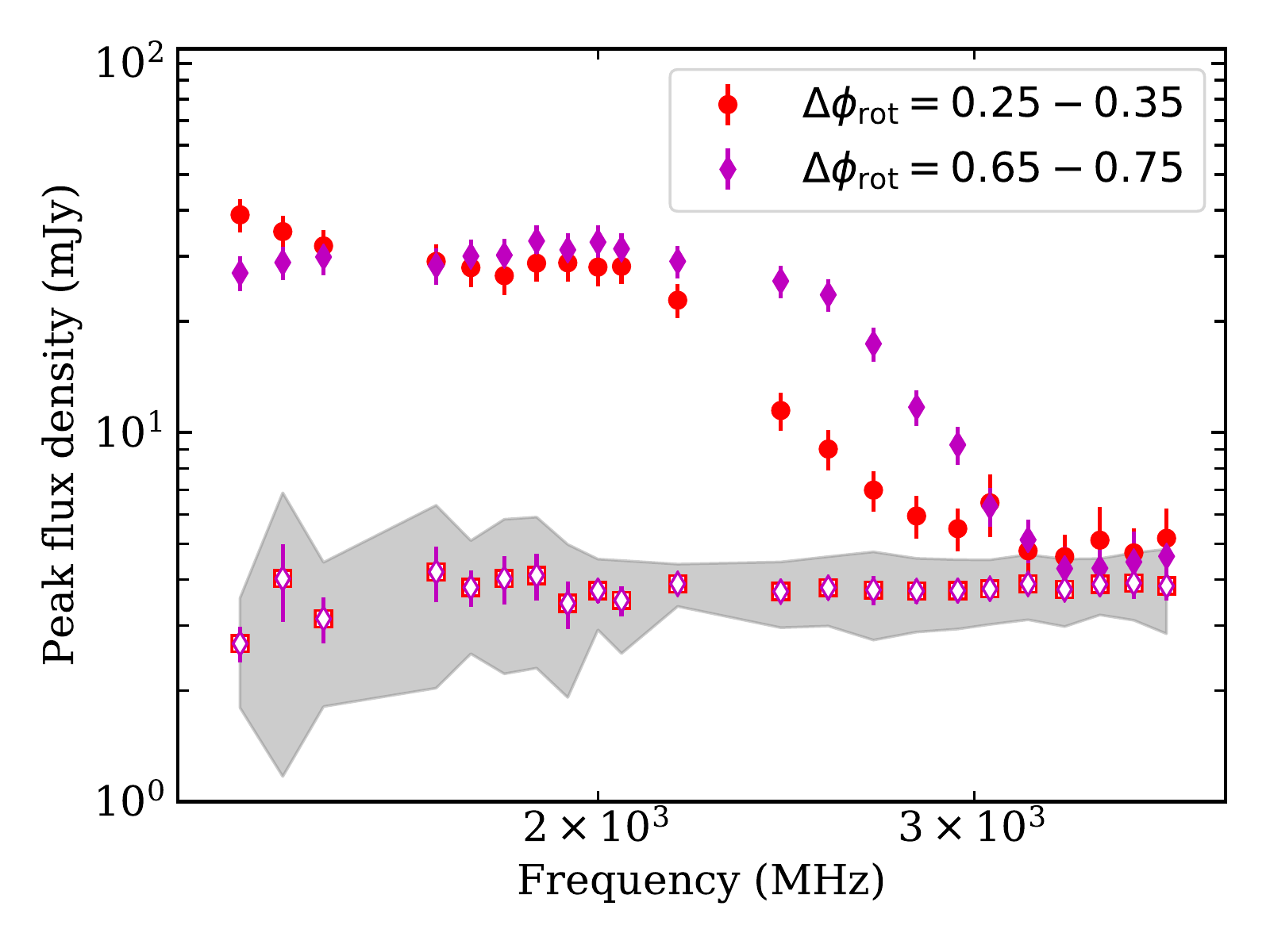}
\includegraphics[width=0.3\textwidth]{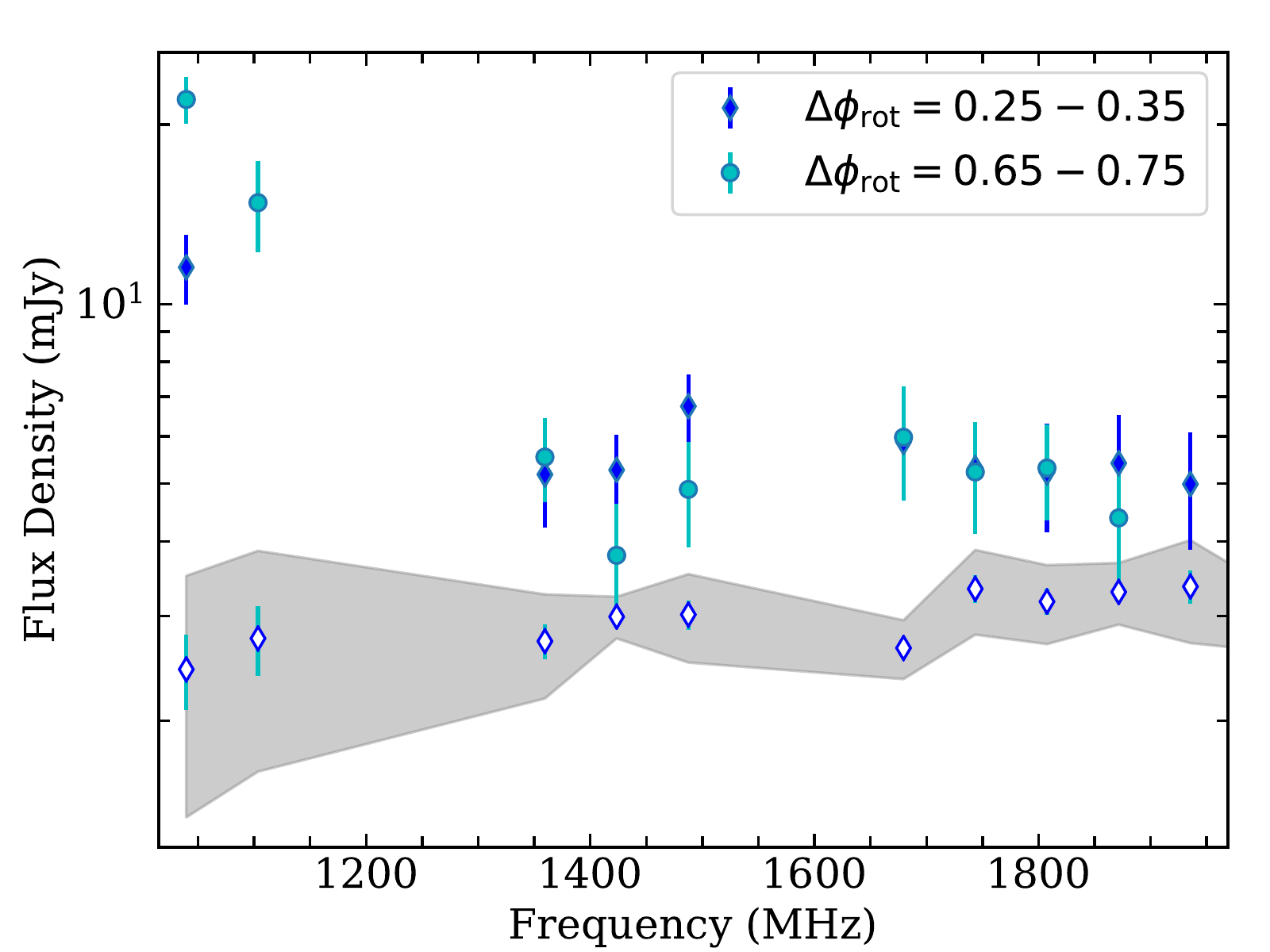}
 \includegraphics[width=0.3\textwidth]{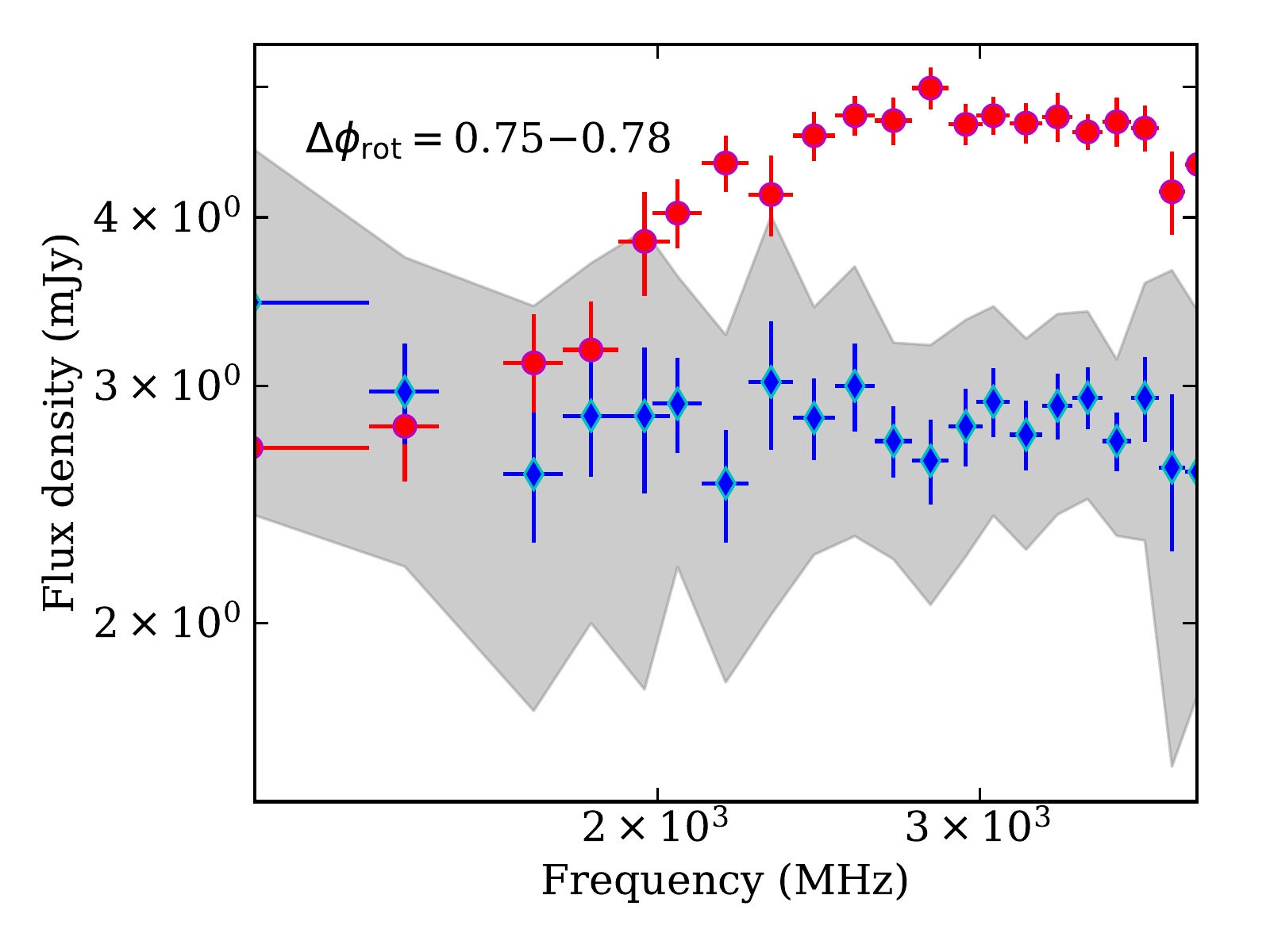}
\caption{\textit{Left:} The spectra (filled markers) corresponding to the peak flux density of the two ECME pulses observed from CU\,Vir. Both pulses, under consideration, are right circularly polarized. The unfilled markers represent basal RCP flux density. The grey region represents $3\times\mathrm{MAD}$, where MAD is the median absolute deviation about the basal flux density (see \S\ref{sec:high_frequenecy_cuv}). These data were acquired on 2019 June 12 with the VLA. $\Delta\phi_\mathrm{rot}$ represents rotational phase range at which an ECME pulse was observed. \textit{Middle:} Same for the LCP pulse, except that the data for the rotational phase range 0.25--0.35 were taken on 2019 July 23. \textit{Right:} The spectrum of the peak flux density of the RCP pulse (red markers), observed only at GHz frequencies between 0.75--85 rotation cycle. Also shown is the spectrum of the peak LCP flux density over the same rotational phase range (blue markers) that acts as a measure of the basal flux density spectrum. The grey shaded region around the LCP flux densities represent $3\sigma$, with $\sigma$ being the errorbars in the basal LCP flux density measurements.\label{fig:L_S_spectra}}
\end{figure*}

In order to locate the upper cut-off frequencies of LCP and RCP pulses (excluding the newly discovered pulse, \S\ref{subsec:additional_RCP_pulse}) with better precision, we extract the lightcurves for every spectral window in our data. Note that each spectral window at L band (1--2 GHz) has a width of 64 MHz, whereas that for the S band is 128 MHz. In Figure \ref{fig:L_S_spectra}, we plot the peak flux density of the ECME pulses against the central frequency of the spectral windows.
Along with that, we also plot the spectrum for the basal flux density. We define the basal flux density to be the median of the flux densities between the rotational phase range 0.4 and 0.6 (where ECME is absent).  The errorbars in the basal flux densities correspond to the median absolute deviation (MAD) over the same range of rotational phases, i.e. 0.4--0.6 \footnote{Except for the case when there is only a single measurement at that range. In that case, the lone measurement and its errorbar are used as the measures of basal flux density and errorbar respectively.}. 
The grey shaded region marks $\leq 3\sigma$ deviation from the basal flux densities, where $\sigma$ represents the associated uncertainty. Similar to \citet{das2020b}, we also define the upper cut-off frequency to be the lowest frequency at which the peak flux density of the ECME pulse merges with the grey shaded region within its error bar.

The above strategy, however, cannot be applied to locate the lower cut-off frequency of the RCP pulse seen near 0.8 rotational phase in S band, since at those frequencies the basal flux density exhibits significant rotational modulation, and the region between 0.4--0.6 rotational phase in fact corresponds to a maximum of the gyrosynchrotron lightcurve with a flux density similar to the peak flux density of the RCP pulse under consideration (Figure \ref{fig:gyrosynchrotron}). We hence use the LCP flux densities within the rotational phase window 0.7--0.8 cycles, that encompasses the RCP pulse, as a measure of the basal flux densities (right of Figure \ref{fig:L_S_spectra}).
The lower cut-off frequency is then defined as the highest frequency (lower than the upper cut-off frequency) at which the peak flux density merges with the grey region.

According to our definition of cut-off frequencies, we obtain the following results:
\begin{enumerate}
    \item The upper cut-off frequency for the RCP ECME pulses, excluding the one visible only at S band, is $3.0\pm 0.06$ GHz.
    \item For the LCP pulses, the upper cut-off frequencies appear slightly different for the pulses observed near the two nulls according to the definition adopted here. For the pulse observed around 0.3 rotational phase, the upper cut-off frequency is $2.00\pm0.03$ GHz, whereas that for the pulse observed around 0.7 rotational phase is $1.42\pm0.03$ GHz.
    \item For the newly discovered RCP pulse (\S\ref{subsec:additional_RCP_pulse}), the lower cut-off frequency comes out to be $2.31\pm0.06$ GHz. Its upper cut-off frequency lies between 5--8.4 GHz.
\end{enumerate}

CU\,Vir is only the second hot magnetic star for which the ECME upper cut-off frequency is precisely located. The other such star, HD\,133880, has also been found to have different upper cut-off frequencies for its RCP and LCP ECME pulses \citep{das2020b}.

\subsection{Rotational modulation of gyrosynchrotron emission}\label{subsec:gyrosyncrotron_modulation} 
The basal flux density of CU\,Vir is known to be due to gyrosynchrotron \citep{trigilio2000,leto2006}.
As the frequency increases, the rotational modulation of the gyrosynchrotron emission becomes more and more prominent \citep[e.g. see Figure 5 of ][]{leto2020}. From Figure \ref{fig:L_S_lightcurves}, we find that the basal flux density is nearly flat at 1.5 GHz, but shows clear rotational modulation at 3.3 GHz and 3.8 GHz. 

\begin{figure}
\centering
\includegraphics[trim={0cm 0.cm 0cm 0.cm},clip,width=0.45\textwidth]{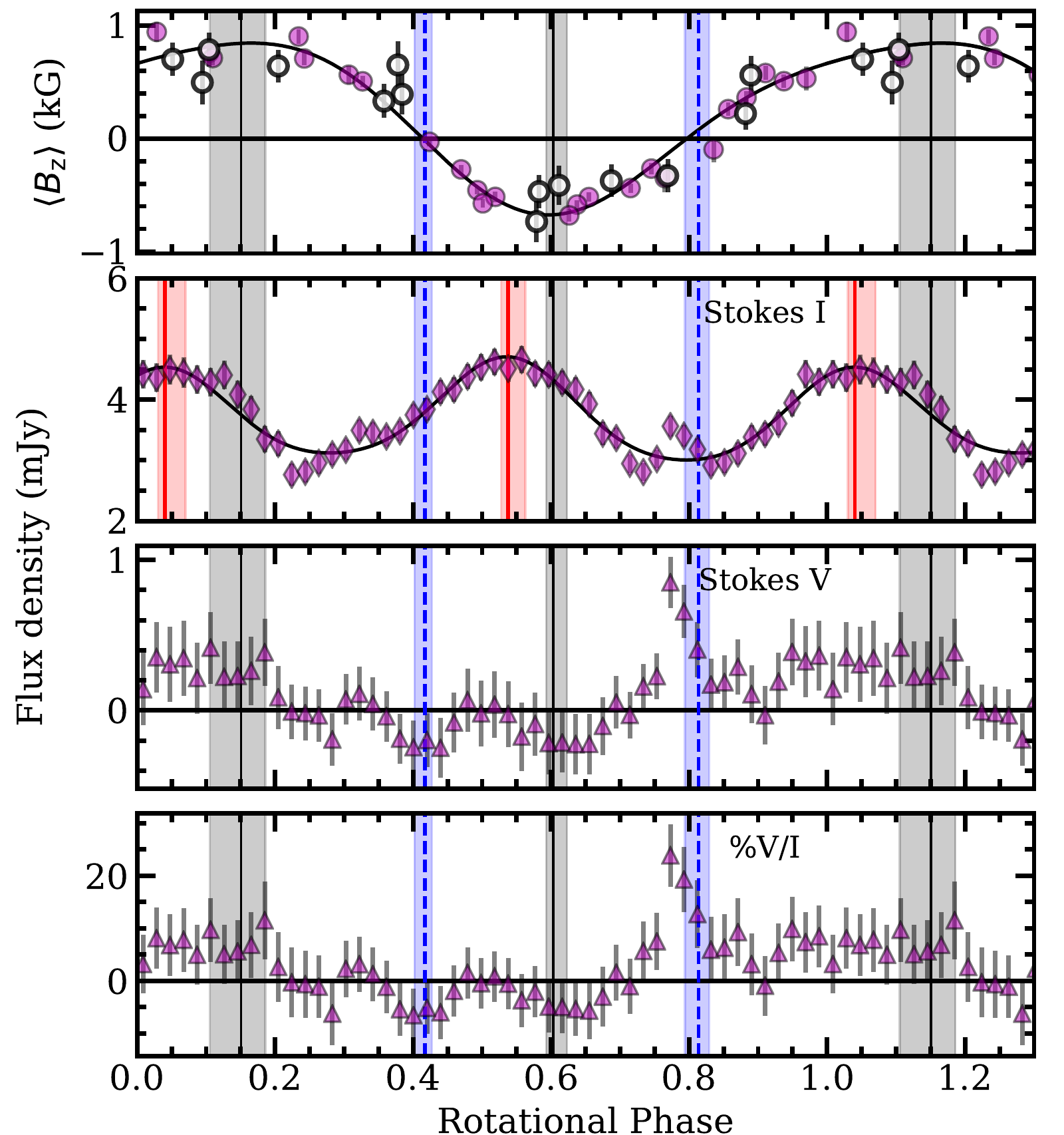}
\caption{\textit{Top:} The rotational modulation of the longitudinal magnetic field \bz of CU\,Vir. The \bz~data are taken from \citet{kochukhov2014}. The filled circles represent their own measurements, whereas the unfilled circles represent data from \citet{borra1980}. \textit{Second to fourth panels:} The rotational modulation of the total intensity Stokes I (second panel), circular polarization Stokes V (third panel), and the percentage circular polarization ($100\times$V/I, bottom panel) at 3.8 GHz, where I=(RCP+LCP)/2 and V=(RCP-LCP)/2. These are obtained after averaging over three data points in the original lightcurves (Figure \ref{fig:L_S_lightcurves}). The effective time resolution is 15 minutes. The solid vertical lines and the surrounding grey shaded regions mark the rotational phases corresponding to the extrema of the \bz~curve and the associated uncertainties respectively. The blue shaded regions mark the nulls of the \bz. The red shaded regions in the middle panel mark the maxima (and associated uncertainties) of the total intensity lightcurve.
See \S\ref{sec:high_frequenecy_cuv} for details.\label{fig:gyrosynchrotron}}
\end{figure}

In Figure \ref{fig:gyrosynchrotron}, we show the lightcurves for Stokes I, i.e. average of RCP and LCP flux densities and the percentage circular polarization at a central frequency of 3.76 GHz (3.56--3.95 GHz), along with the \bz~modulation. The latter was reported by \citet{kochukhov2014}. While obtaining these radio lightcurves, we averaged every three data points in the original lightcurves (time resolution was 5 minutes) which resulted in an effective time resolution of 15 minutes so as to further reduce the errorbars in the flux density measurements. Following \citet{kochukhov2014}, we fit a dipolar+quadrupolar model to the \bz~data resulting in the solid curve shown in the top panel of Figure \ref{fig:gyrosynchrotron}. We find the rotational phases corresponding to the maximum and the minimum of the fitted curve to be $0.15^{+0.04}_{-0.04}$ and $0.60^{+0.02}_{-0.01}$ respectively (shown by vertical black solid lines surrounded by grey shaded regions in the top panel of Figure \ref{fig:gyrosynchrotron}). The null phases come out to be $0.42^{+0.01}_{-0.02}$ and $0.81\pm0.02$ (shown by vertical blue dashed lines surrounded by blue shaded regions in the top panel of Figure \ref{fig:gyrosynchrotron}). In case of the total intensity lightcurve (middle panel of Figure \ref{fig:gyrosynchrotron}), we find the (primary) maxima to lie at $0.04^{+0.03}_{-0.01}$ and $0.54^{+0.02}_{-0.01}$ rotational phases. Thus, unlike the ideal case, the maxima of the gyrosynchrotron lightcurve are not aligned with the extrema of the stellar \bz~curve. 
This discrepancy however vanishes once we take the phase offset of $\approx 0.05-0.06$ cycles, observed between the ECME pulses observed in the year 2010 \citep[radio data near-simultaneous to \bz~data,][]{trigilio2011,kochukhov2014} and those reported in this work (year 2019), into account. This is interesting since the photospheric magnetic field of CU\,Vir is known to deviate significantly from a dipolar topology \citep{kochukhov2014}, whereas at the height of radio emission, we expect the field to be predominantly dipolar. Note that
\citet{leto2020} also observed a rotational phase offset of more than 0.1 cycle between the maxima of the gyrosynchrotron lightcurves over 5.5--21.2 GHz and the extrema of the photospheric \bz~for the star $\rho\,\mathrm{Oph\,A}$ \citep[their radio data were acquired in the year 2019 and the most of the \bz~data were acquired in the years 2017--2018,][]{leto2020, pillitteri2018}. The offset was attributed to the contribution of non-dipolar component to the photospheric \bz~measurements.


\section{Results from sub-GHz observations}\label{sec:results_low_freq}
\begin{figure}
    \centering
    \includegraphics[width=0.48\textwidth]{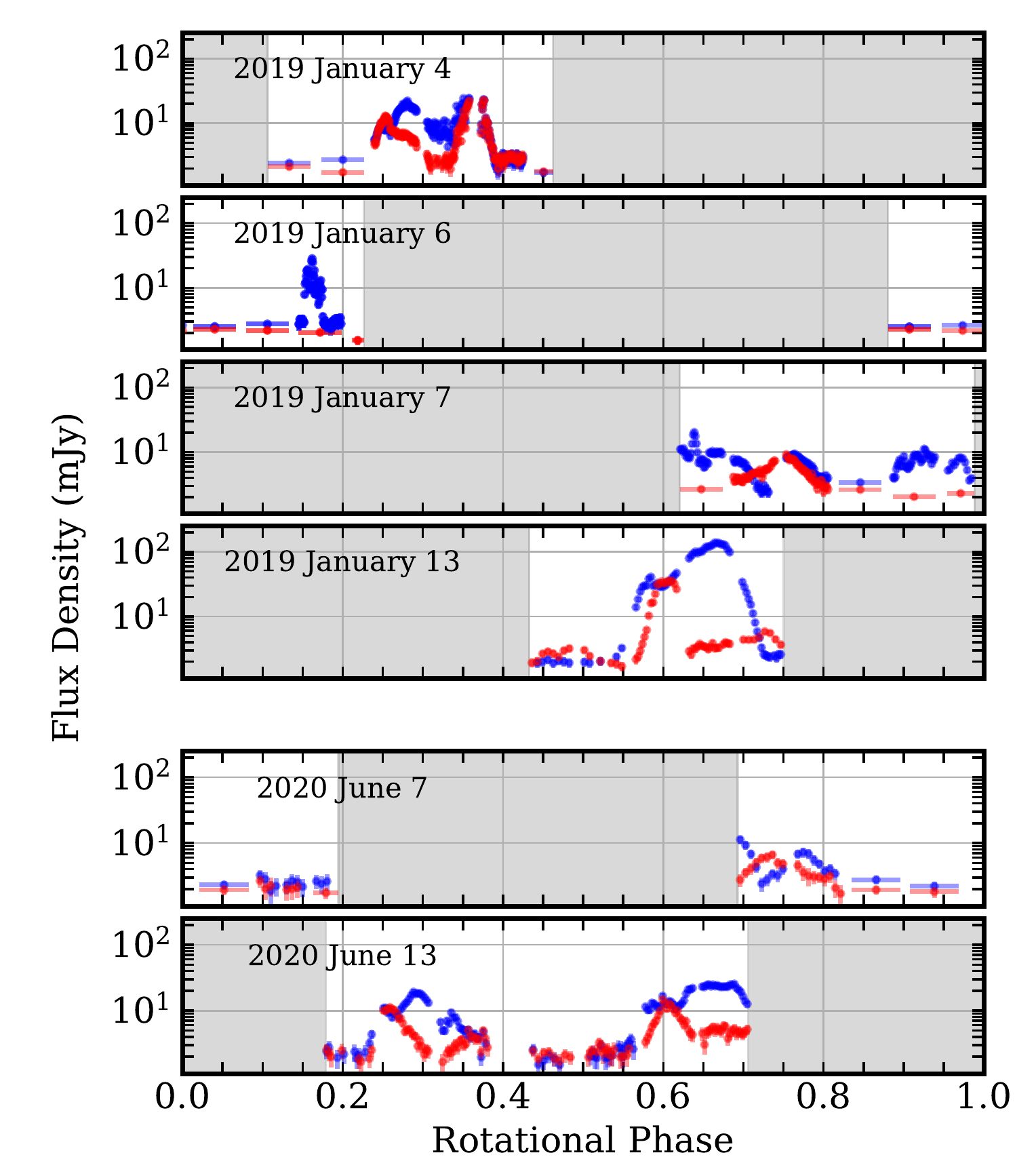}
    \caption{The lightcurves of CU\,Vir in band 4 of the uGMRT obtained during epoch 1 (top four panels) and epoch 2 (bottom two panels). For the definition of epoch 1 and epoch 2, refer to Table \ref{tab:obs}. Red and blue markers represent RCP and LCP respectively which are in accordance with the IEEE convention. The grey shaded regions in each panel indicate unavailability of data in those rotational phase ranges on that day. Note that the Y axes are in log scale.}
    \label{fig:band4_epoch1_2}
\end{figure}

The lower frequency lightcurves along with the higher frequency ones are shown in Figure \ref{fig:band3_band4_L_S}. Before discussing all these lightcurves together, we first present the results obtained for the lower frequencies starting with band 4 (550--800 MHz) followed by that for band 3 (330--461 MHz).

\subsection{Lightcurves in band 4 at the two epochs}\label{subsec:band4}
\begin{figure*}
    \centering
    \includegraphics[width=0.8\textwidth]{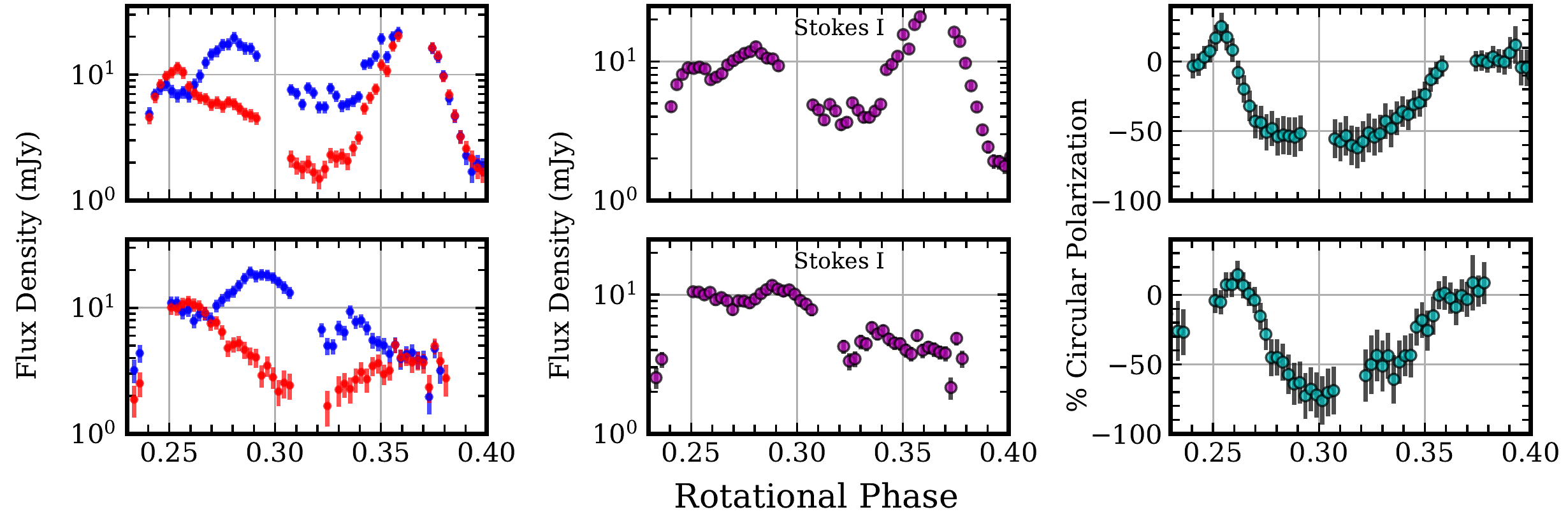}
    \caption{The enhancements observed from CU\,Vir in band 4 (570--804 MHz) of the uGMRT between phases 0.20--0.40 on different days of observation. \textit{Left:} RCP (red) and LCP (blue) lightcurves, \textit{Middle:} Stokes I (average of RCP and LCP) lightcurves, \textit{Right:} lightcurves for the percentage circular polarization, given by $100\times$(RCP--LCP)/(RCP+LCP). The top panel corresponds to data acquired at epoch 1 (2019 January 4) and the bottom panel corresponds to data taken at epoch 2 (2020 June 13). The fact that these enhancements are observed at both epochs show that they are persistent pulses. See \S\ref{subsubsec:band4_persistent_pulses} for details.}
    \label{fig:band4_4jan13jun}
\end{figure*}

\begin{figure*}
    \centering
    \includegraphics[width=0.8\textwidth]{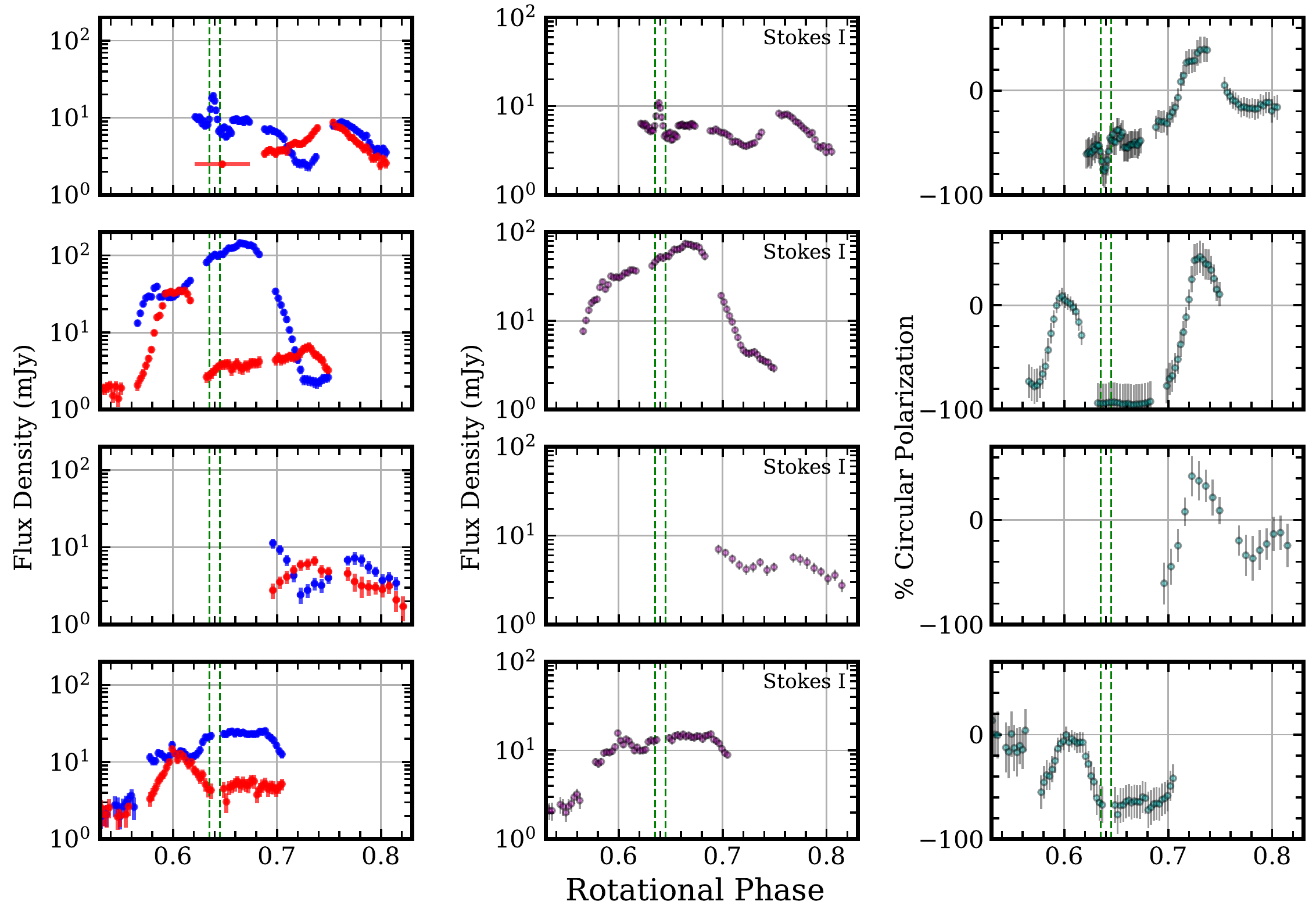}
    \caption{Another set of persistent enhancements observed from CU\,Vir in band 4 of the uGMRT that lie between phases 0.55--0.85. The four rows correspond to four different days of observation. \textit{Left:} RCP (red) and LCP (blue) lightcurves, \textit{Middle:} Stokes I lightcurves, \textit{Right:} lightcurves for the percentage circular polarization. The top two panels correspond to data acquired at epoch 1 (topmost: 2019 January 7, second from the top: 2019 January 13) and the bottom two panels correspond to data acquired at epoch 2 (third from the top: 2020 June 7, fourth from the top: 2020 June 13). The vertical dashed lines mark the narrow LCP enhancement observed only on 2019 January 7 (topmost panel). See \S\ref{subsubsec:band4_persistent_pulses} for details.}
    \label{fig:band4_7jan13jan_7jun13jun}
\end{figure*}

The lightcurves in band 4 on different days of observations are shown in Figure \ref{fig:band4_epoch1_2}. The top four panels correspond to epoch 1 (data acquired in the year 2019) and the bottom two correspond to epoch 2 (data acquired in the year 2020). 
Unlike their higher frequency counterparts, these lightcurves are full of enhancements. There are two types of enhancements: persistent and intermittent/transient. The persistent enhancements are observed over 0.20--0.40 and 0.55--0.80 rotation phases (Figures \ref{fig:band4_4jan13jun} and \ref{fig:band4_7jan13jan_7jun13jun}). The intermittent/transient enhancements are observed on three days at different rotational phases: 0.14--0.18 (2019 January 6), 0.635--0.645 (2019 January 7) and 0.88--1.00 (2019 January 7) rotation phases (Figure \ref{fig:flare_lightcurve_6_7jan2019}). Interestingly, the two rotational phase windows, within which the persistent enhancements are seen, are the same within which the ECME pulses above 1 GHz are seen (Figure \ref{fig:L_S_lightcurves}).

In the subsequent subsections, we discuss these two types of enhancements observed at different rotational phases. 

\subsubsection{Persistent enhancements}\label{subsubsec:band4_persistent_pulses}
As mentioned already, the persistent pulses in band 4 are observed over 0.20--0.40 and 0.55--0.80 rotation cycles. Below we describe their characteristics.

The rotational phases between 0.2--0.4 cycle were covered on two days: 2019 January 4 and 2020 June 13 (Figure \ref{fig:band4_epoch1_2}). The zoomed lightcurves are shown in Figure \ref{fig:band4_4jan13jun}. Though the profiles of the enhancements observed in RCP and LCP appear different, especially in the range 0.32--0.40, they appear similar in terms of the rotational modulation of the percentage circular polarization. The significance of this property is discussed in \S\ref{subsec:a_giant_pulse}.

One curious observation that we made is the offset of $\approx 0.01$ rotation cycle (7.5 minutes) between the LCP enhancements between 0.20--0.32 rotation cycles at the two epochs (see the left panels of Figure \ref{fig:band4_4jan13jun}). It is not clear if there is any offset between the respective RCP pulses as well, since it is not adequately sampled near the peak at epoch 2. As will be seen subsequently, such offsets are not observed for the remaining pulses proving that the observed offset, in this case, is not related to any kind of rotation period evolution, but could be due to `drifting pulse emission region' \citep{ravi2010}.

For the other set of persistent pulses, that lies between 0.55--0.85 rotation cycles (Figure \ref{fig:band4_7jan13jan_7jun13jun}), we have observation from four days: 2019 January 7, 13 (epoch 1), and 2020 June 7, 13 (epoch 2). Between the different days, we observe order of magnitude variation in the flux density over the same rotational phases (e.g. see second and fourth panels of Figure \ref{fig:band4_7jan13jan_7jun13jun}). Though ECME pulses are known to vary in amplitude \citep[e.g.][]{trigilio2011}, variation by an order of magnitude has never been observed before. In fact, our observation of a pulse of strength $> 100\,\mathrm{mJy}$ (second panel of Figure \ref{fig:band4_7jan13jan_7jun13jun}) corresponds to the strongest pulse ever observed from any MRP. Despite such dramatic fluctuation in amplitude of the pulses at different circular polarizations, the profiles for the percentage circular polarization remain essentially identical at all epochs, except for the data shown in the top panel of Figure \ref{fig:band4_7jan13jan_7jun13jun}, where a persistent pulse is contaminated by a non-persistent enhancement between 0.63--0.65 rotation cycles (marked by green dashed lines). 
The latter is  described in the next subsection along with the other non-persistent features observed in band 4.
The two aspects, i.e. the extreme enhancement observed, and the stability of the circular polarization profiles, are further discussed in \S\ref{subsec:a_giant_pulse}.

\subsubsection{Non-persistent features}\label{subsubsec:band4_non_persistent}
\begin{figure*}
    \centering
    \includegraphics[width=0.33\textwidth]{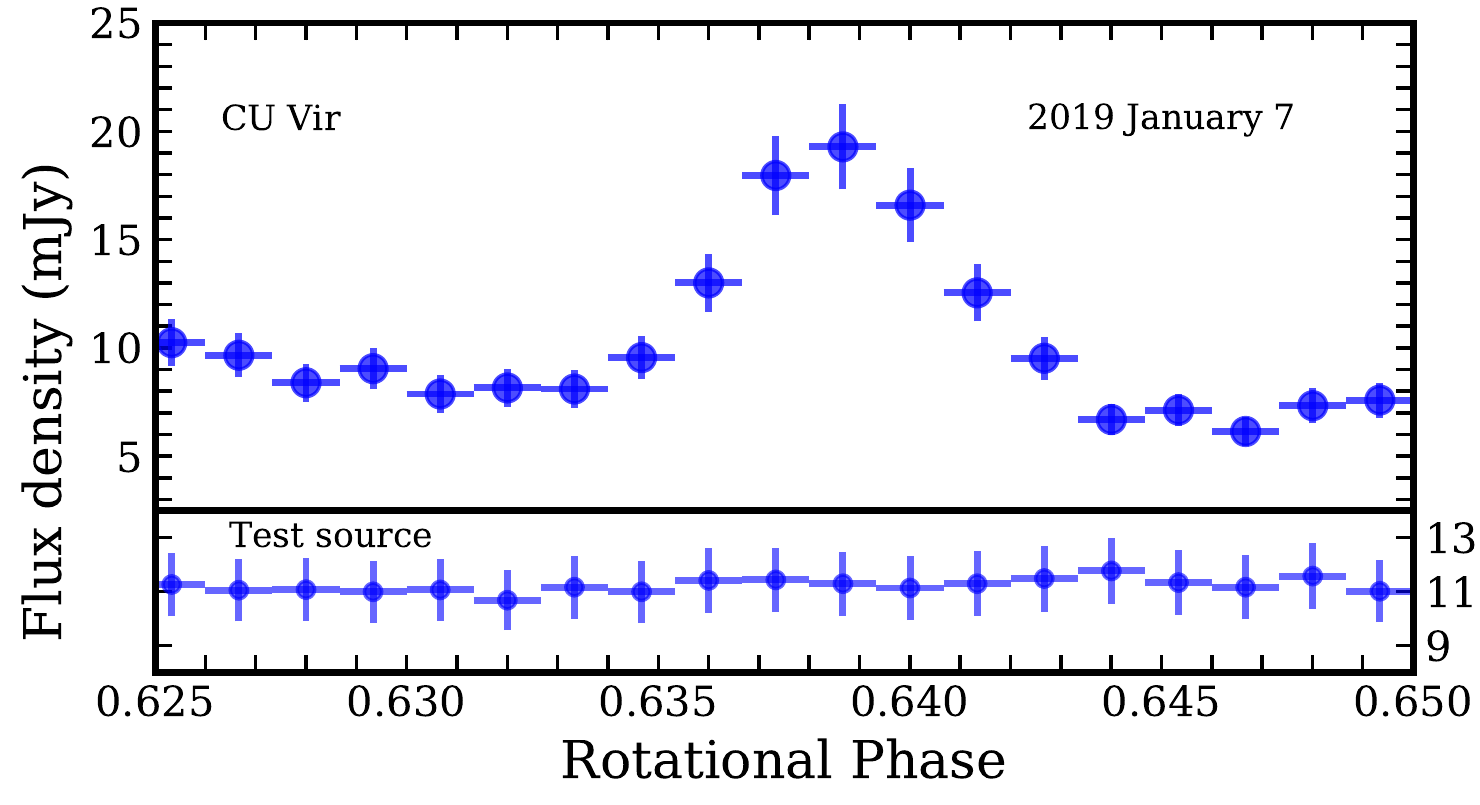}
     \includegraphics[width=0.33\textwidth]{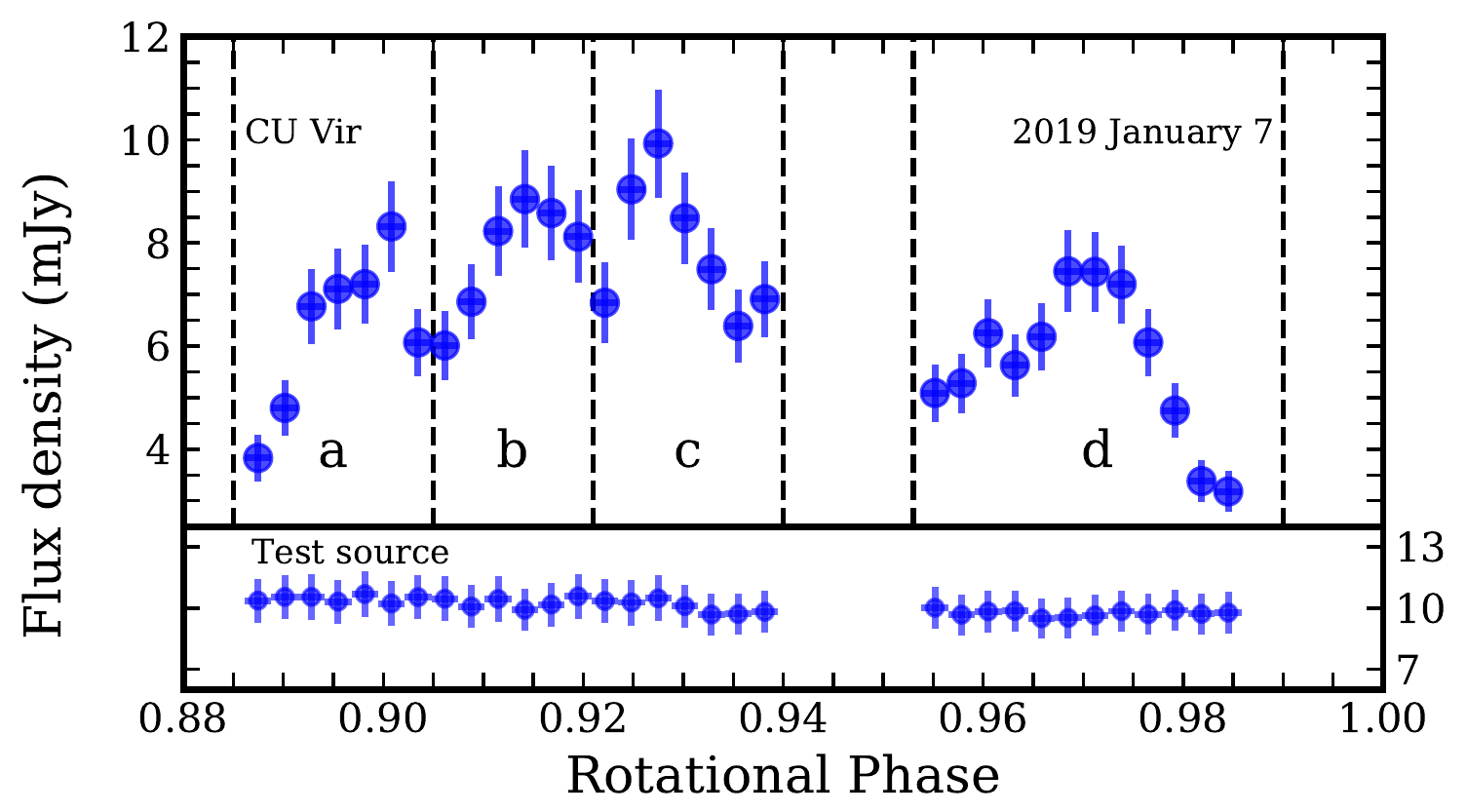}
      \includegraphics[width=0.32\textwidth]{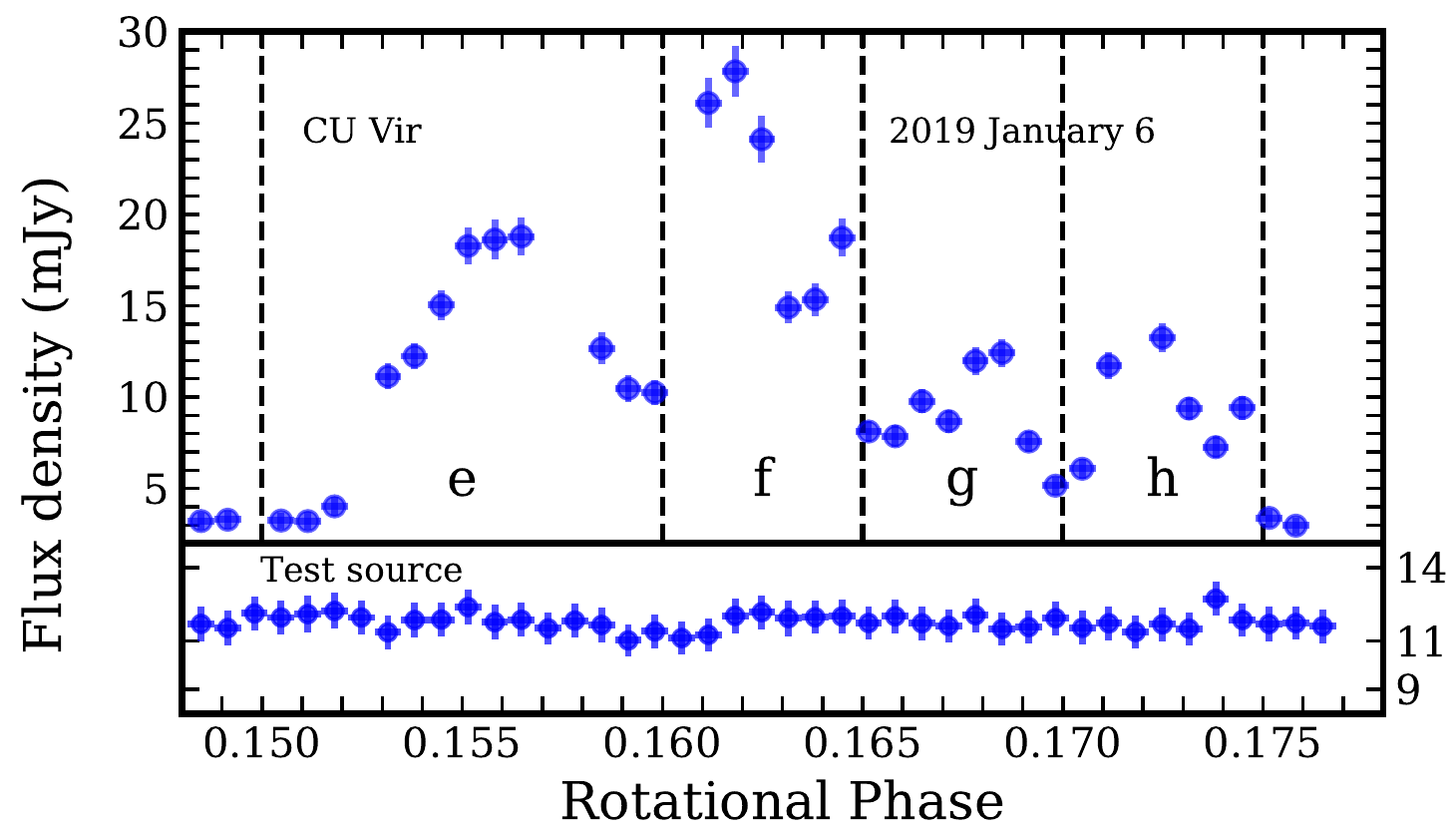}
     \caption{\textit{Left and middle:} The lightcurves for the LCP non-persistent enhancement seen from CU\,Vir on 2019 January 7, along with the corresponding lightcurves for a test source (\S\ref{sec:test_source_check}).  \textit{Right:} The lightcurve for the LCP flare seen from CU\,Vir on 2019 January 6, along with the corresponding lightcurve for a test source.}
    \label{fig:flare_lightcurve_6_7jan2019}
\end{figure*}

Non-persistent features refer to the enhancements that were observed only once despite multiple observations of the corresponding rotational phases. There are three such features, all left circularly polarized, observed on 2019 January 6 and 7. On the latter day, two such features were observed. The first was an extremely narrow feature between 0.635--0.645 cycles, that appeared on top of a persistent LCP pulse (see the region between the vertical dashed lines in the topmost panels of Figure \ref{fig:band4_7jan13jan_7jun13jun}). 
On the same day, we observed another non-persistent enhancement between 0.88--1.0 rotational phases. This and the one observed on the previous day (2019 January 6) between 0.14--0.18 rotational phases have multiple components of smaller width (Figure \ref{fig:flare_lightcurve_6_7jan2019}). The sub-components observed on the same day have similar widths, but those observed on different days, have different widths. The spectral properties of the different components are described in \S\ref{sec:flare_spectra}. The cause of these features are discussed in \S\ref{subsec:flares}.


\subsection{Lightcurves in band 3 at the two epochs}\label{subsec:band3}

\begin{figure*}
    \centering
    \includegraphics[width=0.95\textwidth]{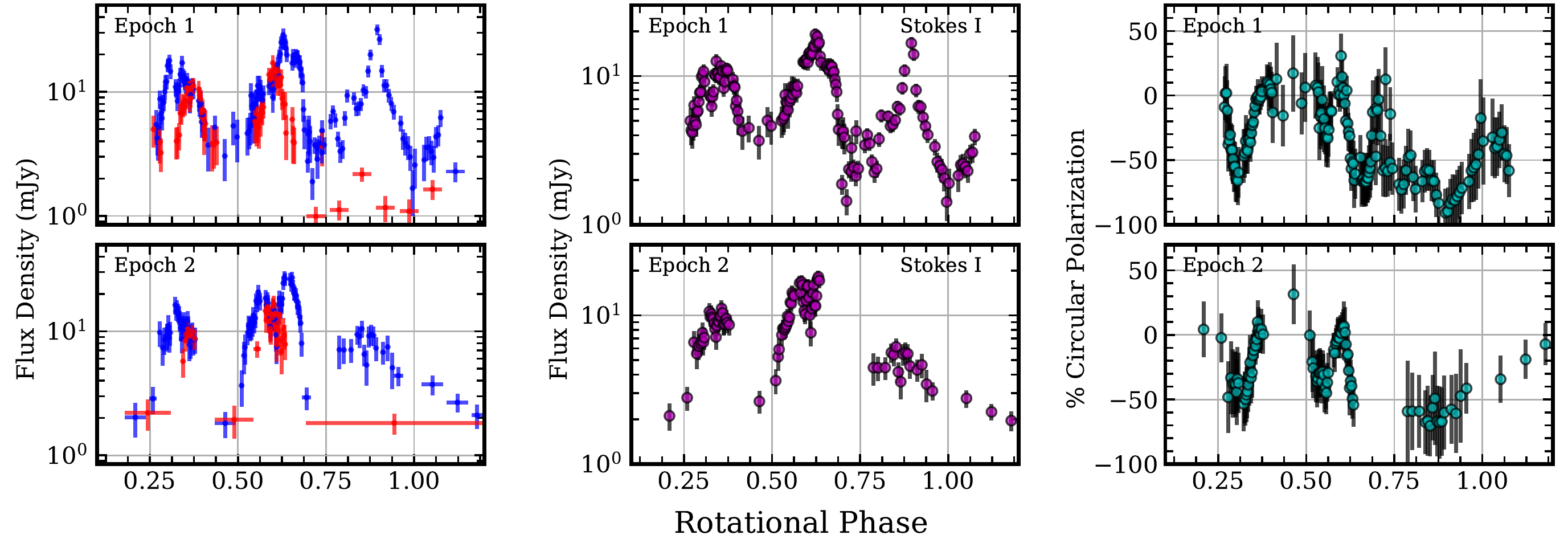}
    \caption{The lightcurves of CU\,Vir in band 3 (279--461 MHz) of the uGMRT obtained during epoch 1 (top panels) and epoch 2 (bottom panels). For the definition of epoch 1 and epoch 2, see Table \ref{tab:obs}. The left panels show the lightcurves for RCP (red markers) and LCP (blue markers), which are in accordance with the IEEE convention. The middle panel shows the same for the total intensity and the right panel show the lightcurves for the percentage circular polarization.}
    \label{fig:band3_epoch1_2}
\end{figure*}

\begin{figure*}
    \centering
    \includegraphics[width=0.4\textwidth]{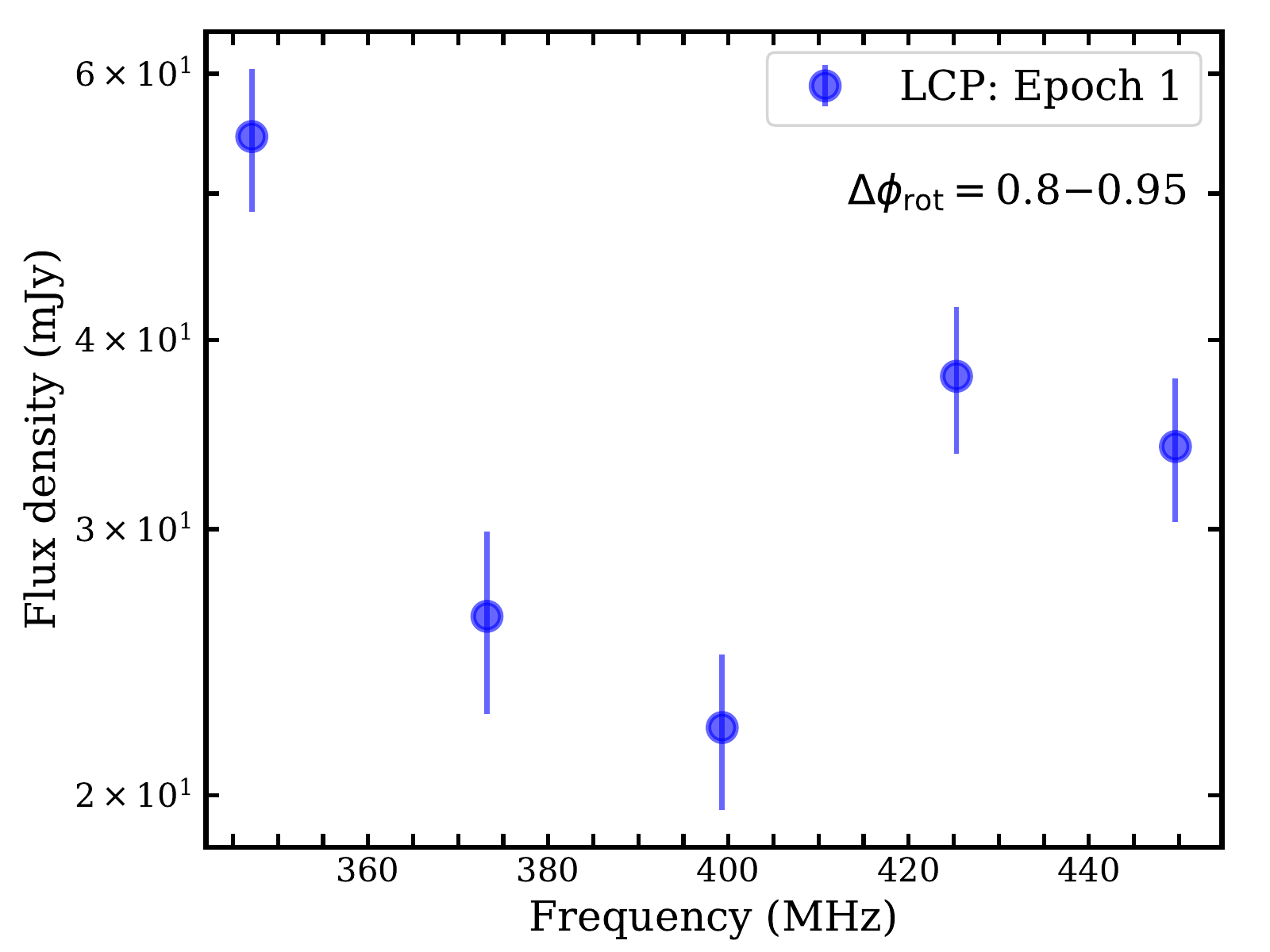}
    \includegraphics[width=0.45\textwidth]{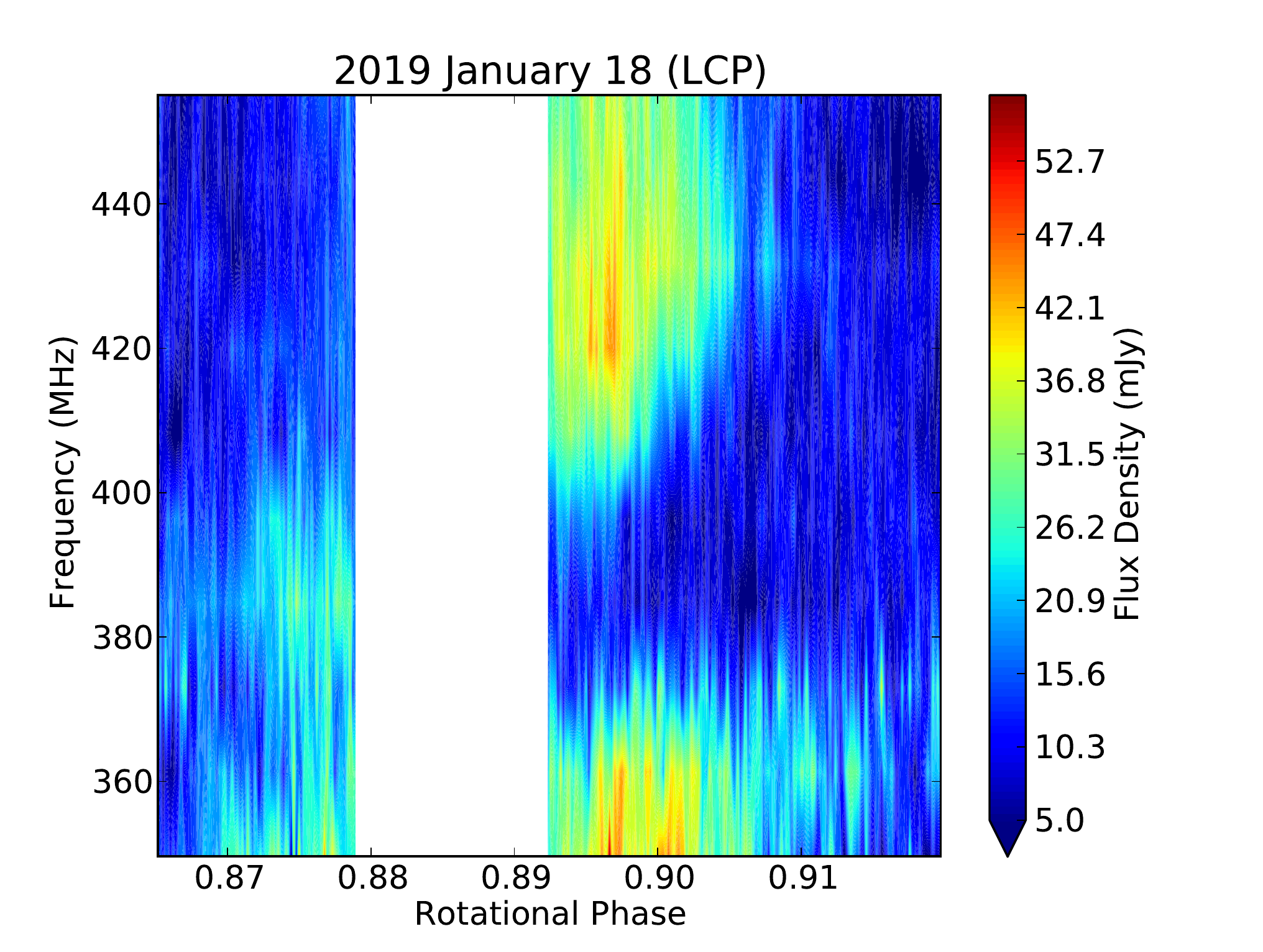}
    \caption{\textit{Left:} The spectrum corresponding to the peak flux density of the LCP enhancement observed on 2019 January 18 (epoch 1) over 0.80--0.95 rotation phases. No RCP enhancement is present in this rotational phase range. The averaging time for each point is 2 minutes. \textit{Right:} The dynamic spectrum for the pulse, frequency resolution is 12 MHz, time resolution is 8 second.}
    \label{fig:spectrum_18jan2019}
\end{figure*}

The lightcurves in band 3 (330--461 MHz) obtained at the two epochs are shown in Figure \ref{fig:band3_epoch1_2}. At each epoch, the full rotational phase coverage was obtained by observing on two days, with very little overlap in rotational phase. 

Similar to what we found for band 4, here also the lightcurves consist of enhancements in both circular polarizations. Interestingly, all of these features appear to be present at both epochs implying that all of them are persistent structures of the lightcurves. Regarding the time evolution of the height of the enhancements, we find that for the inner two sets of RCP and LCP pulses (those lying over 0.2--0.7 rotation cycle), the pulse strength does not evolve significantly with time, indicating a highly stable underlying phenomenon.

For the remaining enhancement (between 0.75--1.10 rotational phase), that is entirely LCP, we find that the pulse at epoch 2 is much weaker than that at epoch 1. Upon examining the spectrum of the maximum observed flux density (using the epoch 1 data only), we find that the peak flux density first decreases upto a certain frequency, and then increases with further increase in frequency (left of Figure \ref{fig:spectrum_18jan2019}). To investigate it further, we extracted the dynamic spectrum around the peak of the enhancement (right of left of Figure \ref{fig:spectrum_18jan2019}).
By examining the dynamic spectrum (flux density on the time-frequency plane), we find that this enhancement is actually the superposition of two enhancements. Between 0.89--0.91 rotational phases, there are two enhancements occurring at the same time, but over disconnected range of frequencies. We could not examine the dynamic spectrum for the enhancement observed at epoch 2 since the signal there is much weaker and the data were much noisier. Thus it is not clear whether both enhancements seen at epoch 1 have counterparts at epoch 2 or not.

Interestingly, one of the non-persistent band 4 enhancements (LCP) was observed over the rotational phase range of 0.88--0.98 rotation cycle (Figure \ref{fig:flare_lightcurve_6_7jan2019}) which is roughly the same over which the LCP enhancement under consideration was observed in band 3. Current observations are however inadequate to comment on whether there is any connection between the two enhancements.

Thus, the difference between the strength of the enhancements over 0.7--1.1 cycle observed at the two epochs in band 3 could either be due to intrinsic variability of the pulse-height, or due to the occurrence of a flare at epoch 1 over the same range of rotation cycles in addition to the persistent pulse.

To summarize, the results in band 3 are similar to those obtained in band 4 in the sense that in both bands, enhancements are observed in both RCP and LCP, and these are not confined to the rotational phases where the GHz ECME pulses are seen. However the properties at the two bands differ in two important aspects. Unlike band 4, no non-persistent enhancement was observed at band 3. Secondly, the pulses observed between 0.2--0.75 rotational phases appear to be extremely stable (in terms of pulse-height) in band 3, which is not the case in band 4.



\section{CU\,Vir over 0.4--4 GHz}\label{sec:combined_results}
\begin{figure*}
    \centering
    \includegraphics[width=0.9\textwidth]{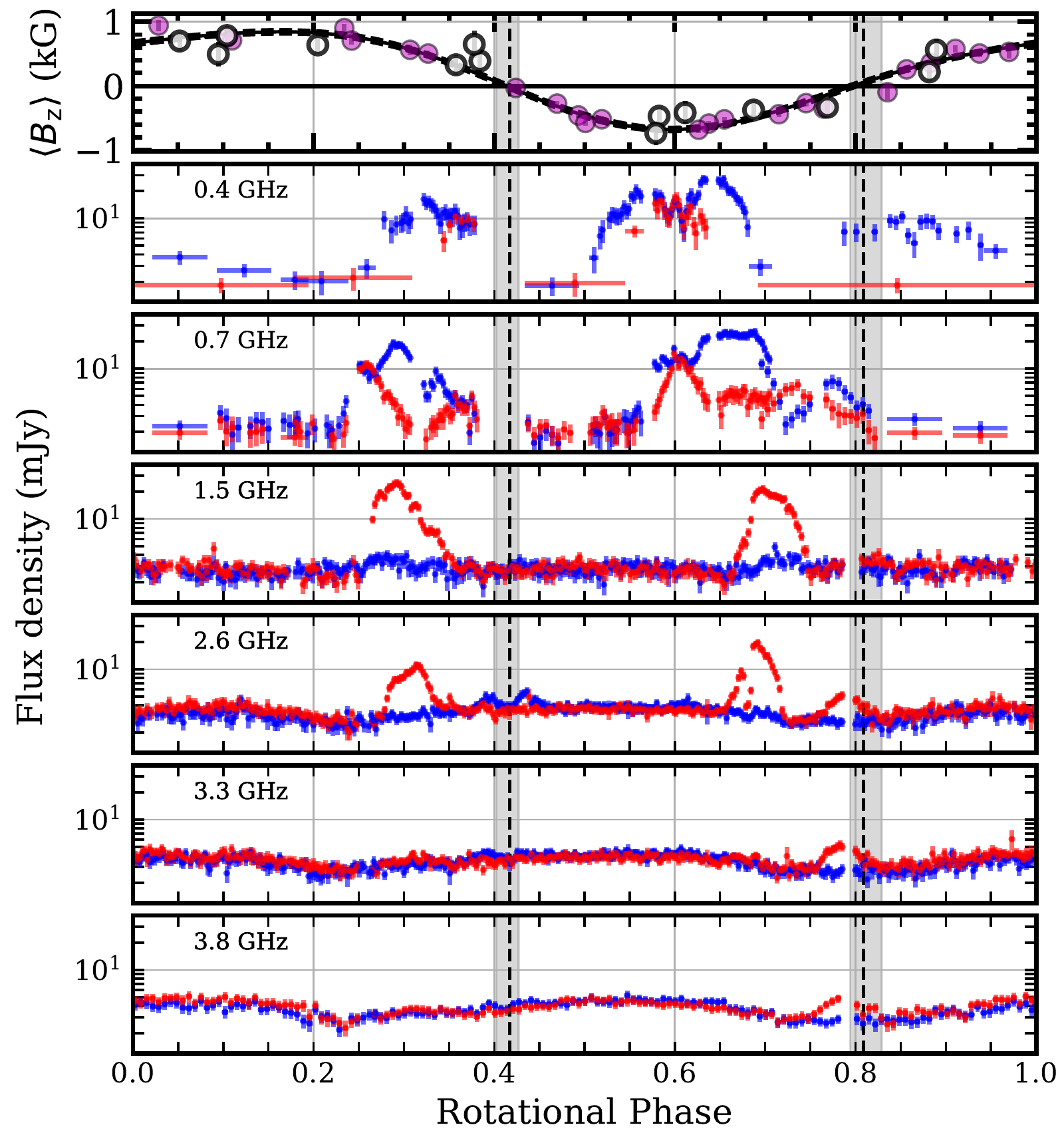}   
    \caption{The combined uGMRT lightcurves at epoch 2 and the higher frequency lightcurves along with the rotational modulation of the stellar \bz. The vertical dashed lines and the grey shaded regions around them represent the nulls of the \bz and their errorbars. The colored arrows are used to mark the pulses that are very likely counterparts to one another. Pink arrows are used for LCP and yellow arrows are used for RCP.}
    \label{fig:band3_band4_L_S}
\end{figure*}

In Figure \ref{fig:band3_band4_L_S}, we show the lightcurves for one complete rotation cycle of CU\,Vir at different frequency bands along with the stellar \bz. For the lightcurves at sub-GHz frequencies, we have used the epoch 2 data which does not have any non-persistent features. It can be clearly seen that while at GHz frequencies, the pulses are confined to a relatively narrow rotational phase range, they are much more spread out at sub-GHz frequencies. We also do not find the pulses to lie around the magnetic nulls (shown by the vertical dashed lines in Figure \ref{fig:band3_band4_L_S}) of the photospheric \bz. This is probably due to the fact that the \bz~at the height of the radio emission and the photospheric \bz~are not identical due to the contribution from the higher order magnetic moments.

\begin{figure*}
    \centering
    \includegraphics[width=0.45\textwidth]{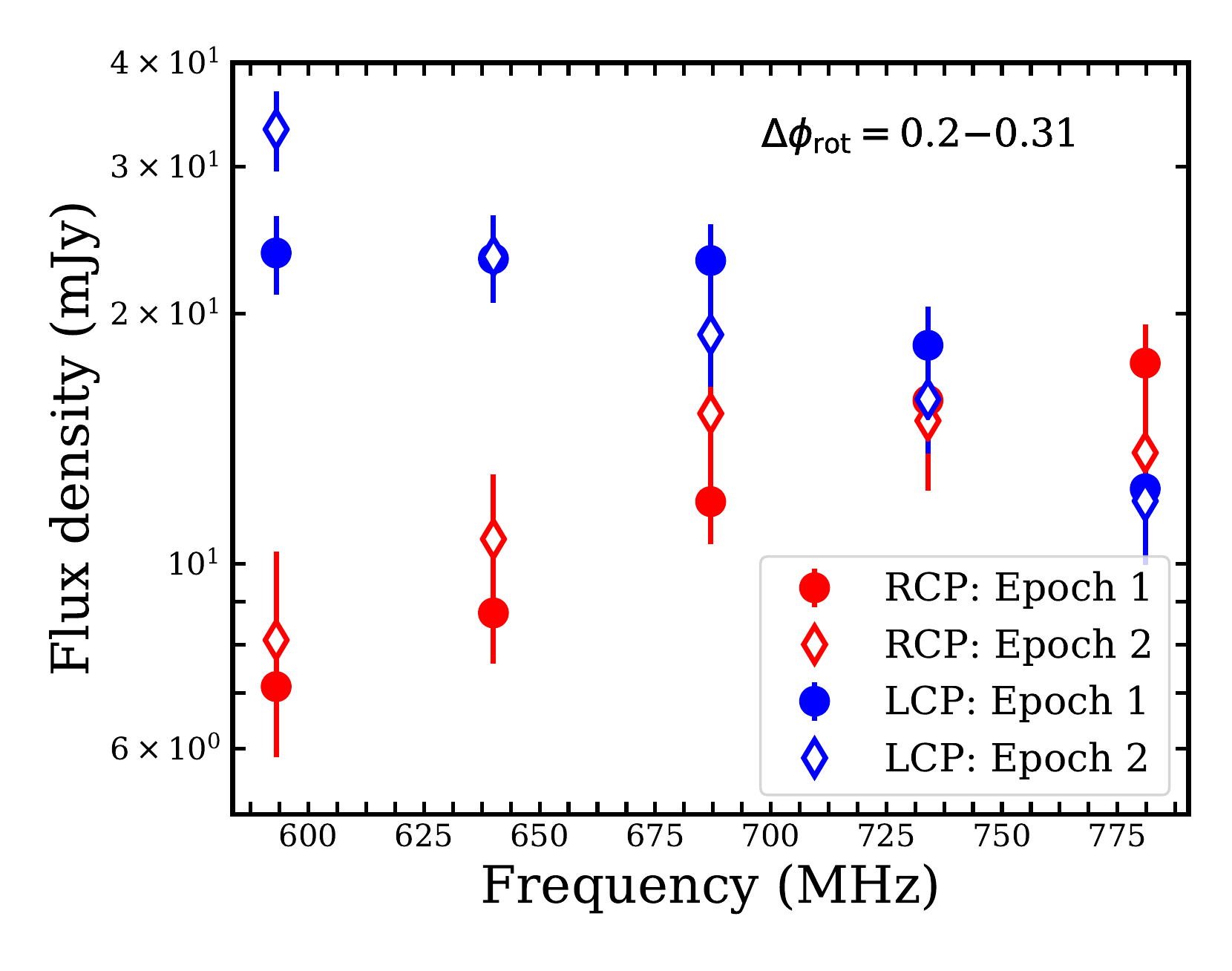}
    \includegraphics[width=0.46\textwidth]{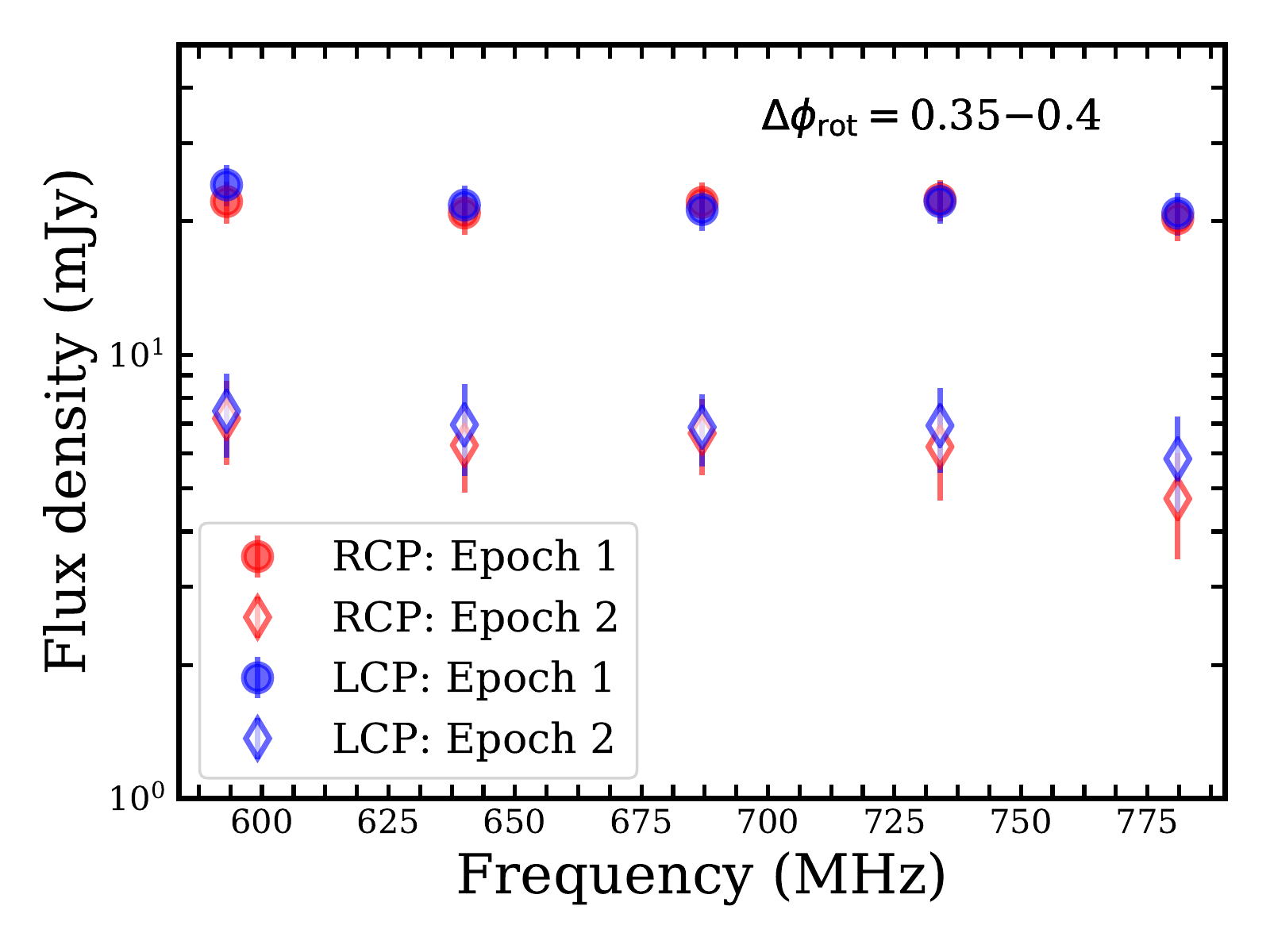}
    \includegraphics[width=0.43\textwidth]{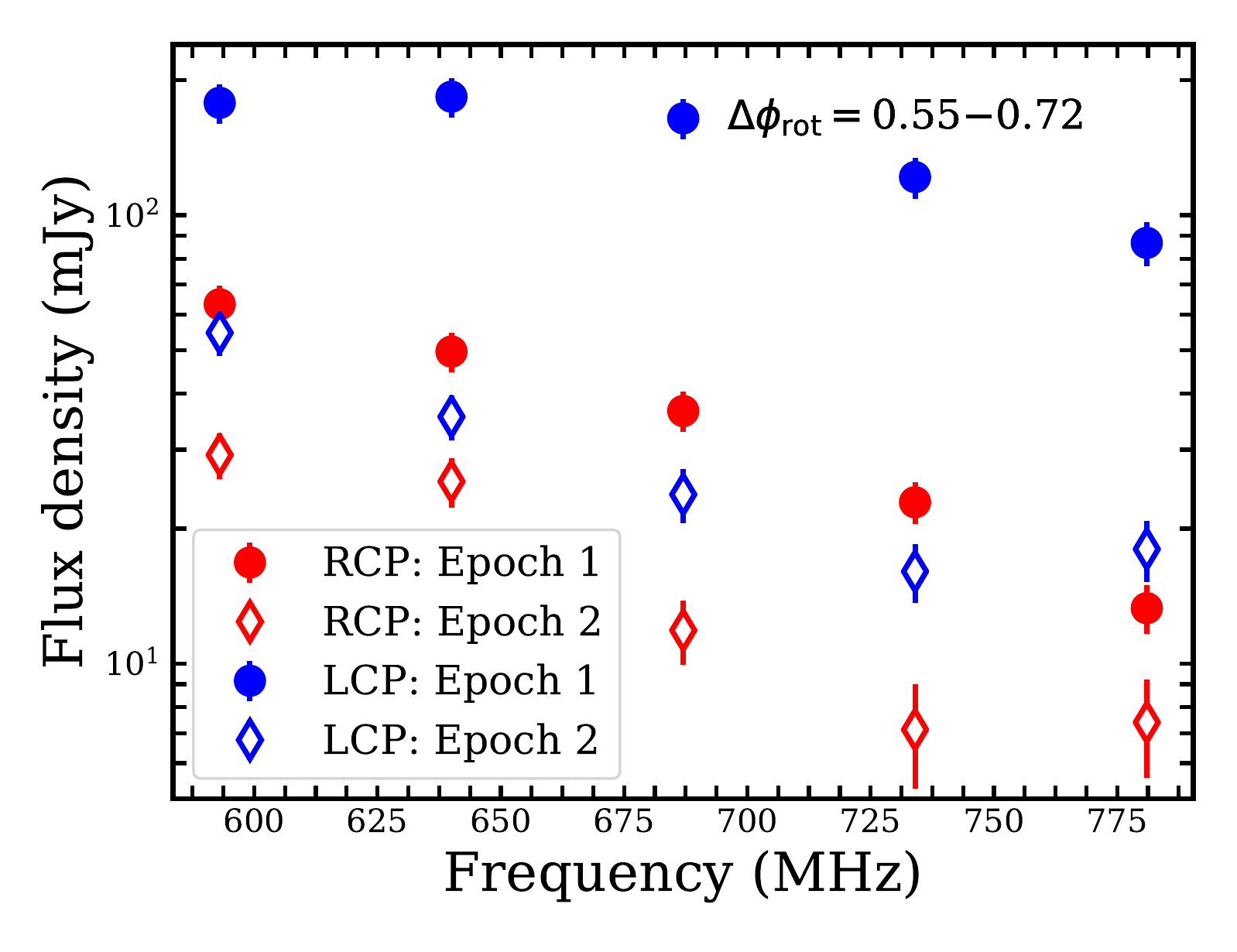}
    \includegraphics[width=0.45\textwidth]{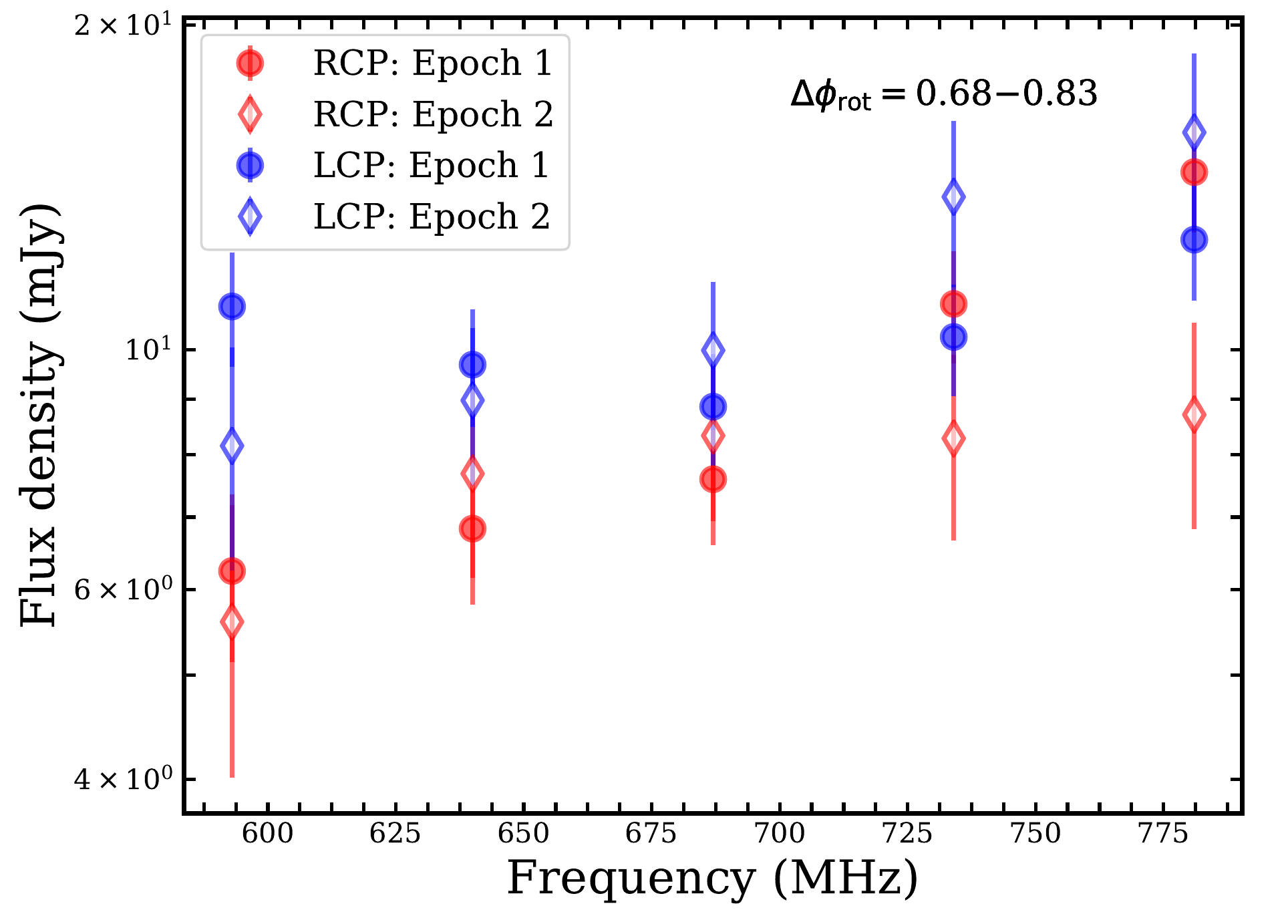}
    \caption{\textit{Top feft:} The spectra over 550--800 MHz corresponding to the peak flux densities of the persistent RCP and LCP enhancement observed from CU\,Vir between 0.20--0.32 rotation cycle on 2019 January 4 (epoch 1) and 2020 June 13 (epoch 2). \textit{Top right:} The spectra corresponding to the maximum flux densities observed between 0.35--0.40 rotation cycles at the two epochs. These data were acquired on 2019 January 4 (epoch 1) and 2020 June 13 (epoch 2). \textit{Bottom left:} Peak flux density spectra of the persistent RCP and LCP pulses observed between 0.55--0.72 rotational cycle on 2019 January 13 (epoch 1) and 2020 June 13 (epoch 2). \textit{Bottom right:} Peak flux density spectra for the pulses observed between 0.68--0.83. Epoch 1 correspond to data acquired on 2019 January 7 and epoch 2 corresponds to data acquired on 2020 June 7. The averaging time for each point is 2 minutes. RCP and LCP are represented by red and blue markers respectively.}
    \label{fig:peak_spectra_band4_persistent_pulses}
\end{figure*}

The goal of this section is to connect the pulses observed at different frequencies to derive a more comprehensive picture. 
In order to achieve that, we examine the intraband spectra of the persistent pulses in band 4, since this is the waveband that bridges the previously unexplored sub-GHz behaviour with the relatively well-characterized properties of the star at GHz frequencies. Note that due to the highly directed nature of the mechanism, the observed spectrum also depends on the visibility conditions of the emission beam at different frequencies, and hence does not necessarily reflect the intrinsic levels of emission. Nevertheless, as will be shown subsequently, these spectra become important while connecting results obtained at different wavebands. Figure \ref{fig:peak_spectra_band4_persistent_pulses} shows the observed spectra of the persistent pulses in band 4. It clearly demonstrates that though the pulses themselves are persistent, their spectra may vary with time. The most robust property of these spectra appears to be the sign of the spectral indices which remain unchanged. 
Table \ref{tab:band4_pers_spectral_properties} lists the spectral indices of these pulses using data of epoch 1.

\begin{deluxetable}{cc|c|c}
\tablecaption{The intraband spectral indices $\alpha$ ($S\propto \nu^\alpha$, with $S$ being the flux density at a frequency $\nu$) of the persistent pulses observed over 570--804 MHz (band 4 of the uGMRT) at different rotational phase ($\Delta\phi_\mathrm{rot}$) ranges. Note that the calculations are based on data taken at epoch 1 (Table \ref{tab:obs}). The last column indicates whether the spectra observed at the two epochs are similar (in terms of both spectral indices and flux densities) or not. For details see \S \ref{subsubsec:band4_persistent_pulses}.\label{tab:band4_pers_spectral_properties}}
\tablehead{
\hline
$\Delta \phi_\mathrm{rot}$ & \multicolumn{2}{c}{$\alpha$} &  comment\\
\hline
&  RCP &  LCP & 
}
\startdata
\hline
0.20--0.31 & $3\pm1$ & $-5\pm1$ & \\
0.35--0.40 & 0 & 0 & \\ 
0.55--0.72\tablenotemark{a} & $-8\pm1$ & $-5\pm1$ & Variable spectra\\
0.68--0.83 & $5\pm1$\tablenotemark{a} & 0 & Variable spectra \\
\hline
\enddata
\tablenotetext{a}{ Between 687--781 MHz}
\end{deluxetable}

In the following subsections, we first connect the results obtained in band 3 and 4, followed by comparison of the results obtained in band 4 and those at GHz frequencies. In the next section, we will discuss the coherent picture that emerges from all these results.

\subsection{Connecting band 3 (0.4 GHz) with band 4 (0.7 GHz)}\label{subsec:connection_band3_4}
\begin{figure}
    \centering
    \includegraphics[width=0.45\textwidth]{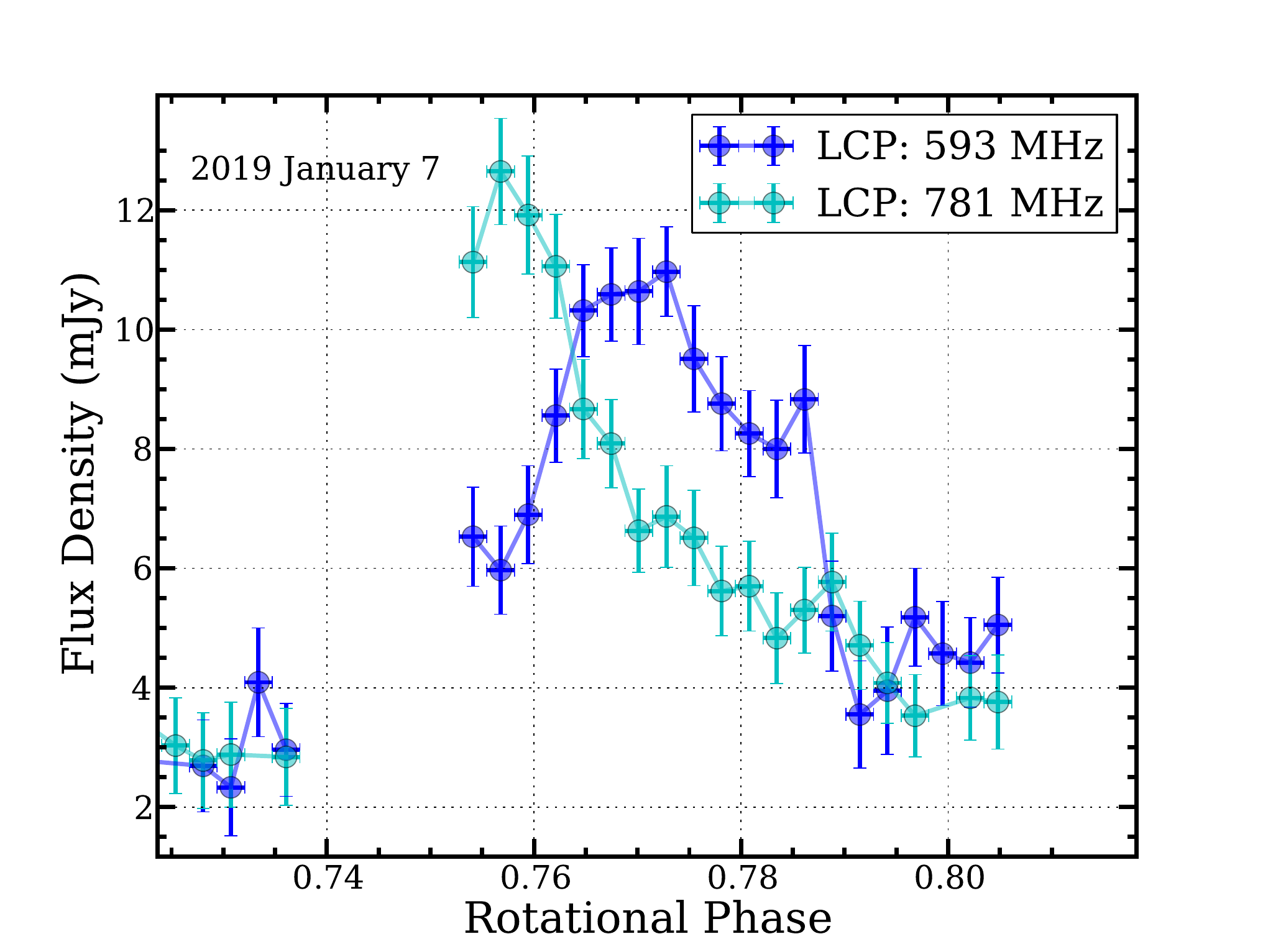}
    \caption{Figure showing the drift of the LCP pulse at 0.7 GHz over 0.72--0.82 cycle within the frequency range 593--804 MHz. The data corresponds to epoch 1 (Table \ref{tab:obs}).}
    \label{fig:2_freq_7jan2019_LCP}
\end{figure}

Upon comparing the second and third panels of Figure \ref{fig:band3_band4_L_S}, we find that the persistent LCP pulse observed over 0.2--0.4 rotation phase at 0.4 GHz (band 3) and 0.7 GHz (band 4) are likely the counterparts to each other. This association implies that this pulse drifts from high to low frequencies with time since the onset of the pulse in band 4 occurs at $\approx 0.04\pm0.01$ cycle ahead of that in band 3. The approximate drift rate is -$161\pm 40$ kHz/s. Between band 3 and band 4, the spectrum corresponding to the peak flux density is flat. 

There are two more distinct LCP pulses at 0.4 GHz (second panel of Figure \ref{fig:band3_band4_L_S}): over 0.50--0.70 cycles and over 0.75--1.10 cycles. One possibility is that these two are respectively the counterparts of the LCP pulses seen over 0.55--0.72 cycles and 0.72--0.82 cycles at 0.7 GHz (third panel of Figure \ref{fig:band3_band4_L_S}). If this association holds, it implies that the LCP pulse over 0.5--0.7 cycle (at 0.4 GHz) drifts at a rate of $129\pm 26$ kHz/s.
In that case, we find that the two LCP pulses of band 3, lying between 0.2--0.4 and 0.5--0.7 rotational phases, drift with the same rate on the time-frequency plane, but in opposite directions.
Using the epoch 2 data, where we employed subarray mode observation, the spectral index for the peak flux density of the LCP pulse between 0.55--0.72 comes out to be zero within error bars, again a characteristic similar to the LCP pulse between 0.2--0.4 rotational phases.

The other association between the LCP pulses over 0.75--1.10 cycle at 0.4 GHz and 0.72--0.82 cycle at 0.7 GHz, implies a significantly slower drift rate. In such a case, we would expect to see the drift of the pulse within a single frequency band as well. Unfortunately, at epoch 1, where the sensitivity is much higher, the corresponding LCP pulse at 0.4 GHz is contaminated by a transient burst candidate (\S\ref{subsec:band3}). We hence examine the intraband drift at 0.7 GHz at epoch 1. In Figure \ref{fig:2_freq_7jan2019_LCP}, we show the LCP pulses at 593 MHz and 781 MHz over 0.72--0.82 rotation cycle (as observed on 2019 January 7). It clearly shows that the pulse drifts to lower frequencies as time progresses. This is consistent with the fact that the LCP pulse over 0.72--0.82 rotation cycle at 0.7 GHz is the counterpart of the LCP pulse over 0.75--1.10 rotation cycle at 0.4 GHz. In this case, the spectral index between band 3 and band 4 (using the epoch 2 data) comes out to be $-0.6\pm0.4$.

The number of RCP pulses at 0.4 GHz is two per rotation cycle (second panel of Figure \ref{fig:band3_band4_L_S}). For the one over 0.3--0.4 cycle, the most likely counterpart at 0.7 GHz is the one that lies over the same range of rotation cycle (third panel of Figure \ref{fig:band3_band4_L_S}). The adjacent RCP pulse over 0.2--0.3 cycle at 0.7 GHz is ruled out since it shows a steep decline in flux density below 0.69 GHz at both epochs (top left panel of Figure \ref{fig:peak_spectra_band4_persistent_pulses}), making it unlikely to give rise to an enhancement with a peak flux density of 10 mJy at 0.4 GHz. 
From the epoch 2 data (where observations at the two bands were recorded simultaneously employing subarray mode), we find that this RCP pulse (between 0.3--0.4 rotational phase) exhibits a spectrum with a spectral index of $\approx -1.3\pm0.4$ (between band 3 and band 4) corresponding to the peak flux density. 

Using an argument similar to the one used in the preceding paragraph, we can rule out the RCP pulse over 0.70--0.85 cycle at 0.7 GHz for being the counterpart of the RCP pulse at 0.4 GHz observed over 0.55--0.65 rotation cycle (see bottom right panel of Figure \ref{fig:peak_spectra_band4_persistent_pulses} for the spectrum of the RCP pulse over 0.70--0.85 cycle at the band 4 frequency range centred at 0.7 GHz). Thus the most probable counterpart for the RCP pulse over 0.55--0.65 cycle at 0.4 GHz is the one observed over 0.55--0.70 rotation cycle at 0.7 GHz. 
Using the data from epoch 2, we find the spectrum (for the peak flux density) to be flat between 0.4 and 0.7 GHz.

Thus the only pulses that do not have counterparts at both frequency bands are the two RCP pulses at 0.7 GHz, seen over 0.2--0.3 cycle and 0.7--0.8 rotation cycle.

\subsection{Connecting band 4 (0.7 GHz) with GHz frequencies}\label{subsec:connection_band4_GHz}
Here we attempt to identify counterparts of the ECME pulses that are observed at frequencies $\gtrsim1$ GHz. The RCP ECME pulse observed around 0.8 rotational phase (bottom three panels of Figure \ref{fig:band3_band4_L_S}) will not be considered here as we already showed that its lower cut-off frequency lies around 2.3 GHz (\S\ref{subsec:gyrosyncrotron_modulation}).

From Figure \ref{fig:band3_band4_L_S}, one can readily see that there are multiple candidates for counterparts of higher-frequency ECME pulses at 0.7 GHz. We assume that the circular polarization of ECME pulses does not change with frequency \citep[this need not be true,][]{das2019a}. 
Let us consider the RCP pulse observed between 0.2--0.4 rotation cycle at 1--3 GHz. There are two possible counterparts for this pulse at 0.7 GHz. For the latter counterpart at 0.7 GHz (over 0.3--0.4 rotation cycle), we inferred in the preceding subsection that it has a counterpart at 0.4 GHz lying over the same range of rotational phases, which indicates a very high drift rate ($>36$ MHz/s, with 8 second time resolution). In addition, it has a spectral index of -1 between 397--687 MHz (\S \ref{subsec:connection_band3_4}). Both these facts make it unlikely to be the counterpart of the high-frequency ECME pulses. The other RCP pulse (over 0.2--0.3 rotation cycle, second panel of Figure \ref{fig:band3_band4_L_S}) has a positive spectral index (Table \ref{tab:band4_pers_spectral_properties}). It also arrives ahead of the pulses at higher frequencies, thus preserves the sequence of arrival of the ECME pulses seen between 1--3 GHz. Therefore, the RCP pulse observed between 0.2--0.3 rotational phases at 0.7 GHz (band 4) is the most likely counterpart of the ECME pulse of same polarization observed between 0.2--0.4 rotation cycle above 1 GHz. 

We next consider the RCP ECME pulse that is observed between 0.6--0.8 rotation cycle at higher frequencies. Again there are two possible counterparts at 687 MHz (band 4): the RCP pulse over 0.55--0.65 rotation cycle and the RCP pulse over 0.7--0.8 rotation cycle (third panel of Figure \ref{fig:band3_band4_L_S}). The first candidate, i.e. the one over 0.55--0.65 rotation cycle is unlikely to be the counterpart since, it has been found to exhibit a spectrum with steeply falling flux densities with increasing frequencies (Figure \ref{fig:peak_spectra_band4_persistent_pulses}, Table \ref{tab:band4_pers_spectral_properties}). 
The RCP pulse over 0.7--0.8 rotation cycle shows nearly flat spectrum, with a hint of a positive spectral index at the higher end of band 4 (Figure \ref{fig:peak_spectra_band4_persistent_pulses}, Table \ref{tab:band4_pers_spectral_properties}). Thus this pulse is favoured over the preceding RCP pulse for being the counterparts of ECME seen at higher frequencies.

If our association is correct, it will imply that the lower cut-off frequency corresponding to the higher frequency RCP ECME pulses lies between the frequency range of band 3 and band 4, i.e. over 461--570 MHz, since none of the two RCP pulse at 687 MHz which are the likely counterparts of higher frequency ECME, has any counterpart at band 3 (\S \ref{subsec:connection_band3_4}).

Similar to the RCP pulse, the weak LCP pulse observed at GHz frequencies also have multiple possible counterparts at sub-GHz frequencies. Two of the persistent LCP pulses in band 4 exhibit very steep spectra and the other two exhibit flat spectra (Table \ref{tab:band4_pers_spectral_properties}). The pulses with steep spectra are unlikely to be of strength $\sim 10$ mJy at 1 GHz unless we consider the super-bright pulse observed on 2019 January 13 (top panel of Figure \ref{fig:giant_pulse}), or variable spectral index at different epochs/different frequency ranges. On the other hand, either of the two LCP pulses (in band 4) with nearly flat spectra (Figure \ref{fig:peak_spectra_band4_persistent_pulses}, Table \ref{tab:band4_pers_spectral_properties}) can give rise to the enhancement seen at 1 GHz. The drift direction within the L band, that would have allowed us to predict the more likely counterparts at lower frequencies, could not be determined due to the weak strength of the pulse. We thus cannot pinpoint the most-likely counterparts of the L band LCP pulse at 0.7 GHz. As will be explained in the following section, all the LCP enhancements seen at band 4 are probably connected to the LCP enhancements seen at GHz frequencies.

\section{A combined look at the star}\label{sec:all_data_combined}
From the last section, the immediate picture that emerges is that CU\,Vir produces `regular' ECME of only one circular polarization (RCP, that corresponds to only one of the magnetic poles) over the frequency range of $\lessapprox 593$ MHz till $\approx 3$ GHz ($\gtrsim 5\,\mathrm{GHz}$ if we also include the ECME pulse at 0.8 phase). Between 0.8--1 GHz, some other mechanism turns on that gives rise to the persistent LCP and RCP pulses that are observable down to at least 360 MHz. The properties of these pulses (peak flux density, percentage circular polarization, pulse-width etc.) that are exclusively observable at sub-GHz frequencies, are similar to those of the pulses identified as the ECME counterparts at sub-GHz frequencies. Hence if the latter are of coherent origin, it is most likely that the same is the case for the former. In that case, the most favourable mechanism to produce such regular pulses is ECME. 
This picture thus requires two different channels to produce ECME in the star, which is hard to reconcile with the existing scenario regarding the magnetosphere around hot magnetic stars.

We can reconcile the above facts if we consider the aforementioned characteristics to reflect the frequency dependence of ECME pulse-profile, and polarization dependence of ECME cut-off frequencies. In \citet{das2020b}, it was reported that the upper cut-off frequencies of the RCP and LCP pulses observed from the hot magnetic star HD\,133880 are different (2 GHz and 3.2 GHz for RCP and LCP respectively near null 1). The situation here is similar to that, but the difference is much higher. For the LCP pulses, the upper cut-off frequency is around 1.5 GHz, whereas for the RCP pulses (including the one at 0.8 phase), the upper cut-off frequency is $\gtrsim 5$ GHz. In this context, we would like to mention that the field strength of the two magnetic poles of CU\,Vir differ by a factor of 2--4 \citep{kochukhov2014}, which suggests that the different upper cut-off frequencies for the RCP and LCP pulses can be a manifestation of the asymmetry between the two magnetic hemispheres in terms of polar field strength. Note that this assumes that the RCP pulse, that has a higher upper cut-off frequency, is produced at the south magnetic polar regions, which is the one with stronger field strength, and vice-versa \citep[ordinary mode emission, e.g. ][]{leto2019}. This explanation is unsatisfactory in one aspect, which is that the star HD\,133880 has an even higher difference between the maximum strength of its two magnetic hemispheres \citep{kochukhov2017}, but the upper cut-off frequencies of RCP and LCP pulses differ by a factor smaller than that observed in CU\,Vir. It is however to be noted that the upper cut-off frequencies for both stars are smaller than the electron-gyrofrequencies corresponding to the maximum stellar magnetic fields, and hence the premature cut-off has been hypothesized to be produced due to the presence of very high density plasma close to the stellar surface \citep{leto2019}. This simply implies that the high-density plasma cloud surrounding the star is not spherically symmetric, which is not unexpected given that these stars do not have an axi-symmetric dipole, and/or aligned magnetic-rotation axes.

The different upper cut-off frequencies for the pulses coming from opposite magnetic hemispheres, can also be related to the star's non-dipolar magnetic field, the effect of which is stronger at higher frequencies (radiation produced closer to the star). Due to this effect, the intrinsic direction of emission could be significantly different for the radiation produced at opposite magnetic polar regions, and one of them could be such that the emission beam never crosses the line of sight up to a certain frequency. At sub-GHz frequencies, the magnetic field of CU\,Vir is probably very close to being a dipole, and hence we can observe pulses of both circular polarization as is expected for a star with dipolar magnetic field. Note that the difference in the intrinsic direction of emission will also affect the rotational phase offsets between pulses at different frequencies.

The frequency dependence of ECME pulse-profile has also been observed in HD\,133880 at a much milder level \citep{das2020b}, and in HD\,142990 (Das et al., in preparation). Besides, \citet{das2020a} showed via simulation that due to propagation effects in the inner magnetosphere of the star, secondary pulses may appear that are visible only over a small frequency window \citep[as compared to the total bandwidth of ECME, see Figure 10 and 11 of ][]{das2020a}. This can be understood by remembering that the observed ECME pulse at a given frequency get contribution from different parts (that can be marked by a unique magnetic azimuth) of the relevant auroral ring above the magnetic poles \citep[see Figure 4 of][]{trigilio2011}. The observed pulse-profile depends on the fractions of the auroral ring that contribute to the pulse and the differences in the directions of the contributing radiation once it is outside the dense stellar magnetosphere \citep[e.g.][]{das2020a}. The pulse-profile will be broad, and in the extreme case, will consist of separate sub-pulses if the difference between the ray direction corresponding to different parts of the contributing magnetic azimuths is too high. This effect is different at different frequencies since different frequencies traverse through different parts of stellar magnetosphere and hence may experience very different kinds of density profiles \citep[e.g.][]{das2020a}. Thus our observation of strong frequency dependence of ECME pulse-profiles suggests that the magnetosphere of CU\,Vir is filled with plasma that has strong density gradients \citep[stronger than a $1/r$ profile assumed for this star, ][]{leto2006} in radial as well as in azimuthal direction (especially because the properties near the two nulls are not identical).


Last but not the least, we discuss a scenario, which we deemed unlikely at the very beginning of this section based on the existing ideas of hot star magnetospheres, but definitely a possibility in view of latest wide-band measurements. As per this scenario, the ECME observed at GHz and sub-GHz frequencies correspond to two different types of `engines'. For ECME observed from planets, in-situ observations proved that there can be multiple channels for production of auroral emission at the same time, \citep[e.g. ECME from Jupiter,][etc.]{Ladreiter1992,zarka2004}. Similar suggestions have been made to explain the auroral radio emission properties of some UCDs \citep[e.g.][]{leto2017,pineda2017}. In almost all such scenarios, one of the channels is satellite-induced ECME \citep[like the Jupitor-Io case,][]{bigg1964}. The consequence of such an interaction is that a stellar magnetic `flux-tube' passing through the satellite participate in the production of ECME (instead of having all the field lines at a given magnetic latitude, like those shown in Figure \ref{fig:auroral_rings}). \citet{kuznetsov2012} also considered the scenario where ECME is produced along a fixed set (sets) of magnetic field lines, called `active longitudes'. In case of hot magnetic stars, the observed ECME characteristics, until now, had been qualitatively consistent with the emission arising in auroral rings formed in a large-scale dipolar magnetosphere. CU\,Vir thus becomes the first star for which the possibility of multi-channel ECME is considered. In that case, the different components of ECME are linked to magnetic field lines anchored at different stellar magnetic latitudes. Given that the magnetic field topology of CU\,Vir is not a simple dipole, the field topology is likely to vary as a function of magnetic latitudes, which will then result into difference in the intrinsic direction of emission. This will cause the rotational phase offsets between different ECME components in the lightcurves. Moreover, since in cases like that involving active longitudes, ECME is directed along the surface of a hollow cone centred at the field lines (belonging to the group of active longitudes), the observed ECME lightcurve can be very different from that expected for ECME directed tangential to the auroral rings \citep[e.g.][]{kuznetsov2012}. This provides an explanation to the unusual sub-GHz features without the requirement of significant propagation effects. Moreover, the hollow-cone beaming pattern also influences the spectral cut-off as recently discussed by \citet{davis2021}.

With the current data, we cannot distinguish between the different scenarios that can be responsible behind the unique ECME characteristics of CU\,Vir. One way to acquire more insight in this regard will be to observe more MRPs over a wide range of frequencies spanning their full rotation cycles so as to examine whether CU\,Vir is a unique case. If we find that the seemingly `anomalous' behavior is actually common among MRPs (may be at different frequencies), propagation effects or/and difference in the intrinsic direction of emission, will be a more favoured candidate since it is a more general phenomenon than either the presence of a satellite, or set(s) of active longitudes. On the other hand, if CU Vir remains unique in this aspect, a more exotic explanation, like the contribution from active longitudes, or a satellite-induced ECME, will have to be considered.

Since our observation suggests that propagation effects might play an important role in shaping the lightcurves at different frequencies, it remains no longer straight-forward to associate the ECME pulses to the different magnetic polar regions \citep[e.g. see the third panels of Figure 8 and 10 in][]{das2020a}. Thus despite detecting pulses of both circular polarizations, we are not able to infer whether the emission is in extra-ordinary or in ordinary mode. 

\section{The transient phenomena observed from CU\,Vir}\label{sec:transients}
\begin{figure}
    \centering
    \includegraphics[trim={0.5cm 7cm 0.5cm 2cm}, clip, width=0.4\textwidth]{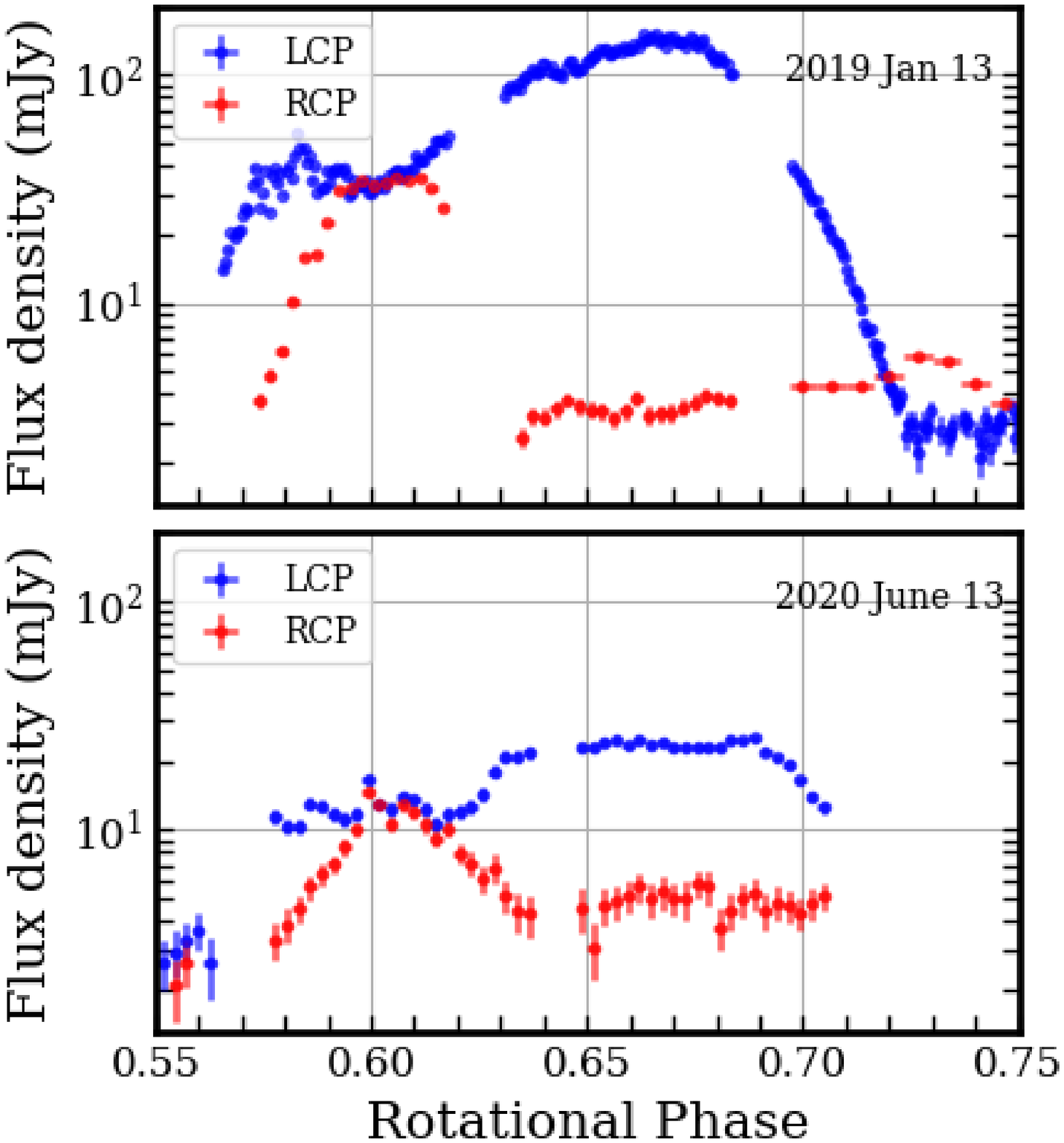}
    \caption{The giant LCP pulse (upper panel) observed at 687 MHz on 2019 January 13. For comparison, the pulses observed over the same rotational phase range at epoch 2 are also shown in the bottom panel.}
    \label{fig:giant_pulse}
\end{figure}

One of the biggest surprises that we obtained from our wideband study is observation of seemingly transient phenomena from this star that is known to have a highly stable magnetic field. At first sight, the only transient event that we observe is the non-persistent features observed in band 4, and possibly also in band 3 (\S\ref{sec:results_low_freq}). However, we also witness another event that decidedly belongs to the category of transient phenomenon. In the next subsection, we describe this event. This will be followed by a discussion about the cause behind the non-persistent low frequency events.

\subsection{A `giant pulse'}\label{subsec:a_giant_pulse}
This event, under consideration, is the observation of an LCP enhancement on 2019 January 13 at 687 MHz, that is approximately an order of magnitude brighter than the typical pulses seen from this star at any radio frequencies (top panel of Figure \ref{fig:giant_pulse}). Although the corresponding rotational phases were covered on only three days, we still suggest that such an enhancement is rare, considering that even if we combine the observation of all the other pulses that we observed, and also the pulses observed at higher frequencies, there is no other report of observation of a $>100$ mJy pulse from CU\,Vir.
By taking the analogy from pulsars, we name this anomalously bright pulse as `giant pulse'.

Interestingly, not only the LCP pulse was enhanced, but also the RCP pulse immediately preceding it, was enhanced on that day (Figure \ref{fig:giant_pulse}). This suggests that the two pulses, though oppositely circularly polarized, have a common source of energy supply. Since we argued that they are produced by ECME, the oppositely circularly polarized pulses correspond to opposite magnetic hemispheres. In that case, the only thing that is common to pulses produced from both magnetic hemispheres is the current sheet at the middle magnetosphere (where plasma from the two magnetic hemispheres meet) that acts as the energy supplier for the radio emission \citep[e.g.][]{trigilio2004}. Thus, the observed correlated enhancement for pulses produced at opposite magnetic hemisphere points to an event that has happened at the magnetic equatorial region. Such an event can be brought about by `centrifugal breakout' (CBO) of plasma trapped in the inner magnetosphere. Below we provide a brief description regarding CBO and their importance in the magnetosphere of a CU\,Vir like star.

The idea of CBO was conceived as an answer to the question of what happens when the magnetosphere around a hot magnetic star is over-filled with stellar wind plasma \citep[relevant only for stars with centrifugal magnetospheres that allow plasma accumulation, see Figure 2 of][]{petit2013}. \citet{townsend2005} proposed that when the mass of the accumulated plasma exceeds a critical value, the whole magnetosphere will be emptied out via a breakout of plasma. This scenario was however challenged due to the lack of observational evidence of such a violent phenomenon \citep{townsend2013}. Most recently, \citet{shultz2020} showed through study of $\mathrm{H}\alpha$ emission that above a certain luminosity, the CBO scenario is the most favourable mechanism for plasma transport in stars with centrifugal magnetospheres. It has been proposed that the reason for not detecting any signature of it in the various diagnostics is that CBO events happen at all times, but as small-scale events distributed over the magnetic azimuths instead of a single large-scale event \citep{shultz2020,owocki2020}. As the current stellar diagnostics (e.g. $\mathrm{H}\alpha$, photometry) are not sensitive to changes that happen at small-scales (due to spatial averaging), they cannot retain any signature of CBOs. ECME, on the other hand, is a directed emission. Thus the spatial averaging involved is negligible in this case which makes it the only probe that has the potential to provide observational evidence of the existence of CBO. The observed correlated change in height of the ECME pulses coming from opposite magnetic hemispheres could be a manifestation of that. There is another significance of this observation. The study of \citet{shultz2020} does not tell what mechanism is favoured for stars that have centrifugal magnetospheres but do not produce detectable $\mathrm{H}\alpha$ emission, such as CU\,Vir . If ECME is indeed established to be a probe for CBOs, it will also enable us to understand how the magnetospheric plasma escapes in lower luminosity stars that are not bright in $\mathrm{H}\alpha$.

Since CBOs are hypothesized to be confined to small groups of magnetic azimuths at a time, the attribution of the pulse-height variation to the CBO events provide a natural explanation as to why only the RCP pulse preceding the giant pulse was enhanced, but the one succeeding it remain unaffected (compare second and the fourth panels of Figure \ref{fig:band4_7jan13jan_7jun13jun}). This can happen when the magnetic azimuths corresponding to the emission sites of the succeeding RCP pulse are significantly different than those for the giant (LCP) pulse. On the other hand, the emission sites of the giant (LCP) pulse and the preceding RCP pulse are most likely connected by common/neighbouring magnetic azimuths.

In this context, we would like to point out that ECME pulses are generally known to exhibit variable height that has previously been proposed to be due to instability at the emission site \citep{trigilio2011}. In our sub-GHz observation, apart from the giant pulse, we observed variable flux density for only one pair of pulses (among the persistent enhancements) lying between 0.32--0.40 rotation cycle (Figure \ref{fig:band4_4jan13jun}). In that case, we find the lightcurves for the percentage circular polarization identical between the two epochs of observation (right of Figure \ref{fig:band4_4jan13jun}). This suggests that here also the cause of the variation lies at the site common to both magnetic hemispheres, i.e. the magnetic equatorial region. Thus CBOs could be the reason (or, one of the reasons) behind the variable pulse-heights exhibited by ECME from hot magnetic stars. 
The observation of the giant pulse and its companion RCP pulse is a potential signature of a stronger than usual (since giant pulse is likely a rare phenomenon) CBO event. 

\subsection{Non-persistent enhancements observed from CU\,Vir}\label{subsec:flares}
The other type of transients that we observed are the non-persistent enhancements observed at sub-GHz frequencies. 
Here we consider only the confirmed such features.
Besides, the apparently solitary event observed on 2019 January 7 over 0.635--0.645 rotational phases (\S\ref{subsubsec:band4_non_persistent}) will not be considered here as it lies on top of a persistent LCP pulse (top panel of Figure \ref{fig:band4_7jan13jan_7jun13jun}) and it is not possible to disentangle their properties. 

The events under consideration could either be flares or intermittent pulses. The two classes differs in terms of the extent to which one can predict their appearance in the lightcurves. By `flare', we refer to the events that are totally unpredictable in all aspects. But if the events are classified as `intermittent', it means that we can predict the rotational phases of their observation, but cannot predict whether in a given rotation cycle, the pulse will appear or not (i.e. the site of origin is spatially stable, but the phenomenon is temporally unstable). We will consider both possibilities while discussing the emission mechanism behind these events.

For the event observed on 2019 January 6, the highest observed circular polarization is $\approx 90\%$ at the lowest spectral bin, i.e. at 593 MHz (within band), but the fraction drops to around 60\% at the highest spectral bin (781 MHz). The brightness temperature corresponding to the peak flux density ($28$ mJy) comes out to be $> 10^{15}$ K (since the area of emission is unlikely to be as large as the stellar size), indicating a coherent emission mechanism behind. Similarly, for the non-persistent enhancement seen on 2019 January 7 between 0.88--1.0 rotational phases, the maximum observed circular polarization is 67\% and the brightness temperature comes out to be $>10^{14}$ K, again proving that the emission is coherent in nature. 

Before we discuss the emission mechanisms, we would first like to draw attention to certain interesting facts.
The first one is that on both days, the events are composed of multiple `sub-event's that have durations $\sim 4-8$ minutes (for those seen on 2019 January 6) and $\sim 15-30$ minutes (for those seen on 2019 January 7). 
Interestingly, these sub-events are nearly equispaced in time, especially on 2019 January 6 (Figure \ref{fig:flare_lightcurve_6_7jan2019}). Quasi-periodic oscillation in the radio lightcurve was seen from the sun in which the period was associated with the timescale in which magnetic stress is built up and then relieved through reconnection \citep[e.g.][]{mohan2019}.

Another important point to note here is the rotational phase of arrival of the events. They were observed at around 0.16 and 0.93 rotational phase, which is very close to the rotational phase 0.04 corresponding to the maximum of the stellar longitudinal magnetic field at the height of radio emission (\S\ref{subsec:gyrosyncrotron_modulation}), and also close to the maximum of the stellar \bz. Since we found the events to be left-circularly polarized, it implies that the magneto-ionic mode of emission is ordinary (O-mode). There is a caveat here, which arises due to the transient nature of the events. Since they must involve some sudden irregular changes in the magnetosphere, the relevant magnetic field at the site of emission need not match the stable topology of the stellar magnetic field.

We now consider the two possible emission mechanisms which are plasma emission and ECME. In case of the former, emission should be at the fundamental plasma frequency since the flares have very high degree ($\approx 90\%$) of polarization. This will produce O-mode emission which is favoured by our observation (left circular polarization from the northern magnetic hemispheres). The corresponding plasma density then comes out to be $\approx 6\times 10^9\,\mathrm{cm^{-3}}$. This is quite a large value considering that the estimated plasma density inside the star's the inner magnetosphere, which is supposed to be the densest part of the stellar magnetosphere, is $\sim10^9\,\mathrm{cm^{-3}}$ \citep{leto2006,trigilio2011}. In addition, the emission site has to be sufficiently far away from the star so that the electron gyrofrequency at the emission site is less than the plasma frequency, i.e. the local magnetic field must be less than 245 G. Thus we need a temporary density enhancement at a region relatively far away from the star \citep[the maximum surface magnetic field strength is 4 kG,][]{kochukhov2014}. A possible site that meets this criterion is a region close to the current sheet at the middle magnetosphere in which continuous reconnection happens. Sudden ejection of mass from the inner magnetosphere can cause both enhanced reconnection rate and particle density. The separation between the small events that we observed might be related to the timescale associated with the mass ejection. Thus, once again, the observation of these transient events (radio flares) might be associated with the CBO scenario.
However, the nature of CBO is such that it is present at all times as multiple small-scale events \citep{owocki2020} distributed over the magnetic azimuths, so that if it is associated with radio emission, that should be observable at all rotational phases. Therefore, the radio events under consideration cannot be the result of regular CBO events, but might be associated with occasional stronger than usual CBO phenomenon (similar to the case of the giant pulse). Note that this scenario implies that the non-persistent events are indeed flares.


In case of ECME, we know that the emission direction makes a large angle with the local magnetic field direction (close to $90^\circ$). The fact that we see the pulse when the \bz~is close to being maximum, implies that the site of emission is far away from the magnetic polar regions (unlike the regular ECME that is seen from this type of stars), since otherwise the direction of emission cannot be parallel to the line of sight at that rotational phase. 
If the radio events are flares, they will require a temporary magnetic mirroring set-up in the stellar magnetosphere so as to produce the inverted particle distribution. The magnetic field required is 245 G for the fundamental and 123 G for the second harmonic emission. It is very difficult to understand how, far away from the polar region, a temporary magnetic mirroring will form, which will give rise to distinct ECME flares with a few minutes separation. From that perspective, plasma emission is favoured over ECME for being the reason behind the enhancements if they belong to the flare category.

An alternate possibility is that the enhancements, under consideration, are intermittent ECME pulses. There are two observations in support of this scenario. The first is that CU\,Vir is known to produce intermittent ECME pulses at 2.3 GHz \citep{ravi2010} that does not have any satisfactory explanation yet. Secondly, we observed a potentially transient enhancement in band 3 over the same range of rotational phases as that for the second non-persistent enhancement observed on 2019 January 7 (0.88--1.0 rotational phase, \S\ref{subsec:band3}). In such a case, the huge offset of these `intermittent pulses' from the persistent enhancements could arise due to propagation effects \citep{das2020a}, or intrinsic difference in direction of emission (e.g. if they are emitted at different harmonics of the electron gyrofrequency than that for the primary ECME pulses), or a combination of both. The reason for their intermittency could be related to the CBO events, that enable sufficient growth of a particular harmonic emission, which is otherwise, not achieved. However, we would like to emphasize that based on the current observations, it is not possible to conclude whether these enhancements are indeed flares, or they are intermittent ECME pulses.


There is one more possibility which cannot be ignored, which is the presence of a companion. In the past, the magnetic Bp star $\sigma\,\mathrm{Ori\,E}$ was reported to flare in X-ray \citep{groote2004}. However a close companion \citep[a K or an M dwarf,][]{mullan2009} was discovered by \citet{bouy2009} which cast serious doubt on the previous claim of flares from the Bp star.

We would like to mention that flares in optical bands have been claimed from A and B type stars \citep{balona2013,balona2020}. If these flares are confirmed to arise at the early-type stars, one will need to revise the understanding about production of stellar flares which might also be applicable to flaring activities in radio bands.

\section{Summary}\label{sec:summary}
In this paper, we present the first ultra-wideband (0.4--4 GHz) study conducted for the hot magnetic star CU\,Vir. The sub-GHz observations were carried out with the uGMRT and the observations over 1--4 GHz were carried out with the VLA. We detected ECME of both right and left circular polarizations below 1 GHz, down to our lowest observing frequency. Despite detecting pulses of both circular polarizations, we could not determine the magneto-ionic mode of the emission as our results suggest strong influence of the magnetospheric propagation effects on the observed properties of the pulses. Several new results have come out of this study. The new results from our observation at GHz frequencies are listed below:

\begin{enumerate}
    \item \textit{Detection of LCP pulse above 1 GHz:} For the first time, we detected LCP pulses above 1 GHz from CU\,Vir, in addition to RCP pulse. The LCP pulses are absent at S band (2--4 GHz).
    
    \item \textit{Previously unreported ECME pulse at GHz frequencies:} We discovered an additional ECME pulse, also in RCP, that has a lower cut-off frequency at 2.3 GHz and upper cut-off frequency between 5--8.4 GHz. This corresponds to the highest frequency detection of an ECME pulse from a hot magnetic star. The observation of this pulse might be related to the non-dipolar stellar magnetic field, or ECME produced in magnetic field lines anchored at different magnetic latitudes (invoked for planetary and UCD auroral radio emission).
     
    \item \textit{Upper cut-off frequency of `regular' ECME:} Using the continuous frequency coverage over 1--4 GHz, we located the upper cut-off frequency of the RCP ECME pulse (until now, thought to be the only type of pulse produced by CU\,Vir) to lie at $\approx 3$ GHz.
    
    \item \textit{Rotational modulation of the gyrosynchrotron emission:} The basal flux density of CU\,Vir show clear rotational modulation at 2.6 GHz and above. We find that the rotational phases for the maxima of the lightcurve are shifted from the rotational phases of the extrema of the stellar \bz. We attribute this to the effect of non-dipolar components on the photospheric \bz.  
\end{enumerate}

The highlight of this paper is the result of the multi-epoch observation of CU\,Vir at sub-GHz frequencies, that has never been done before. Below we list the main findings at sub-GHz frequencies:

\begin{enumerate}
     \item \textit{Dominance of LCP pulses}: CU\,Vir has been known to produce only right circularly polarized ECME, attributed to the fact that one of its magnetic poles does not produce the emission. We, on the other hand, found that pulses of both circular polarizations are produced in the stellar magnetosphere below 800 MHz, and in fact, LCP pulses are more abundant and stronger than the RCP pulses. 
    
    \item \textit{Constraint on the lower cut-off frequency of ECME:} We show that the star produces ECME down to our lowest frequency of observation, which means that the lower cut-off frequency of ECME lies below 0.4 GHz for both RCP and LCP.
    
    \item \textit{Frequency dependence of ECME pulse-profile:} The profiles of the ECME pulses, both for RCP and LCP, exhibit strong dependence on frequency. In particular, over 0.6--0.8 GHz, each RCP pulse appears to have two sub-structures, one of which vanishes with further decrease in observing frequency. This can be explained by propagation effects in a magnetosphere with strong gradients in plasma density. Alternatively, it can be a result of partial contribution from ECME produced by a different engine that follows a different ECME beaming pattern (hollow-cone beaming instead of tangential plane beaming).

    \item \textit{More than two pairs of ECME pulses per rotation cycle:} For the first time, we observed more than two pairs (one from each polarization) of ECME pulses per stellar rotation (in case of an MRP). This is significantly in contrast to the star's behavior at GHz frequencies, where we mainly observe two RCP pulses (and no LCP at all beyond $\approx 1.5\,\mathrm{GHz}$) per rotation cycle. This could be either due to propagation effects, or an indication of the presence of a different `ECME' engine. 
    
    \item \textit{Discovery of a `giant pulse'}: We report the first observation of a `giant pulse' of which origin could be linked to the centrifugal breakout of magnetically confined plasma. The peak flux density is higher by nearly an order of magnitude than that of the typical pulses seen from this star.
    
    \item \textit{Observation of sub-GHz intermittent pulse/flares:} At one of the epochs (epoch 1), we observed enhancements at 550--800 MHz that were never observed a second time during out observation campaign. Some of these events appear to exhibit quasi-periodicity which might be associated with the time-scale of underlying magnetic reconnection processes. All the confirmed events, under this category, are left circularly polarized.

    \item \textit{Potential observational signature of centrifugal breakout:} A key idea that has emerged from this work is the possibility of being able to use ECME to trace centrifugal breakout (CBO) of plasma confined in the stellar magnetosphere. Based on our observation of relatively stable profiles for circular polarization, we propose that CBO events are responsible for causing correlated variabilities in the amplitude of the ECME pulses of opposite circular polarizations. No other currently used stellar diagnostic is capable of retaining any signature of the CBO events as the latter happen at small-scales.

\end{enumerate}

Our results demonstrate the strong frequency dependence of observed ECME from MRPs, and hence the need to observe such objects at wide range of frequencies. The two-epoch observation shows that several ECME pulse-properties are time variable, yet the profiles for circular polarization remain nearly identical between the two epochs, suggesting for the first time that locally the phenomenon is highly stable, and the changes are due to external events (e.g. the CBOs). This opens a new door to characterize the emission and also to `use' them to characterize other magnetospheric phenomenon.
Complete characterization of its properties via multi-epoch, wideband radio observations of stars known to produce such emission will be essential for achieving that goal.

\section*{Acknowledgements}
We sincerely than the referee for the constructive comments that help us to improve the paper immensely. We acknowledge support of the Department of Atomic Energy, Government of India, under project no. 12-R\&D-TFR-5.02-0700. BD thanks Z. Mikul{\'a}{\v{s}}ek for providing useful information regarding the rotation period evolution of CU\,Vir. PC  acknowledges support from the Department of Science and Technology via 
SwarnaJayanti Fellowship awards (DST/SJF/PSA-01/2014-15). 
We thank the staff of the GMRT and the National Radio Astronomy Observatory (NRAO) that made our observations
possible. The GMRT is run by the National Centre for Radio Astrophysics of the Tata Institute of Fundamental Research. The National Radio Astronomy Observatory is a facility of the National Science Foundation operated under cooperative agreement by Associated Universities, Inc. This research has made use of NASA’s Astrophysics Data System.




\bibliography{das} 

\begin{thebibliography}{}
\expandafter\ifx\csname natexlab\endcsname\relax\def\natexlab#1{#1}\fi
\providecommand{\url}[1]{\href{#1}{#1}}
\providecommand{\dodoi}[1]{doi:~\href{http://doi.org/#1}{\nolinkurl{#1}}}
\providecommand{\doeprint}[1]{\href{http://ascl.net/#1}{\nolinkurl{http://ascl.net/#1}}}
\providecommand{\doarXiv}[1]{\href{https://arxiv.org/abs/#1}{\nolinkurl{https://arxiv.org/abs/#1}}}

\bibitem[{{Andre} {et~al.}(1988){Andre}, {Montmerle}, {Feigelson}, {Stine}, \&
  {Klein}}]{andre1988}
{Andre}, P., {Montmerle}, T., {Feigelson}, E.~D., {Stine}, P.~C., \& {Klein},
  K.-L. 1988, \apj, 335, 940, \dodoi{10.1086/166979}

\bibitem[{{Balona}(2013)}]{balona2013}
{Balona}, L.~A. 2013, \mnras, 431, 2240, \dodoi{10.1093/mnras/stt322}

\bibitem[{{Balona}(2020)}]{balona2020}
---. 2020, arXiv e-prints, arXiv:2008.06305.
\newblock \doarXiv{2008.06305}

\bibitem[{{Bigg}(1964)}]{bigg1964}
{Bigg}, E.~K. 1964, \nat, 203, 1008, \dodoi{10.1038/2031008a0}

\bibitem[{{Borra} \& {Landstreet}(1980)}]{borra1980}
{Borra}, E.~F., \& {Landstreet}, J.~D. 1980, \apjs, 42, 421,
  \dodoi{10.1086/190656}

\bibitem[{{Bouy} {et~al.}(2009){Bouy}, {Hu{\'e}lamo}, {Mart{\'\i}n}, {Marchis},
  {Barrado Y Navascu{\'e}s}, {Kolb}, {Marchetti}, {Petr-Gotzens}, {Sterzik},
  {Ivanov}, {K{\"o}hler}, \& {N{\"u}rnberger}}]{bouy2009}
{Bouy}, H., {Hu{\'e}lamo}, N., {Mart{\'\i}n}, E.~L., {et~al.} 2009, \aap, 493,
  931, \dodoi{10.1051/0004-6361:200810267}

\bibitem[{{Chandra} {et~al.}(2015){Chandra}, {Wade}, {Sundqvist}, {Oberoi},
  {Grunhut}, {ud-Doula}, {Petit}, {Cohen}, {Oksala}, \&
  {David-Uraz}}]{chandra2015}
{Chandra}, P., {Wade}, G.~A., {Sundqvist}, J.~O., {et~al.} 2015, \mnras, 452,
  1245, \dodoi{10.1093/mnras/stv1378}

\bibitem[{{Das} {et~al.}(2019{\natexlab{a}}){Das}, {Chandra}, {Shultz}, \&
  {Wade}}]{das2019a}
{Das}, B., {Chandra}, P., {Shultz}, M.~E., \& {Wade}, G.~A. 2019{\natexlab{a}},
  \apj, 877, 123, \dodoi{10.3847/1538-4357/ab1b12}

\bibitem[{{Das} {et~al.}(2019{\natexlab{b}}){Das}, {Chandra}, {Shultz}, \&
  {Wade}}]{das2019b}
---. 2019{\natexlab{b}}, \mnras, L133, \dodoi{10.1093/mnrasl/slz137}

\bibitem[{{Das} {et~al.}(2018){Das}, {Chandra}, \& {Wade}}]{das2018}
{Das}, B., {Chandra}, P., \& {Wade}, G.~A. 2018, \mnras, 474, L61,
  \dodoi{10.1093/mnrasl/slx193}

\bibitem[{{Das} {et~al.}(2020{\natexlab{a}}){Das}, {Chandra}, \&
  {Wade}}]{das2020b}
---. 2020{\natexlab{a}}, \mnras, 499, 702, \dodoi{10.1093/mnras/staa2499}

\bibitem[{{Das} {et~al.}(2020{\natexlab{b}}){Das}, {Kudale}, {Chandra},
  {Bhattacharya}, {Roy}, \& {Gupta}}]{das2020}
{Das}, B., {Kudale}, S., {Chandra}, P., {et~al.} 2020{\natexlab{b}}, arXiv
  e-prints, arXiv:2004.08542.
\newblock \doarXiv{2004.08542}

\bibitem[{{Das} {et~al.}(2020{\natexlab{c}}){Das}, {Mondal}, \&
  {Chandra}}]{das2020a}
{Das}, B., {Mondal}, S., \& {Chandra}, P. 2020{\natexlab{c}}, \apj, 900, 156,
  \dodoi{10.3847/1538-4357/aba8fd}

\bibitem[{{Davis} {et~al.}(2021){Davis}, {Vedantham}, {Callingham}, {Shimwell},
  {Vidotto}, {Zarka}, {Ray}, \& {Drabent}}]{davis2021}
{Davis}, I., {Vedantham}, H.~K., {Callingham}, J.~R., {et~al.} 2021, arXiv
  e-prints, arXiv:2105.01021.
\newblock \doarXiv{2105.01021}

\bibitem[{{Dulk}(1985)}]{dulk1985}
{Dulk}, G.~A. 1985, \araa, 23, 169, \dodoi{10.1146/annurev.aa.23.090185.001125}

\bibitem[{{Groote} \& {Schmitt}(2004)}]{groote2004}
{Groote}, D., \& {Schmitt}, J.~H.~M.~M. 2004, \aap, 418, 235,
  \dodoi{10.1051/0004-6361:20034300}

\bibitem[{{Hallinan} {et~al.}(2006){Hallinan}, {Antonova}, {Doyle}, {Bourke},
  {Brisken}, \& {Golden}}]{hallinan2006}
{Hallinan}, G., {Antonova}, A., {Doyle}, J.~G., {et~al.} 2006, \apj, 653, 690,
  \dodoi{10.1086/508678}

\bibitem[{{Kochukhov} {et~al.}(2014){Kochukhov}, {L{\"u}ftinger}, {Neiner},
  {Alecian}, \& {MiMeS Collaboration}}]{kochukhov2014}
{Kochukhov}, O., {L{\"u}ftinger}, T., {Neiner}, C., {Alecian}, E., \& {MiMeS
  Collaboration}. 2014, \aap, 565, A83, \dodoi{10.1051/0004-6361/201423472}

\bibitem[{{Kochukhov} {et~al.}(2017){Kochukhov}, {Silvester}, {Bailey}, {Land
  street}, \& {Wade}}]{kochukhov2017}
{Kochukhov}, O., {Silvester}, J., {Bailey}, J.~D., {Land street}, J.~D., \&
  {Wade}, G.~A. 2017, \aap, 605, A13, \dodoi{10.1051/0004-6361/201730919}

\bibitem[{{Krticka, J.} {et~al.}(2019){Krticka, J.}, {Mikul\'asek, Z.}, {Henry,
  G. W.}, {Jan\'{\i}k, J.}, {Kochukhov, O.}, {Pigulski, A.}, {Leto, P.},
  {Trigilio, C.}, {Krtickov\'a, I.}, {L\"uftinger, T.}, {Prv\'ak, M.}, \&
  {Tich\'y, A.}}]{krticka2019}
{Krticka, J.}, {Mikul\'asek, Z.}, {Henry, G. W.}, {et~al.} 2019, A\&A, 625,
  A34, \dodoi{10.1051/0004-6361/201834937}

\bibitem[{{Kuznetsov} {et~al.}(2012){Kuznetsov}, {Doyle}, {Yu}, {Hallinan},
  {Antonova}, \& {Golden}}]{kuznetsov2012}
{Kuznetsov}, A.~A., {Doyle}, J.~G., {Yu}, S., {et~al.} 2012, \apj, 746, 99,
  \dodoi{10.1088/0004-637X/746/1/99}

\bibitem[{{Ladreiter} \& {Leblanc}(1992)}]{Ladreiter1992}
{Ladreiter}, H.~P., \& {Leblanc}, Y. 1992, in Planetary Radio Emissions III,
  45--67

\bibitem[{{Lenc} {et~al.}(2018){Lenc}, {Murphy}, {Lynch}, {Kaplan}, \&
  {Zhang}}]{lenc2018}
{Lenc}, E., {Murphy}, T., {Lynch}, C.~R., {Kaplan}, D.~L., \& {Zhang}, S.~N.
  2018, \mnras, 478, 2835, \dodoi{10.1093/mnras/sty1304}

\bibitem[{{Leto} {et~al.}(2016){Leto}, {Trigilio}, {Buemi}, {Umana},
  {Ingallinera}, \& {Cerrigone}}]{leto2016}
{Leto}, P., {Trigilio}, C., {Buemi}, C.~S., {et~al.} 2016, \mnras, 459, 1159,
  \dodoi{10.1093/mnras/stw639}

\bibitem[{{Leto} {et~al.}(2006){Leto}, {Trigilio}, {Buemi}, {Umana}, \&
  {Leone}}]{leto2006}
{Leto}, P., {Trigilio}, C., {Buemi}, C.~S., {Umana}, G., \& {Leone}, F. 2006,
  \aap, 458, 831, \dodoi{10.1051/0004-6361:20054511}

\bibitem[{{Leto} {et~al.}(2017){Leto}, {Trigilio}, {Oskinova}, {Ignace},
  {Buemi}, {Umana}, {Ingallinera}, {Todt}, \& {Leone}}]{leto2017}
{Leto}, P., {Trigilio}, C., {Oskinova}, L., {et~al.} 2017, \mnras, 467, 2820,
  \dodoi{10.1093/mnras/stx267}

\bibitem[{{Leto} {et~al.}(2019){Leto}, {Trigilio}, {Oskinova}, {Ignace},
  {Buemi}, {Umana}, {Cavallaro}, {Ingallinera}, {Bufano}, {Phillips},
  {Agliozzo}, {Cerrigone}, {Todt}, {Riggi}, \& {Leone}}]{leto2019}
{Leto}, P., {Trigilio}, C., {Oskinova}, L.~M., {et~al.} 2019, \mnras, 482, L4,
  \dodoi{10.1093/mnrasl/sly179}

\bibitem[{{Leto} {et~al.}(2020{\natexlab{a}}){Leto}, {Trigilio}, {Leone},
  {Pillitteri}, {Buemi}, {Fossati}, {Cavallaro}, {Oskinova}, {Ignace},
  {Krti{\v{c}}ka}, {Umana}, {Catanzaro}, {Ingallinera}, {Bufano}, {Agliozzo},
  {Phillips}, {Cerrigone}, {Riggi}, {Loru}, {Munari}, {Gangi}, {Giarrusso}, \&
  {Robrade}}]{leto2020}
{Leto}, P., {Trigilio}, C., {Leone}, F., {et~al.} 2020{\natexlab{a}}, \mnras,
  493, 4657, \dodoi{10.1093/mnras/staa587}

\bibitem[{{Leto} {et~al.}(2020{\natexlab{b}}){Leto}, {Trigilio}, {Buemi},
  {Leone}, {Pillitteri}, {Fossati}, {Cavallaro}, {Oskinova}, {Ignace},
  {Krti{\v{c}}ka}, {Umana}, {Catanzaro}, {Ingallinera}, {Bufano}, {Riggi},
  {Cerrigone}, {Loru}, {Schillir{\'o}}, {Agliozzo}, {Phillips}, {Giarrusso}, \&
  {Robrade}}]{leto2020b}
{Leto}, P., {Trigilio}, C., {Buemi}, C.~S., {et~al.} 2020{\natexlab{b}},
  \mnras, 499, L72, \dodoi{10.1093/mnrasl/slaa157}

\bibitem[{{Lo} {et~al.}(2012){Lo}, {Bray}, {Hobbs}, {Murphy}, {Gaensler},
  {Melrose}, {Ravi}, {Manchester}, \& {Keith}}]{lo2012}
{Lo}, K.~K., {Bray}, J.~D., {Hobbs}, G., {et~al.} 2012, \mnras, 421, 3316,
  \dodoi{10.1111/j.1365-2966.2012.20555.x}

\bibitem[{{McMullin} {et~al.}(2007){McMullin}, {Waters}, {Schiebel}, {Young},
  \& {Golap}}]{mcmullin2007}
{McMullin}, J.~P., {Waters}, B., {Schiebel}, D., {Young}, W., \& {Golap}, K.
  2007, in Astronomical Society of the Pacific Conference Series, Vol. 376,
  Astronomical Data Analysis Software and Systems XVI, ed. R.~A. {Shaw},
  F.~{Hill}, \& D.~J. {Bell}, 127

\bibitem[{{Melrose} \& {Dulk}(1982)}]{melrose1982}
{Melrose}, D.~B., \& {Dulk}, G.~A. 1982, \apj, 259, 844, \dodoi{10.1086/160219}

\bibitem[{{Mikul{\'a}{\v{s}}ek} {et~al.}(2019){Mikul{\'a}{\v{s}}ek},
  {Krti{\v{c}}ka}, {Henry}, {Jan{\'\i}k}, \& {Pigulski}}]{mikulasek2019}
{Mikul{\'a}{\v{s}}ek}, Z., {Krti{\v{c}}ka}, J., {Henry}, G.~W., {Jan{\'\i}k},
  J., \& {Pigulski}, A. 2019, in Astronomical Society of the Pacific Conference
  Series, Vol. 518, Physics of Magnetic Stars, ed. D.~O. {Kudryavtsev}, I.~I.
  {Romanyuk}, \& I.~A. {Yakunin}, 125

\bibitem[{{Mikul{\'a}{\v{s}}ek} {et~al.}(2011){Mikul{\'a}{\v{s}}ek},
  {Krti{\v{c}}ka}, {Henry}, {Jan{\'\i}k}, {Zverko},
  {{\v{Z}}i{\v{z}}{\v{n}}ovsk{\'y}}, {Zejda}, {Li{\v{s}}ka},
  {Zv{\v{e}}{\v{r}}ina}, {Kudrjavtsev}, {Romanyuk}, {Sokolov}, {L{\"u}ftinger},
  {Trigilio}, {Neiner}, \& {de Villiers}}]{mikulasek2011}
{Mikul{\'a}{\v{s}}ek}, Z., {Krti{\v{c}}ka}, J., {Henry}, G.~W., {et~al.} 2011,
  \aap, 534, L5, \dodoi{10.1051/0004-6361/201117784}

\bibitem[{{Mohan} {et~al.}(2019){Mohan}, {McCauley}, {Oberoi}, \&
  {Mastrano}}]{mohan2019}
{Mohan}, A., {McCauley}, P.~I., {Oberoi}, D., \& {Mastrano}, A. 2019, \apj,
  883, 45, \dodoi{10.3847/1538-4357/ab3a94}

\bibitem[{{Mullan}(2009)}]{mullan2009}
{Mullan}, D.~J. 2009, \apj, 702, 759, \dodoi{10.1088/0004-637X/702/1/759}

\bibitem[{{Mutel} {et~al.}(2008){Mutel}, {Christopher}, \&
  {Pickett}}]{mutel2008}
{Mutel}, R.~L., {Christopher}, I.~W., \& {Pickett}, J.~S. 2008, \grl, 35,
  L07104, \dodoi{10.1029/2008GL033377}

\bibitem[{{Owocki} {et~al.}(2020){Owocki}, {Shultz}, {ud-Doula}, {Sundqvist},
  {Townsend}, \& {Cranmer}}]{owocki2020}
{Owocki}, S.~P., {Shultz}, M.~E., {ud-Doula}, A., {et~al.} 2020, \mnras, 499,
  5366, \dodoi{10.1093/mnras/staa2325}

\bibitem[{{Petit} {et~al.}(2013){Petit}, {Owocki}, {Wade}, {Cohen},
  {Sundqvist}, {Gagn{\'e}}, {Ma{\'\i}z Apell{\'a}niz}, {Oksala}, {Bohlender},
  {Rivinius}, {Henrichs}, {Alecian}, {Townsend}, {ud-Doula}, \& {MiMeS
  Collaboration}}]{petit2013}
{Petit}, V., {Owocki}, S.~P., {Wade}, G.~A., {et~al.} 2013, \mnras, 429, 398,
  \dodoi{10.1093/mnras/sts344}

\bibitem[{{Pillitteri} {et~al.}(2018){Pillitteri}, {Fossati}, {Castro
  Rodriguez}, {Oskinova}, \& {Wolk}}]{pillitteri2018}
{Pillitteri}, I., {Fossati}, L., {Castro Rodriguez}, N., {Oskinova}, L., \&
  {Wolk}, S.~J. 2018, \aap, 610, L3, \dodoi{10.1051/0004-6361/201732078}

\bibitem[{{Pineda} {et~al.}(2017){Pineda}, {Hallinan}, \& {Kao}}]{pineda2017}
{Pineda}, J.~S., {Hallinan}, G., \& {Kao}, M.~M. 2017, \apj, 846, 75,
  \dodoi{10.3847/1538-4357/aa8596}

\bibitem[{{Ravi} {et~al.}(2010){Ravi}, {Hobbs}, {Wickramasinghe}, {Champion},
  {Keith}, {Manchester}, {Norris}, {Bray}, {Ferrario}, \& {Melrose}}]{ravi2010}
{Ravi}, V., {Hobbs}, G., {Wickramasinghe}, D., {et~al.} 2010, \mnras, 408, L99,
  \dodoi{10.1111/j.1745-3933.2010.00939.x}

\bibitem[{{Robrade} {et~al.}(2018){Robrade}, {Oskinova}, {Schmitt}, {Leto}, \&
  {Trigilio}}]{robrade2018}
{Robrade}, J., {Oskinova}, L.~M., {Schmitt}, J.~H.~M.~M., {Leto}, P., \&
  {Trigilio}, C. 2018, \aap, 619, A33, \dodoi{10.1051/0004-6361/201833492}

\bibitem[{{Sharma} \& {Vlahos}(1984)}]{sharma1984}
{Sharma}, R.~R., \& {Vlahos}, L. 1984, \apj, 280, 405, \dodoi{10.1086/162006}

\bibitem[{{Shultz} {et~al.}(2020){Shultz}, {Owocki}, {Rivinius}, {Wade},
  {Neiner}, {Alecian}, {Kochukhov}, {Bohlender}, {ud-Doula}, {Landstreet},
  {Sikora}, {David-Uraz}, {Petit}, {Cerraho{\u{g}}lu}, {Fine}, {Henson}, {MiMeS
  Collaboration}, \& {BinaMIcS Collaboration}}]{shultz2020}
{Shultz}, M.~E., {Owocki}, S., {Rivinius}, T., {et~al.} 2020, \mnras, 499,
  5379, \dodoi{10.1093/mnras/staa3102}

\bibitem[{{Stevens} \& {George}(2010)}]{stevens2010}
{Stevens}, I.~R., \& {George}, S.~J. 2010, in Astronomical Society of the
  Pacific Conference Series, Vol. 422, High Energy Phenomena in Massive Stars,
  ed. J.~{Mart{\'\i}}, P.~L. {Luque-Escamilla}, \& J.~A. {Combi}, 135

\bibitem[{{Townsend} \& {Owocki}(2005)}]{townsend2005}
{Townsend}, R.~H.~D., \& {Owocki}, S.~P. 2005, \mnras, 357, 251,
  \dodoi{10.1111/j.1365-2966.2005.08642.x}

\bibitem[{{Townsend} {et~al.}(2013){Townsend}, {Rivinius}, {Rowe}, {Moffat},
  {Matthews}, {Bohlender}, {Neiner}, {Telting}, {Guenther}, {Kallinger},
  {Kuschnig}, {Rucinski}, {Sasselov}, \& {Weiss}}]{townsend2013}
{Townsend}, R.~H.~D., {Rivinius}, T., {Rowe}, J.~F., {et~al.} 2013, \apj, 769,
  33, \dodoi{10.1088/0004-637X/769/1/33}

\bibitem[{{Treumann}(2006)}]{treumann2006}
{Treumann}, R.~A. 2006, \aapr, 13, 229, \dodoi{10.1007/s00159-006-0001-y}

\bibitem[{{Trigilio} {et~al.}(2000){Trigilio}, {Leto}, {Leone}, {Umana}, \&
  {Buemi}}]{trigilio2000}
{Trigilio}, C., {Leto}, P., {Leone}, F., {Umana}, G., \& {Buemi}, C. 2000,
  \aap, 362, 281.
\newblock \doarXiv{astro-ph/0007097}

\bibitem[{{Trigilio} {et~al.}(2008){Trigilio}, {Leto}, {Umana}, {Buemi}, \&
  {Leone}}]{trigilio2008}
{Trigilio}, C., {Leto}, P., {Umana}, G., {Buemi}, C.~S., \& {Leone}, F. 2008,
  \mnras, 384, 1437, \dodoi{10.1111/j.1365-2966.2007.12749.x}

\bibitem[{{Trigilio} {et~al.}(2011){Trigilio}, {Leto}, {Umana}, {Buemi}, \&
  {Leone}}]{trigilio2011}
---. 2011, \apjl, 739, L10, \dodoi{10.1088/2041-8205/739/1/L10}

\bibitem[{{Trigilio} {et~al.}(2004){Trigilio}, {Leto}, {Umana}, {Leone}, \&
  {Buemi}}]{trigilio2004}
{Trigilio}, C., {Leto}, P., {Umana}, G., {Leone}, F., \& {Buemi}, C.~S. 2004,
  \aap, 418, 593, \dodoi{10.1051/0004-6361:20040060}

\bibitem[{{Zarka}(2004)}]{zarka2004}
{Zarka}, P. 2004, Advances in Space Research, 33, 2045,
  \dodoi{10.1016/j.asr.2003.07.055}

\end{thebibliography}

\appendix

\section{Test source lightcurves}\label{sec:test_source_check}
\begin{figure}
    \centering
    \includegraphics[width=0.45\textwidth]{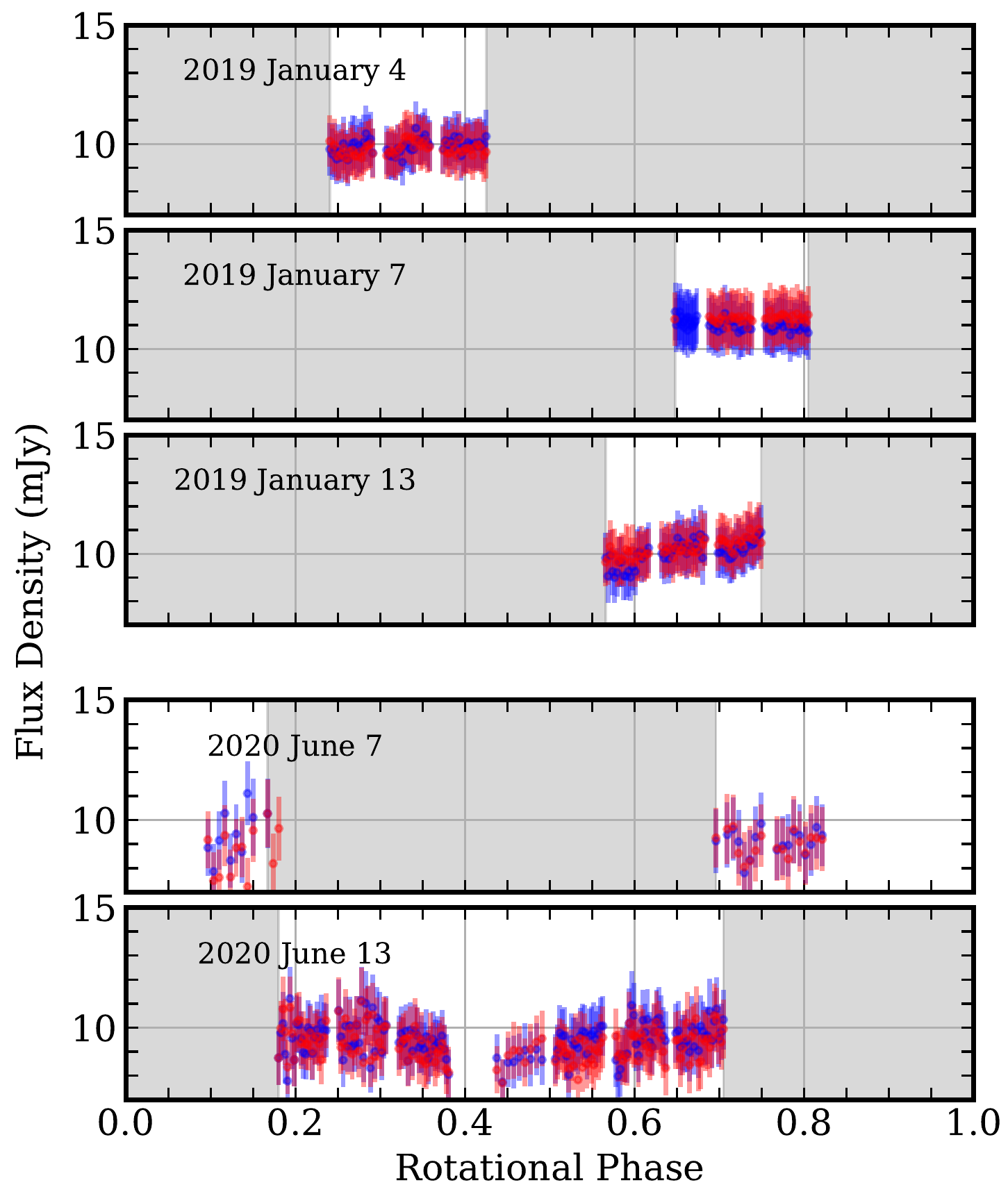}
    \includegraphics[width=0.45\textwidth]{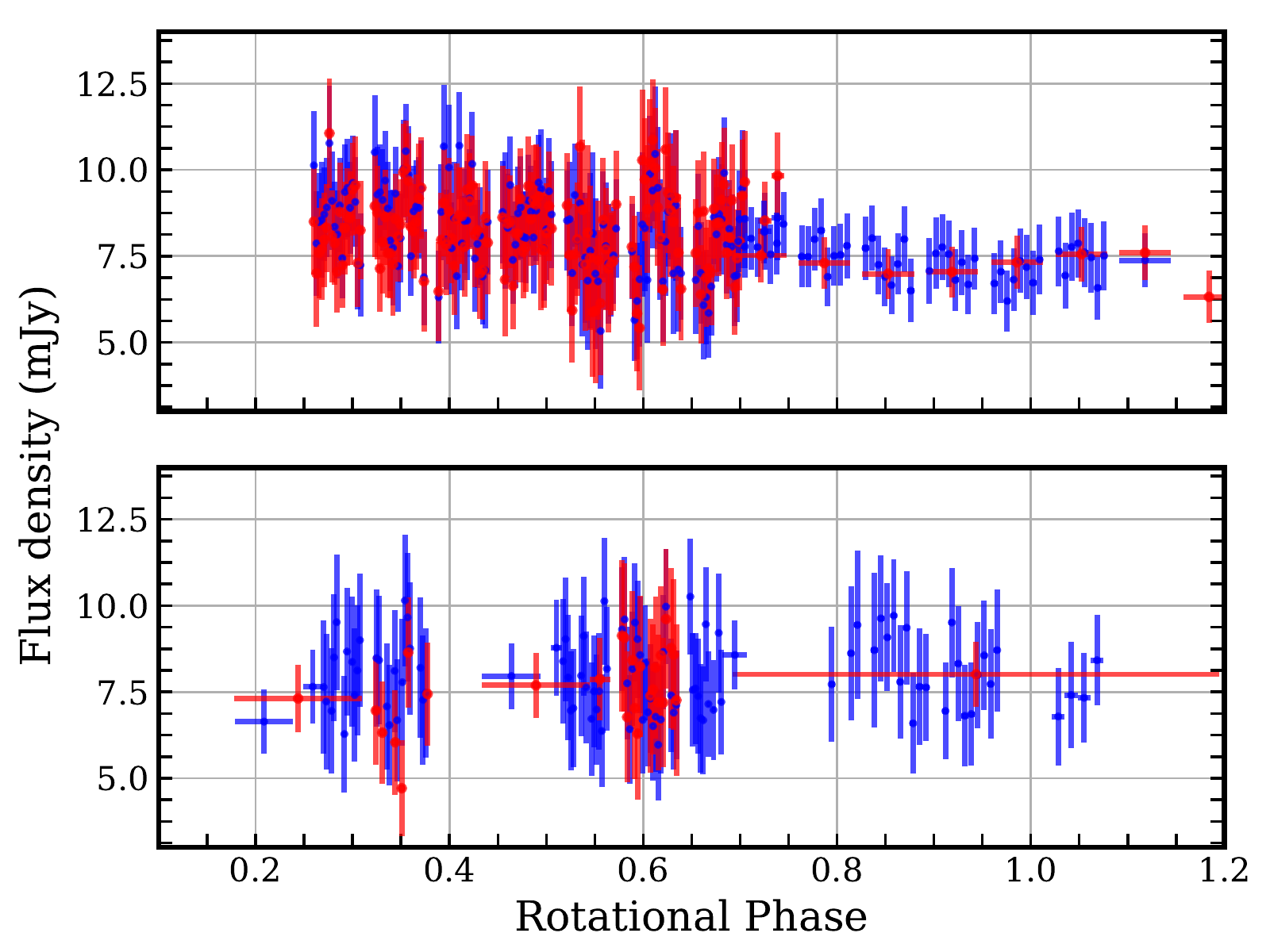}
    \caption{The lightcurves of a test source (J2000 coordinates are RA: $14^\mathrm{h}11^\mathrm{m}13^\mathrm{s}.8619$, Dec: $+02^\mathrm{d}31^{'}35^{''}.92$) \textit{Left:} in band 4 of the uGMRT at the rotational phases where persistent enhancements from CU\,Vir was observed; \textit{Right:} in band 3 of the uGMRT, top and bottom panels correspond to epoch 1 and 2 respectively (for the definition of epoch 1 and 2, see Table \ref{tab:obs}).}
    \label{fig:test_source_lightcurve_band4_band3}
\end{figure}

In order to check if the observed variation in flux density of the target is real, we examined the sub-GHz lightcurves of a test source (J2000 coordinates are RA: $14^\mathrm{h}11^\mathrm{m}13^\mathrm{s}.8619$, Dec: $+02^\mathrm{d}31^{'}35^{''}.92$, $\approx 17'$ away from CU\,Vir), at the rotational phases where we observed significant variation in the flux density of CU\,Vir. For the non-persistent enhancements observed in band 4, we have already shown the lightcurves of the test source in the main body itself (Figure \ref{fig:flare_lightcurve_6_7jan2019}). In the left panel of Figure \ref{fig:test_source_lightcurve_band4_band3}, we show the lightcurves corresponding to the persistent enhancements observed in band 4. Similarly, in the right panel, we show the same for band 3. At both epochs, the lightcurves for the test source do not show any systematic variation at either band. Also the LCP and RCP flux densities are similar. This implies that any systematic variation in flux density observed for the target (CU\,Vir) is intrinsic to the star.

\section{Spectra of the non-persistent features observed in band 4 (550--800 MHz) of the uGMRT}\label{sec:flare_spectra}

\begin{figure*}
    \centering
     \includegraphics[width=0.3\textwidth]{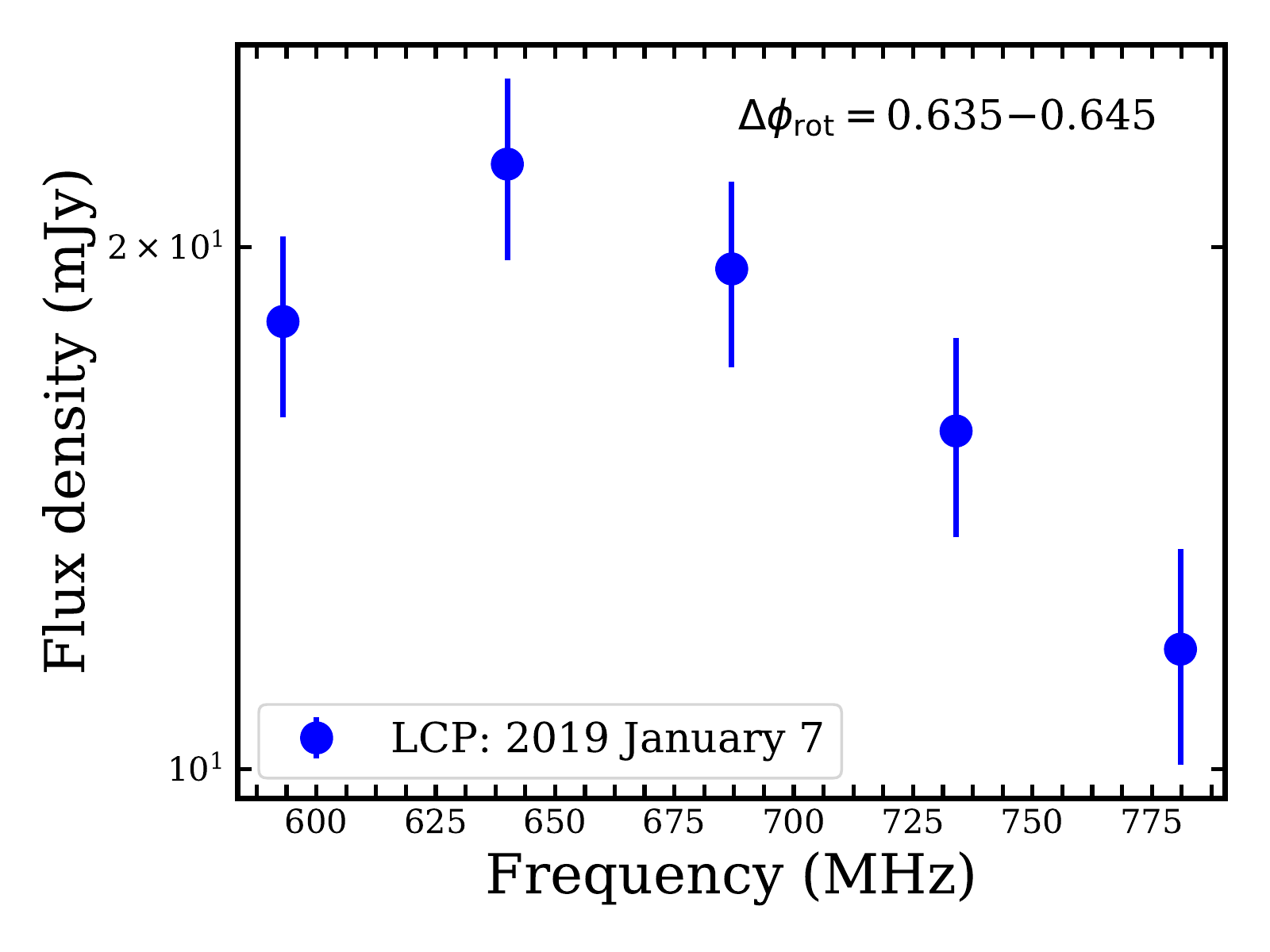}
      \includegraphics[width=0.3\textwidth]{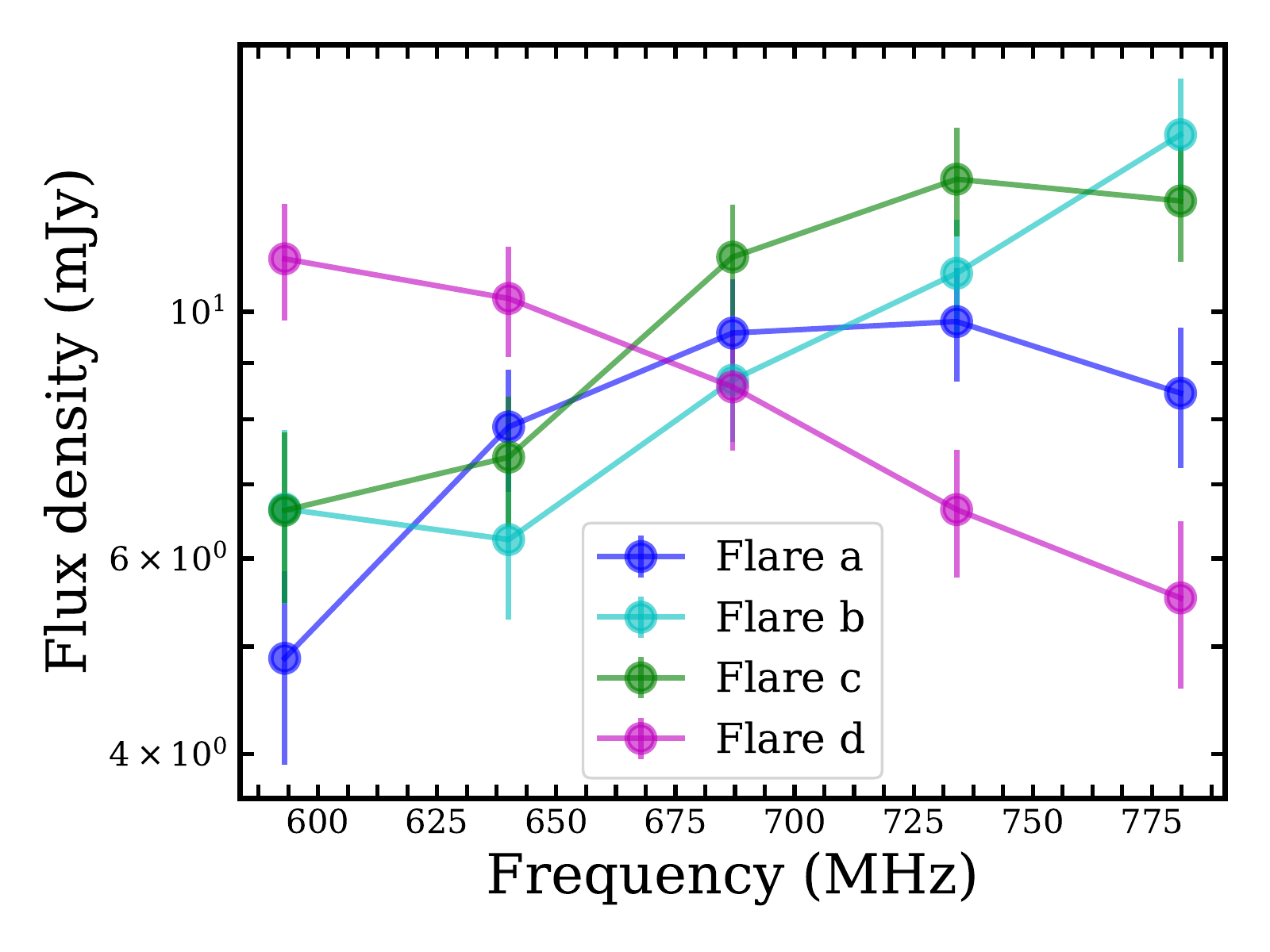}
     \includegraphics[width=0.3\textwidth]{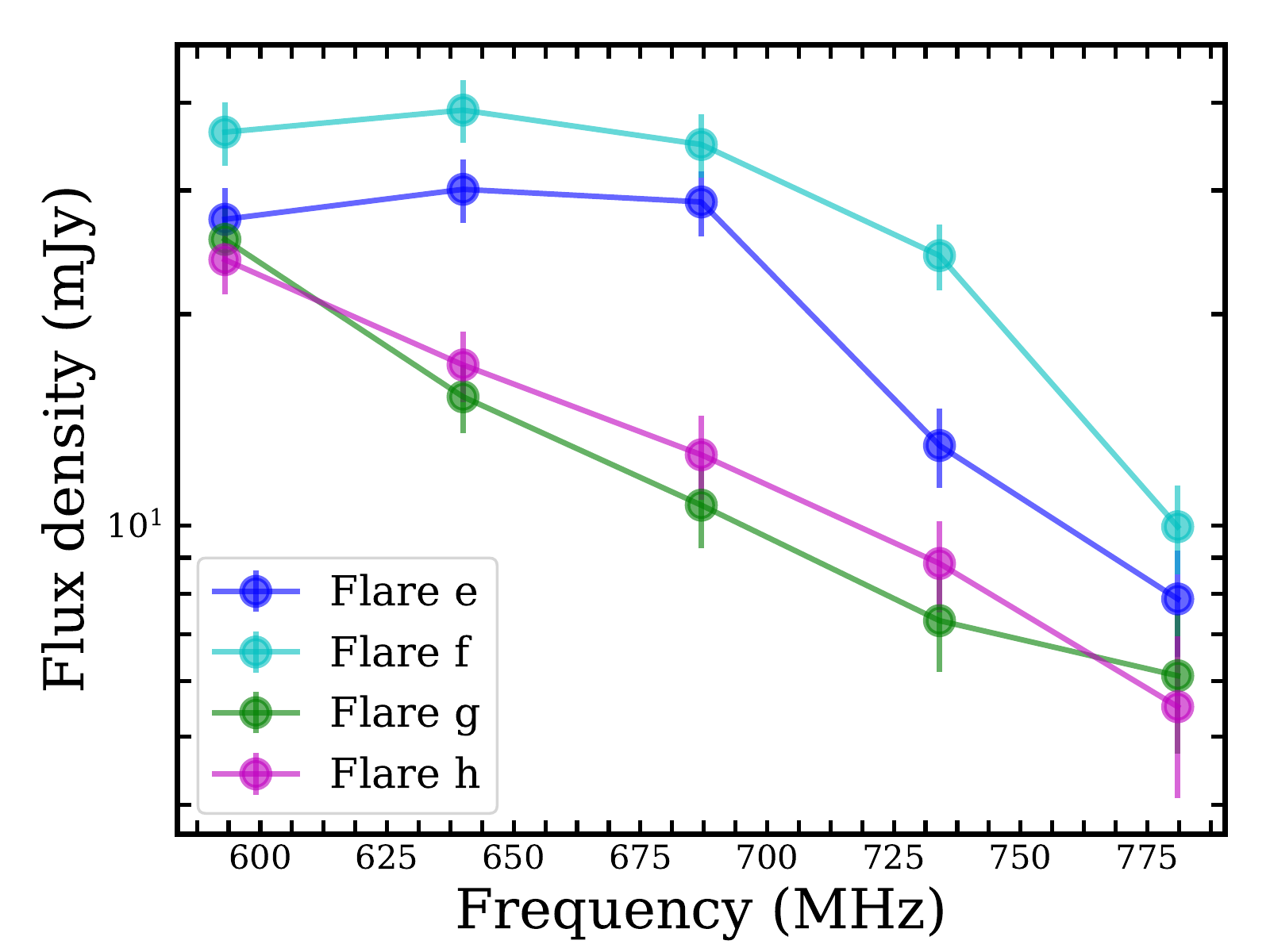}
    \caption{The peak flux density spectra for the non-persistent enhancements observed from CU\,Vir at 570--804 MHz. Al of them are left circular polarized and hence only the LCP flux densities are plotted here. \textit{Left:} The enhancement observed on 2019 January 7 between 0.635--0.645 cycle (see top panels of Figure \ref{fig:band4_7jan13jan_7jun13jun}). The averaging time is 1 minute. \textit{Middle:} The set of enhancements observed between 0.88--1.00 rotation cycle on 2019 January 7. The different labels correspond to different rotational phases of observation: 0.88--0.905 (`a'), 0.905--0.921 (`b'), 0.921--0.94 (`c') and 0.95--0.99 (`d') rotation cycles (see left of Figure \ref{fig:flare_lightcurve_6_7jan2019}). The averaging time for each point is 2 minutes. \textit{Right:} The set of enhancements observed on 2019 January 6 over the rotational phase range 0.14--0.18 cycles. Here also, the labels correspond to the rotational phases of observation of a particular feature: 0.15--0.16 (`e'), 0.16--0.165 (`f'), 0.165--0.17 (`g') and 0.17--0.175 (`h') rotation cycles (see right of Figure \ref{fig:flare_lightcurve_6_7jan2019}). The averaging time for each point is 2 minutes.
}
    \label{fig:flare_spectra}
\end{figure*}

The spectra corresponding to the maximum flux density of the non-persistent features observed in band 4 of the uGMRT (570--804 MHz) are shown in Figure \ref{fig:flare_spectra}. Note that all these enhancements were observed only in LCP, and hence the plotted flux densities are for LCP only. For the feature observed on 2019 January 7 between 0.635--0.645 rotation cycles (left of Figure \ref{fig:flare_spectra}), we find only a solitary enhancement. However, this was observed on top of a persistent LCP pulse and hence it is not entirely clear if it has associated components (like the ones observed for the other non-persistent features) or not. The observed spectrum for the peak flux
density shows a turn over at 640 MHz, and after that the flux density falls with a spectral index of $-3.9\pm1.5$. 

For the feature observed over 0.88--1.00 rotation cycles (middle panel of Figure \ref{fig:flare_spectra}), we find that the three enhancements over 0.88--0.94 rotation cycle (`a', `b' and `c' in Figure \ref{fig:flare_spectra}, also see the left plot in Figure \ref{fig:flare_lightcurve_6_7jan2019}) show similar spectra, whereas the spectrum for the flare over 0.95--0.99 rotation cycle (`d' in Figure \ref{fig:flare_spectra}) is strikingly different (opposite sign of spectral index). Thus, the features `a' to `c' and feature `d' are most probably have different emission sites but happen to lie adjacent to each other in time. As we could not cover the rotational phase range connecting the events `c' and `d', there is also a possibility of evolution of the spectrum (for the same enhancement) with time. However we do not see any sign of spectral evolution between the events `a' and `c' which makes this possibility unlikely.

The non-persistent enhancement observed between 0.14--0.18 rotation cycle on 2019 January 6 also consists of multiple bursts of smaller duration (see the right plot in Figure \ref{fig:flare_lightcurve_6_7jan2019}, \S\ref{subsubsec:band4_non_persistent}). There are four clear components to this enhancement over the following rotational phase range: 0.15--0.16, 0.160--0.165, 0.165--0.170 and 0.170--0.175 (labelled as `e', `f', `g' and `h' respectively in Figure \ref{fig:flare_lightcurve_6_7jan2019}). The spectra for these four sub-structures (corresponding to the peak flux densities) are shown in the right of Figure \ref{fig:flare_spectra}. They all exhibit very steep spectra (negative spectral indices) beyond 687 MHz, indicating that the upper cut-off frequency is close to the high-frequency end of the observing band. The spectra for the events `e' and `f' are similar, characterized by a flat part between 593--687 MHz followed by a steep decline in peak flux density above 687 MHz. In case of the events over 0.165--0.170 (`g' in Figure \ref{fig:flare_spectra}) and 0.170--0.175 (`h' in Figure \ref{fig:flare_spectra}) rotation cycles, the spectra exhibit negative spectral indices throughout the observing band. This could either imply that the events observed in the two rotational phase ranges have different emission sites, or, that the turn-over frequency of the spectrum got shifted to lower frequencies due to change in physical conditions at the emission site.

\newpage

\begin{deluxetable}{ccccccc}
\tablecaption{The flux density measurements at uGMRT band 3 (300--500 MHz). HJD stands for Heliocentric Julian Day, HJD1 and HJD2 are respectively the beginning and the end of the timerange corresponding to the measurements. Error implies uncertainty in the flux density and it includes the fitting error, image rms and the uncertainty involved in absolute flux density calibration. Pol stands for polarization. $\nu_0$ corresponds to the central frequency of a band with a bandwidth of $\Delta\nu$.}
\label{tab:band3_data}
\tablehead{
\hline
HJD1 & HJD2 & Flux density & Error & Pol & $\nu_0$ & $\Delta\nu$\\
& & (mJy) & (mJy) & & (MHz) & (MHz)
}
\startdata
\hline
2458494.417703 &	2458494.419092	& 4.99	& 1.42	& RCP	& 397 &	128\\
2458494.427426 &	2458494.428815	& 3.95	& 1.30	& RCP	& 397 &	128\\
2458494.428815 & 	2458494.430204	& 3.23	& 0.96	& RCP	& 397 &	128\\
2458494.451827 & 	2458494.453216	& 4.01	& 1.15	& RCP	& 397 &	128\\
2458494.453216 & 	2458494.454605	& 4.04	& 1.16	& RCP	& 397 &	128\\
\hline
\enddata
\tablecomments{Table \ref{tab:band3_data} is published in its entirety in the machine-readable format.
      A portion is shown here for guidance regarding its form and content.}
\end{deluxetable}

\begin{deluxetable}{ccccccc}
\tablecaption{The flux density measurements at uGMRT band 4 (550--900 MHz). HJD stands for Heliocentric Julian Day, HJD1 and HJD2 are respectively the beginning and the end of the timerange corresponding to the measurements. Error implies uncertainty in the flux density and it includes the fitting error, image rms and the uncertainty involved in absolute flux density calibration. Pol stands for polarization. $\nu_0$ corresponds to the central frequency of a band with a bandwidth of $\Delta\nu$.}
\label{tab:band4_data}
\tablehead{
\hline
HJD1 & HJD2 & Flux density & Error & Pol & $\nu_0$ & $\Delta\nu$\\
& & (mJy) & (mJy) & & (MHz) & (MHz)
}
\startdata
\hline
2458487.569470 &	2458487.597250 & 	2.13 & 	0.24 & 	RCP &	687 &	234\\
2458487.604340 & 	2458487.632120 &	1.75 & 	0.20 &	RCP &	687 &	234\\
2458487.638540 & 	2458487.638890 & 	5.05 & 	0.68 &	RCP &	687 &	234\\
2458487.638890 & 	2458487.639230 & 	4.58 & 	0.64 &	RCP	&   687 &	234\\
2458487.639230 & 	2458487.639580 &	4.85 &	0.64 &	RCP	&   687 &	234\\
\hline
\enddata
\tablecomments{Table \ref{tab:band4_data} is published in its entirety in the machine-readable format.
      A portion is shown here for guidance regarding its form and content.}
\end{deluxetable}

\begin{deluxetable}{ccccccc}
\tablecaption{The flux density measurements at VLA L band (1000--2000 MHz). HJD stands for Heliocentric Julian Day, HJD1 and HJD2 are respectively the beginning and the end of the timerange corresponding to the measurements. Error implies uncertainty in the flux density and it includes the fitting error, image rms and the uncertainty involved in absolute flux density calibration. Pol stands for polarization. $\nu_0$ corresponds to the central frequency of a band with a bandwidth of $\Delta\nu$.}
\label{tab:Lband_data}
\tablehead{
\hline
HJD1 & HJD2 & Flux density & Error & Pol & $\nu_0$ & $\Delta\nu$\\
& & (mJy) & (mJy) & & (MHz) & (MHz)
}
\startdata
\hline
2458647.503081&	2458647.504470&	9.89 &	1.07&	RCP&	1500&	1000\\
2458647.504470&	2458647.505859&	14.50&	1.51&	RCP&	1500&	1000\\
2458647.505859&	2458647.507247&	17.12&	1.78&	RCP&	1500&	1000\\
2458647.507247&	2458647.508636&	19.49&	1.98&	RCP&	1500&	1000\\
2458647.508636&	2458647.510025&	18.50&	1.90&	RCP&	1500&	1000\\
\hline
\enddata
\tablecomments{Table \ref{tab:Lband_data} is published in its entirety in the machine-readable format.
      A portion is shown here for guidance regarding its form and content.}
\end{deluxetable}

\begin{deluxetable}{ccccccc}
\tablecaption{The flux density measurements at VLA S band (2000--4000 MHz). HJD stands for Heliocentric Julian Day, HJD1 and HJD2 are respectively the beginning and the end of the timerange corresponding to the measurements. Error implies uncertainty in the flux density and it includes the fitting error, image rms and the uncertainty involved in absolute flux density calibration. Pol stands for polarization. $\nu_0$ corresponds to the central frequency of a band with a bandwidth of $\Delta\nu$.}
\label{tab:Sband_data}
\tablehead{
\hline
HJD1 & HJD2 & Flux density & Error & Pol & $\nu_0$ & $\Delta\nu$\\
& & (mJy) & (mJy) & & (MHz) & (MHz)
}
\startdata
\hline
2458647.505410&	2458647.515820&	7.13 &	0.81&	RCP&	2051 &	128\\
2458647.515820&	2458647.526240&	17.20&	1.75&	RCP&    2051 &	128\\
2458647.526240&	2458647.536650&	16.98& 	1.76&	RCP&	2051 &	128\\
2458647.536650&	2458647.547070&	5.21 &	0.57&	RCP&	2051 &	128\\
2458647.547070&	2458647.557490&	4.10 &	0.47&	RCP&	2051 &	128\\
\hline
\enddata
\tablecomments{Table \ref{tab:Sband_data} is published in its entirety in the machine-readable format.
      A portion is shown here for guidance regarding its form and content.}
\end{deluxetable}





\end{document}